\documentclass[a4paper,11pt]{article}
\pdfoutput=1
\usepackage{jheppub}
\usepackage{multicol,bbm,bigcenter}
\usepackage{citesort}
\usepackage{enumerate}
\newcommand\fverb{\setbox\fverbbox=\hbox\bgroup\verb}
\newcommand\fverbdo{\egroup\medskip\noindent%
\fbox{\unhbox\fverbbox}\ }
\newcommand\fverbit{\egroup\item[\fbox{\unhbox\fverbbox}]}

\newcommand{\lsim}{\raisebox{-0.13cm}{~\shortstack{$<$ \\[-0.07cm] $\sim$}}~} 
\newcommand{\gsim}{\raisebox{-0.13cm}{~\shortstack{$>$ \\[-0.07cm] $\sim$}}~} 
\newcommand{\beq}{\begin{eqnarray}} 
\newcommand{\eeq}{\end{eqnarray}} 
\newcommand{\tb}{\tan \beta}
\newcommand{\ra}{\to}
 
\pdfoutput=1

\preprint{\parbox[t]{0.5\textwidth}{\raggedleft
    KCL-PH-TH-2015-35, LPT-15-90}}

\title{Prospects for Higgs physics at energies up to 100 TeV}

\author[a]{Julien Baglio,}
\author[b]{Abdelhak Djouadi}
\author[c]{and J\'er\'emie Quevillon}

\affiliation[a]{Institut f\"{u}r Theoretische Physik,
Eberhard Karls Universit\"{a}t T\"{u}bingen, Auf der Morgenstelle 14,
D--72076 T\"{u}bingen, Germany}
\affiliation[b]{Laboratoire de Physique Th\'eorique, Universit\'e
  Paris--Sud and CNRS, F--91405 Orsay, France}
\affiliation[c]{Theoretical Particle Physics \& Cosmology, Department
  of Physics, King's College London, \\ London, WC2R 2LS, United Kingdom}

\emailAdd{julien.baglio@uni-tuebingen.de} 
\emailAdd{Abdelhak.Djouadi@th.u-psud.fr}
\emailAdd{jeremie.quevillon@kcl.ac.uk}

\abstract{We summarise the prospects for Higgs boson physics at future
  proton--proton colliders with centre of mass (c.m.) energies up to
  100 TeV. We first provide the production cross sections for the Higgs
  boson of the Standard Model from 13 TeV to 100 TeV, in the main
  production mechanisms and  in subleading but important ones such as
  double Higgs production, triple production and associated production
  with two gauge bosons or with a single top quark. We then discuss
  the production of Higgs particles in beyond the Standard Model
  scenarios, starting with the one in the continuum of a pair of
  scalar, fermionic and vector dark matter particles in Higgs--portal
  models in various channels with virtual Higgs exchange. The cross
  sections for the production of the heavier CP--even and CP--odd
  neutral Higgs states and the charged Higgs states in two--Higgs
  doublet models, with a specific study of the case of the Minimal
  Supersymmetric Standard Model, are then given. The sensitivity of a
  100 TeV proton machine to probe the new Higgs states is discussed
  and compared to that of the LHC with a c.m. energy of 14 TeV and at
  high luminosity.}


\begin{document}
\maketitle
\flushbottom 

\renewcommand{\thefootnote}{\arabic{footnote}}
\setcounter{footnote}{0}
\setcounter{page}{2}
\section{Introduction}  

It has been expected for a long time that the probing of the
electroweak symmetry breaking (EWSB) mechanism would be at least a
two--step process. The first step was the search and the (non)
observation of a Higgs--like particle that would confirm (refute) the 
hypothesis of the Standard Model (SM) --- that the electroweak
symmetry is spontaneously broken by a scalar field that develops a
non-zero vacuum expectation value --- and many of its new physics
extensions~\cite{Higgs:1964ia,Higgs:1964pj,Englert:1964et,Guralnik:1964eu};
for a review see Ref.~\cite{Djouadi:2005gi}. This test has been passed
successfully with the historical discovery by the ATLAS and CMS
collaborations~\cite{Aad:2012tfa,Chatrchyan:2012ufa} of a Higgs boson
with a mass of 125 GeV at the CERN LHC. A second step, that is as
important as the first one, is to probe in all its facets the EWSB
mechanism and to assess whether the Higgs sector of the theory is
SM--like or involves new degrees of freedom. Compared to what has been
achieved so far, this latter step needs first a much more precise
determination of the basic properties of the observed Higgs particle,
and second a more complete probe of the TeV scale in order to
directly discover or definitely exclude the light new degrees of
freedom that are expected to appear in almost all extensions of the
SM. This is particularly true in models that allow for a natural Higgs
boson with a mass that is protected against very high scales and,
hence, solve the so-called hierarchy problem.

All the measurements in the Higgs sector performed so
far~\cite{HiggsCombo} need to be significantly improved in order to
better constrain the SM and to probe smaller effects of new
physics. In particular, a more precise measurement of the Higgs
couplings to the fermions and gauge bosons should be performed in the
already probed channels. One thus needs to experimentally determine
more precisely the cross sections in the dominant Higgs production
modes~\cite{Djouadi:2005gi,Dittmaier:2011ti} such as the loop induced
gluon fusion mechanism $gg\to H$ with the most easily accessible decay
channels such as $H\to \gamma\gamma, H \to ZZ^* \to 4\ell, H \to WW^*
\to 2\ell 2\nu$ with $\ell=e,\mu$. Sub-dominant but very important
channels such as vector boson fusion $qq \to Hqq$ with in particular
the $H \to  \tau\tau$ decay channel and Higgs--strahlung $q\bar q \to
HW,HZ$ with $H\to b\bar b$ decays need to be more thoroughly
investigated. The goal would be to reach an accuracy at the percent
level, which is the size of the higher order electroweak corrections
and presumably also that of the potential new physics effects (see for
instance Ref.~\cite{Gupta:2013zza}).

Additional production channels such as associated Higgs production
with top quark pairs, $pp\to t\bar t H$, that would allow for a direct
determination of the top--quark Yukawa coupling and might provide an
unambiguous determination of the Higgs CP--properties,  and rare decay
modes such as $H\to Z\gamma$ which could give complementary
information to the $H\to \gamma\gamma$ decay and $H\to \mu^+ \mu^-$
which would test for the first time Higgs couplings to other fermions
than those of the third generation still need to be probed. Other
higher order processes for single Higgs boson production, such
associated production with two vector bosons or a single top quark
might also provide interesting information.

Of prime importance would be the measurement of the Higgs
self--coupling that allows for the complete determination of the Higgs
potential which is responsible for EWSB.  The Higgs trilinear coupling
can in principle be studied by measuring the rate for double Higgs
production which, unfortunately, is very small at the LHC. The quartic
Higgs couplings can only be accessed in triple Higgs production which
is hopeless at the LHC.

In the beyond the SM context, one needs to search for new Higgs
bosons with mass scales that are larger than those probed so far
and/or to  detect them in modes that have not yet been considered. 
This would be the case, for instance, of the additional neutral and 
charged Higgs bosons predicted in two--Higgs doublet models
(2HDM)~\cite{Gunion:1989we,Branco:2011iw}, as realised in the minimal
supersymmetric extension of the SM
(MSSM)~\cite{Gunion:1989we,Djouadi:2005gj}. In these models, four
additional physical Higgs bosons are present, besides the one already
observed and that we will now  denote by $h$: a heavier CP--even $H$
state, a CP--odd $A$ state and two charged $H^\pm$ bosons. The four
Higgs masses $M_h, M_A, M_H$ and $M_{H^\pm}$, as well as the ratio of
the vacuum expectation values of the two scalar fields
$\tan\beta=v_2/v_1$ and the mixing angle $\alpha$ that diagonalizes
the two CP--even Higgs states, are unrelated in a general 2HDM. In the
MSSM however, supersymmetry imposes strong constraints on the
parameters and, in fact, only two of them (e.g. $\tan \beta$ and
$M_A$) are independent at tree level.

In the MSSM at high $A$ masses, one is in the decoupling regime 
in which the lighter CP--even  $h$ state will have almost SM--like
couplings while the four states $H,A$ and $H^\pm$ become very heavy,
degenerate in mass and decouple from the massive gauge
bosons~\cite{Haber:1995be}. In the 2HDM, to cope naturally 
with the fact that the observed $h$ boson is SM--like, one invokes the
so--called alignment
limit~\cite{Pich:2009sp,Craig:2013hca,Carena:2013ooa,Bernon:2014nxa}
in which only one Higgs doublet gives masses to the $V=W/Z$ bosons. In
this case,  the mixing angle $\alpha$ is such that the $h$ couplings
are the same as the ones of the SM Higgs state,
i.e.~$\alpha=\beta-\frac12 \pi$. In this case too, the $H$ state will
no longer couple to massive gauge bosons as also does the pseudoscalar
$A$ boson in CP--invariant theories. Hence, in both 2HDMs and the
MSSM, the additional Higgs states need to be searched for at
relatively high masses and in channels that do not involve the vector
bosons and, hence, are rather complicated experimentally. Because
there is a sum--rule that makes that the sum of the coupling squared
of all the CP--even Higgs particles to $W/Z$ boson should add up to
the SM--Higgs coupling squared $g_{hVV}^2$~\cite{Gunion:1989we}, most
other Higgs extensions in which the light $h$ state is forced to
SM--like, will have a phenomenology that is similar to that of 2HDMs
and the MSSM. This would be for instance the case of the next-to-MSSM
(NMSSM) in which the additional singlet--like states almost decouple
from fermions and gauge bosons if $h$ is SM--Higgs
like~\cite{Maniatis:2009re,Ellwanger:2009dp}.

Another interesting example of a new physics extension is the
so-called Higgs portal dark matter (DM)
scenario~\cite{Silveira:1985rk,McDonald:1993ex,Burgess:2000yq,Mambrini:2011ik}. One
postulates the existence of a cosmologically stable, massive and
weakly interacting neutral particle that would account for the DM in
the universe and which couples only to the observed Higgs boson. This
particle is thus undetectable and would appear only as missing energy
when produced in association with SM particles. If the DM particle is
light, $M_{\rm DM} \lsim \frac12 M_h$, it will appear in the decays of
the observed Higgs boson, the latter being produced in the usual
production channels, expect for the dominant process $gg\to h$ in
which an additional jet in the final state is required in order to
make the process visible. These processes have been intensively
discussed, either directly by searching for missing energy signals or
indirectly through the measurement of the observed Higgs signal
strengths; see e.g. Ref.~\cite{Djouadi:2013lra} a recent review.  In
turn, if the DM particles are heavier than $\frac12 M_h$, the Higgs
boson should be virtual, $h^* \to$ DM,  making that the production
processes are at higher order in perturbation theory and  have rather
small production rates. Hence, as in the case of the MSSM or 2HDM, the
production cross sections of the new states are too low at present
energies and luminosities to allow for the probing of the entire
parameter space allowed of these portal models.

To have more sensitivity to new physics phenomena, one needs a
significantly larger sample of Higgs bosons that are produced, and two
options are then at hand. The first is a large increase of the
integrated luminosity. This is the high--luminosity LHC option
(HL-LHC) on which there is presently a wide consensus in the
community: collecting up to 3 ab$^{-1}$ of data at a c.m. energy of
$\sqrt s= 14$ TeV should be the next step for the LHC and the particle
physics goal for the next decade. Prospects in the context of Higgs
physics for the HL--LHC option have been discussed in details in
Ref.~\cite{ATLAS:2013hta,CMS:2013xfa}.

However, in many cases and in particular for the probing of high mass
scales (such as the production of new Higgs bosons with TeV masses) 
or very rare processes for the already observed Higgs state
(such as double Higgs production), this high--luminosity option will  not
be sufficient. A more radical option would be a significant increase in the 
c.m. energy. In this context, an upgrade of the LHC to an energy 
about 2--3 times higher has been discussed and, for instance,
detailed studies of the physics of a $\sqrt s=33$ TeV collider have been
performed~\cite{Baur:2002ka}.  More recently, a Future Circular
Collider (FCC-hh), a hadron collider with a c.m. energy of 100 TeV,
has been proposed as a potential follow-up of the LHC at
CERN\footnote{See the web site
  \url{https://fcc.web.cern.ch/Pages/default.aspx}.};
 such a very high energy machine is also under study in China
 \cite{Tang:2015qga}.

The discovery potential of such machines\footnote{These energies have
  been considered in the late 1980's: $\sqrt s=33$ TeV is close to the
  energy that was foreseen for the late Superconducting Super Collider
  (SSC) ($\sqrt s =40$ TeV) and  $\sqrt s=100$ TeV is a factor of two
  lower than the energy of the Eloisatron collider which was proposed
  in Europe at that time~\cite{Basile:1983cz}.} is immense\footnote{A
  review of the physics potential of such a machine in various
  scenarios, with very little overlap with our analysis here, has
  appeared very recently \cite{Arkani-Hamed:2015vfh}.} and, in the
context of new Higgs states for instance, the search reach would be a
factor three to six larger than the LHC depending on the considered
(pair or singly produced) particle. Even in the case of the observed
SM--like Higgs boson, the increase in the production cross sections
allowed by the higher energy would be more than an order of magnitude
compared to the 14 TeV LHC, and high luminosities would provide a
sufficient data sample to probe very difficult channels such as Higgs
pair production that cannot be seriously considered even at the
HL--LHC.
In the already probed Higgs channels, a very high precision  can be
achieved in the determination of the Higgs properties at such
machines.

In fact, with the expected accuracy, major theoretical improvements in
the determination of the Higgs cross sections and decay branching
ratios (besides the recent improvements in the gluon fusion channel
for single and double Higgs production that will be discussed in the
present paper) need to be performed. In particular, more precise
determinations of the gluon and quark structure functions and of the
strong coupling $\alpha_s$ would be required to reduce the
uncertainties in the cross
sections~\cite{Baglio:2010ae,Dittmaier:2011ti} along with a better
determination of the bottom (and eventually charm) quark mass which,
together with $\alpha_s$, are the main source of uncertainties in the
Higgs decay branching ratios
\cite{Djouadi:1995gt,Baglio:2010ae,Denner:2011mq}. Progresses in these
directions are expected to occur in the next decades. However, to
discuss in detail the potential of the high--energy machine, it is
still useful to provide an estimate of the theoretical precision with
which all observables are currently known in order to quantify the
possible improvement in the measurements at these colliders and to
make comparisons with other options such as, for instance, an
electron--positron collider with $\sqrt s \gsim 250$ GeV.

These are the issues that will be analysed in this paper. While
several analyses have addressed some of the
processes~\cite{Fcc-physics,LHC-XS,Alwall:2014hca,Torrielli:2014rqa,Papaefstathiou:2015paa,Chen:2015gva,Fuks:2015hna},
we will attempt to perform a comprehensive analysis of Higgs physics
at a 100 TeV hadron collider, not only for  Higgs production in the
SM, but also in some  well motivated extensions such as 2HDM/MSSM and
Higgs-portal DM scenarios. The paper is organised as follows. In the
next section, we discuss the production of the SM Higgs boson: single
and double production in the main channels and single production in
subleading channels such as associated production with two vector
bosons or single top quarks; in the case of the main channels, an
attempt to estimate the theoretical uncertainties is made.  The
production of DM particles through the already observed SM--like Higgs
boson in Higgs--portal models will be analysed in section 3 and 
numerical results for the production
cross sections in the various channels will be presented
in the case of  spin 0, $\frac12$ and 1 DM states. In section
4, we analyse the main processes for producing the heavier neutral and
charged Higgs particles in 2HDMs like the MSSM and both single and
pair production processes will be addressed; the parameter space that
could be probed at 100 TeV will be delineated.  A short conclusion
will be given in the final section 5.
 
\section{Production of the  SM Higgs boson} 

\subsection{Single Higgs production and higher order corrections} 

\subsubsection{Gluon fusion}

The main production process for a single Higgs boson at hadron
colliders is gluon fusion, $gg \to H$. This process proceeds at the 
quantum level already at leading order (LO)~\cite{Georgi:1977gs} with
triangular top and bottom quark loops from which the $H$ state is
emitted. In the Standard Model, the top quark loop contribution is by
far dominating; the bottom loop contribution is rather small and its
interference with the dominant top loop does not exceed 10\% of the
total contribution. The next--to--leading order (NLO) QCD
corrections were calculated two decades ago, first in the infinite top
quark mass approximation $M_H \ll 2m_t$~\cite{Djouadi:1991tka,Dawson:1990zj}
by applying the low energy theorem which related the $Hgg$ amplitude to the
QCD $\beta$ function~\cite{Ellis:1975ap,Shifman:1979eb} and 
then using the exact quark mass dependence in the
loop~\cite{Spira:1995rr}.  It was shown that if the LO cross section
contains the exact top--quark mass dependence, the two results
approximately agree with each other. This is particularly true in the
Higgs mass range $M_H \lsim 2m_t$ where the LO $Hgg$ amplitude does
not develop an imaginary part.  

The NLO QCD corrections are found to be very large, with a
$K$--factor\footnote{The $K$--factor is defined as the ratio of cross
  sections at the higher order (HO) and the lowest order (LO), $K\!=
  \! \sigma_{\rm HO}/\sigma_{\rm LO}$ when $\alpha_s$ and the PDFs are
  consistently evaluated at the respective perturbative orders.}  of
around 2 for $M_H=125$ GeV at a c.m. energy $\sqrt s=8$ TeV. 
The next--to--next--to--leading order
(NNLO) QCD corrections were computed  in the infinite top mass
limit~\cite{Harlander:2002wh,Anastasiou:2002yz,Ravindran:2003um}, 
leading to  an increase of about 25\% for the total cross section. 
Later the NNLO corrections have been evaluated in a top mass 
expansion~\cite{Marzani:2008az,Harlander:2009mq,Pak:2009dg,Harlander:2009my}
and it was found that the limit $M_H \ll 2m_t$ is again good  for 
$M_H \lsim 350$ GeV. The resummation of the soft gluons at next-to-next-to leading
logarithm (NNLL) \cite{Catani:2003zt}  leads only to a moderate increase of the 
cross section if the central value of the renormalisation and factorisation scales
is chosen to be at $\mu_0 = \mu_R=\mu_F=\frac12 M_H$ which, in passing
improves the convergence behaviour of the perturbation series
\cite{Anastasiou:2008tj}.  Note, however, that the calculation of the
subleading bottom quark loop contribution beyond the NLO
approximation is still lacking as one cannot use the effective field
theory (EFT) approach with an infinitely large loop mass as in the
case of the top contribution.

The gluon fusion process has now
reached an impressive accuracy with the very recent completion of the
calculation of the next--to--next--to--next--to--leading order (N$^3$LO) QCD
corrections~\cite{Anastasiou:2015ema} after a huge amount of  theoretical 
efforts in the past few years~\cite{Moch:2005ky,Ahrens:2010rs,
  Hoschele:2012xc,Anastasiou:2013srw,Buehler:2013fha,Anastasiou:2013mca,
  Kilgore:2013gba,Anastasiou:2014vaa,Ahmed:2014cla,Ahmed:2014uya,
  Li:2014bfa,deFlorian:2014vta,Anastasiou:2014lda,Dulat:2014mda,
  Duhr:2014nda,Li:2014afw}. For the $M_H=125$ GeV Higgs boson, 
the N$^3$LO corrections amount to a $\approx +2\%$ increase of the cross section 
at both $\sqrt s=8$ TeV and $\sqrt s=14$ TeV but with a  
reduction of the scale uncertainty to about 3\% as will be discussed
later. Improved predictions at N$^3$3LO using renormalisation group
equations have also been calculated very recently~\cite{Ahmed:2015sna}
and jet-vetoed cross section has also reached the N$^3$LO+NNLL
accuracy~\cite{Banfi:2015pju}. Soft-gluon resummation has been
recently extended to N$^3$LL order, leading to an increase of $\sim
5\%$ over the fixed-order results, while mass effects up to NLL are
found to be negligible~\cite{Schmidt:2015cea}.

The electroweak (EW) radiative corrections have been computed at NLO,
first in the infinite loop mass limit $m_t, M_V\! \gg \! 2
M_H$~\cite{Djouadi:1994ge,Djouadi:1997rj,Aglietti:2004nj} and, then,
in an exact calculation
\cite{Degrassi:2004mx,Actis:2008ug}. Approximate mixed QCD-EW
corrections at  NNLO are also available ~\cite{Anastasiou:2008tj} in
the EFT approach with the infinite loop mass limit, $M_{H}\ll M_V$ in
this case. Both types of corrections amount to a few percent.

Currently no public release of the tools used to calculate the N$^3$LO
corrections is available. Since the N$^3$LO QCD correction turns out to be quite
small, we will thus stick to the NNLO QCD
order supplemented with the mixed NNLO QCD+EW corrections and use the
program {\tt iHixs}~\cite{Buehler:2012zf} with a central scale $\mu_0
= \frac12 M_H$. This scale choice, which as mentioned earlier accounts for the
next--to--next--to--leading logarithmic (NNLL) increase of the total
cross section~\cite{Catani:2003zt,Anastasiou:2008tj,deFlorian:2009hc}, 
is motivated by the better convergence of the perturbative expansion as 
it was shown in Ref.~ \cite{Anastasiou:2008tj} and stressed again at
N$^3$LO~\cite{Anastasiou:2015ema}.

\subsubsection{Vector boson fusion}

The vector boson fusion (VBF) channel, in which the Higgs boson is 
produced in association with two jets, $qq \to V^* V^* qq \to Hqq$, is a pure 
electroweak process at
LO~\cite{Cahn:1983ip,Dicus:1985zg,Altarelli:1987ue}. The central scale
in this process is usually chosen to be $\mu_0 = Q^*_V$, the momentum
transfer of the fusing weak bosons. For the fully inclusive 
cross section, the NLO QCD corrections have been known for a while in
both the structure function
approach~\cite{Han:1992hr,Spira:1997dg,Djouadi:1999ht} and
exactly~\cite{Figy:2003nv}; they give rise to an~${\cal O}(10\%)$
increase of the total cross section. The NNLO QCD corrections in the
structure function approach were computed a few years ago and found to
be rather small, below the percent
level~\cite{Bolzoni:2010xr,Bolzoni:2011cu} for the inclusive rate,
while they can be much more sizeable when cuts are included and in the
differential distributions as found
recently~\cite{Cacciari:2015jma}. NLO EW corrections are also
available~\cite{Ciccolini:2007ec,Figy:2010ct} and yield a shift of the
order of 5\% in the cross section. The radiative corrections to the
fully inclusive cross section are thus moderate and well under
control.

However, this process is interesting only when some specific cuts are
employed to single out the VBF topology, namely forward jets with high
transverse momenta, a large rapidity gap between the two forward jets,
central Higgs decay products and no jet activity in this central area. 
This is necessary in order to suppress the QCD background and to
minimise the contamination from the contribution at NNLO of the
gluon-fusion process, $gg\to Hjj$, which leads to the same final state
In this case the NLO corrections are also known
exactly~\cite{Figy:2003nv} and lead to the same enhancement of the
cross section than for the inclusive case. However, a recent
calculation of the QCD corrections at NNLO~\cite{Cacciari:2015jma}
shows that their contribution amounts to ${\cal O}(10\%)$ and are thus
non negligible as in the case of the inclusive cross section. More
importantly, the variation of the VBF cross section with the
renormalisation and factorisation scales is much larger than in the
inclusive case.

Nevertheless, in our study, we will stick to the inclusive cross
section and, given the smallness of the NNLO QCD corrections, we will
give our predictions only at the NLO QCD+EW approximation using the
Monte-Carlo program {\tt VBFNLO}~\cite{Arnold:2008rz,Baglio:2014uba}.

\subsubsection{Higgs--strahlung channels}

The Higgs--strahlung processes $q\bar{q}'\to HV$ with $V=W^\pm$ or $Z$
are also pure EW processes at LO. The NLO QCD corrections are actually
pure Drell-Yan  corrections~\cite{Altarelli:1979ub} to the process $q\bar{q}\to
V^*$~\cite{Han:1991ia,Baer:1992vx,Ohnemus:1992bd,Spira:1997dg,Djouadi:1999ht}
and for the associated production with a $W$ boson this extends up to
NNLO~\cite{Brein:2011vx} by the use of the classic NNLO results for
Drell--Yan~\cite{Hamberg:1990np,Harlander:2002wh}. On top of these
contributions there are NNLO QCD corrections where the Higgs is
radiated off the top loops, they amount to~$\sim 1$\% for both
processes~\cite{Brein:2011vx}. In the case of the
associated production with a $Z$ boson there is, in addition to the
previously discussed contributions, the opening at NNLO of the channel
$gg\to HZ$ not going through a virtual weak
boson~\cite{Brein:2003wg} that has been extended to include the real
radiation effects that describe much better the kinematics of this
subprocess~\cite{Hespel:2015zea}. The threshold N$^3$LO corrections
are also available for the hadronic process~\cite{Kumar:2014uwa}.

The NLO+NNLO QCD Drell--Yan type corrections
are moderate, of the order of at most $+35\%$ and the gluon fusion
contributions in $ZH$ production are of the order of 10\% at $\sqrt s=14$
TeV. In the past few years there have been improvements in the
description of the $gg\to HZ$ subprocess with NLO QCD corrections and
soft-gluon resummation that reduce the uncertainties in this
subprocess~\cite{Altenkamp:2012sx,Harlander:2014wda}. The NLO EW
corrections are negative and reduce the cross section by an amount
of~$3-8\%$ at LHC energies~\cite{Ciccolini:2003jy}. They are computed in
the so-called $G_F$ scheme where the electromagnetic coupling constant
$\alpha$ is derived from the Fermi constant $G_F$ such that the
corrections are not sensitive to the value of the light quark
masses. The combination of the EW and NLO corrections follows
Ref.~\cite{Brein:2004ue}:
\begin{align}
\sigma_{WH}^{\rm QCD+EW} = \sigma_{WH}^{\rm QCD} (1+\delta_{WH}^{\rm EW}),\,\,
\sigma_{ZH}^{\rm QCD+EW} = \sigma_{ZH}^{\rm QCD,DY}
(1+\delta_{ZH}^{\rm EW}) + \sigma_{gg\to ZH}.
\end{align}

For the computation of the cross-section we will use the computer
program {\tt vh@nnlo}~\cite{Brein:2012ne} which includes the full NNLO
QCD + NLO EW corrections. We will then neglect the improvement in the
gluon fusion subchannel for $ZH$ production, which is formally a
N$^3$LO contribution to the whole hadronic process.

\subsubsection{Associated production with a heavy quark pair}

Associated Higgs production with top or bottom quark pairs are the
channels which are most affected by QCD backgrounds. The LO cross
section for $t\bar{t} H$ production was computed decades
ago~\cite{Raitio:1978pt,Ng:1983jm,Kunszt:1984ri,Marciano:1991qq} and
NLO QCD corrections are known to be modest provided that the central
scale $\mu_0\!=\!\frac12
M_H+m_t$~\cite{Beenakker:2001rj,Beenakker:2002nc,Dawson:2002tg} is
used. In the past years the NLO EW corrections have been
computed~\cite{Frixione:2014qaa,Yu:2014cka,Frixione:2015zaa} and
soft-gluon resummation is also available~\cite{Kulesza:2015vda} and
has extended to NNLL accuracy and has been used to obtain approximate
NNLO results that are only slightly larger than NLO results but with
an uncertainty band reduced by a factor of
two~\cite{Broggio:2015lya}.

The associated production with $b$--quark
pairs~\cite{Raitio:1978pt,Ng:1983jm}, in contrast, displays a
different behaviour. The NLO QCD corrections can be calculated
in the same way as for $t\bar{t}H$ production but turn out to be
rather large~\cite{Dittmaier:2003ej,Dawson:2003kb}. This is due to the
large logarithms generated by the integration of the transverse
momenta of the final-state $b$--quarks. 

These large logarithms can be resummed by considering the bottom quark
as a massless constituent of the quark and using the Altarelli--Parisi
evolution~\cite{Altarelli:1977zs} of the bottom quark parton
distribution function (PDF) to the scale of the process. Then one
works in a five-flavour scheme (5FS) and the LO process that needs to
be considered is $b\bar{b}\to H$~\cite{Dicus:1988cx}. Requiring a
high-$p_T$ final-state $b$ quark requires the NLO QCD
corrections~\cite{Dicus:1998hs,Balazs:1998sb,Campbell:2002zm,Maltoni:2003pn} 
and the NNLO QCD corrections introduce back the process $gg\to b
\bar{b} H$~\cite{Harlander:2003ai}. It turns out that choosing $\mu_0
= \frac14 M_H$ as the factorisation scale and using the running bottom
quark mass at the scale of the Higgs boson mass greatly improves the
perturbative convergence of the
series~\cite{Maltoni:2003pn,Harlander:2003ai}. The fully exclusive $gg
\to b \bar{b} H$ process, calculated with four active parton flavours
in a four-flavour scheme (4FS), and the 5FS process $b\bar{b}\to H$,
should converge against the same value at higher perturbative
orders. The NLO EW corrections were also calculated in the past decade
and found to be modest~\cite{Dittmaier:2006cz}, we will then neglect
them. We will also not include the very recent N$^3$LO corrections
that are not yet implemented in a public code. They have been found to
reduce further the scale uncertainty of the
predictions~\cite{Ahmed:2014cha,Ahmed:2014era}.

In our numerical analysis of the cross sections, we will use the program 
{\tt bbH@NNLO}~\cite{Harlander:2003ai} for  the associated production
with bottom quarks at NNLO QCD and we will use {\tt MadGraph 5} in the
{\tt aMC@NLO} framework~\cite{Alwall:2014hca} for the NLO QCD
corrected cross sections of the associated production with top
quarks.

\subsection{The cross sections including the theoretical uncertainties}

The production cross sections are affected by theoretical
uncertainties which are in general divided in two categories, with an
additional one in the $gg\to H$ case.

$i)$
{The scale uncertainty, which reflects the dependence of the
    truncated expansion of the cross section at a given perturbative
    order in  $\alpha_s$ on the  renormalisation scale $\mu_R$ that
    defines $\alpha_s$ and on the factorisation scale $\mu_F$ at
    which the matching of the perturbative calculation (the matrix
    element) and the (non perturbative) structure functions is
    made. This is generally estimated by varying the two scales in the
    interval
    \begin{align}
      \label{scale}
      \small{\frac12} \mu_0 \leq \mu_R, \mu_F\leq 2\mu_0
    \end{align}
    with $\mu_0$ being the central scale chosen for a given process and
    with some restrictions on the ratio $\mu_R/\mu_F$ depending on the
    process\footnote{In some cases, like in the gluon fusion process which is 
      subject to large QCD corrections, this domain of variation is sometimes 
      considered to be too small. Other approaches to estimate the scale uncertainty 
      have been proposed in this case; see for instance the approach
      of Ref.~\cite{Cacciari:2011ze,Bagnaschi:2014wea} for which one
      obtains results that are similar to those obtained by extending
      the domain of scale variation to $\frac13 \mu_0 \leq \mu_R,
      \mu_F\leq 3\mu_0$.}.
    For the various processes, we have adopted the following central
    scales in our analysis:
    \begin{align}
      \mu_0^{gg\to H} = \frac12 M_H,\,\, \mu_0^{qq'\to Hqq'} = Q_V^*,\,\,
      \mu_0^{q\bar{q}' \to V H} = M_{V\! H},\,\,  \mu_0^{q\bar{q}/gg\to t
        \bar{t}H} = m_t + \small{\frac12} M_H. 
    \end{align}
    In the case of $b \bar{b} \to H$ production we use the following set-up
    for the central scales:
    \begin{align}
      \mu_R^{b\bar{b}\to H} = M_H,\,\, \mu_F^{b\bar{b}\to H} =  M_H/4.
    \end{align}
    For all processes, the scale uncertainty is derived by a variation of the 
    renormalisation and factorisation scales within a factor of two from the
    central scale, as in Eq.~(\ref{scale}). In the case of the $b\bar{b}\to H$ process
    however, following the LHC Higgs Cross Section Working Group
    framework~\cite{Dittmaier:2011ti} and given the asymmetry in the
    choice of the scales $\mu_R$ and $\mu_F$, we use the following domains of 
    variation: $\frac15 M_H \leq \mu_R \leq 5 M_H$, $\frac{1}{10} M_H
    \leq \mu_F \leq 0.7 M_H$.  {The asymmetrical choice of the central
    renormalisation and factorisation scales for the $b\bar{b}\to H$
    process is justified by the requirement of a good agreement
    between the 4FS and the 5FS~\cite{Dittmaier:2011ti}. This is also
    supported by the scale dependence studied in
    Ref.~\cite{Harlander:2003ai}. This requirement also drives the
    choice of a larger scale variation interval as exemplified in
    Ref.~\cite{Dawson:2005vi}, and by the wish to have a reliable
    estimate of the scale uncertainties at NNLO given the very small
    dependence of the total cross section on the renormalisation scale.}}

$ii)$
{The PDF and $\alpha_s$ uncertainty stemming from the impact of the
    uncertainties in the modelling and the experimental data used in
    the fit that provides the PDF sets and the value of the strong
    coupling constant  $\alpha_s$.  With the cross sections for most
    channels now known at  NNLO and even beyond, thereby reducing the
    scale uncertainty, this is becoming the major source of
    theoretical errors. According to  usual
    practice,  we will consider the 90\% CL correlated
    PDF+$\Delta^{\rm exp} \alpha_s$ uncertainties using the MSTW2008
    PDF sets~\cite{Martin:2009iq,Martin:2009bu} and the following value
    for $\alpha_s$:
    \begin{align}
      \label{alphas}
      \alpha_s(M_Z^2)=  0.1171~^{+0.0014}_{-0.0014}~{\rm (68\%~CL)}~{\rm
        or}~ ^{+0.0032}_{-0.0032}~{\rm (90\%~CL)}~{\rm at~NNLO} 
    \end{align}}

$iii)$
{In the case of the gluon fusion channel, a third uncertainty
    related to the use of the effective field theory approach to
    account for the corrections beyond NLO should also be included as
    done in Refs.~\cite{Baglio:2010um, Baglio:2010ae} for the Tevatron
    and the LHC respectively. Indeed, the infinite quark mass
    limit is not adequate for the bottom loop which generates a
    $\approx 10\%$ contribution when it interferes with the top loop
    contribution. The exact $b$--quark contribution is known at NLO but not at
    NNLO\footnote{According to the authors of
      Ref.~\cite{Anastasiou:2015ema} in which the ``tour de force" of
      deriving the N$^3$LO corrections has been achieved, this
      generates an uncertainty of order 2\% and the NNLO calculation
      with the exact $m_b$ dependence  is now becoming the next big
      challenge in the field of higher order corrections.}. In Ref.~\cite{Baglio:2010ae}, the
    uncertainty generated by the use of the EFT approach in this case
    has been estimated to be of order 3--4\% for $M_H\!=\!125$ GeV
    (when the uncertainty due to the choice of a renormalisation
    scheme for the $b$--quark mass is also included). In addition, the
    mixed QCD--EW corrections at NNLO have been derived in the limit
    $M_W\! \gg \!M_H$ which is clearly not adequate. This leads to an
    additional error of  few percent \cite{Baglio:2010ae} which generates
    a total EFT uncertainty of order of 5--7\% for $M_H\!=\!125$ GeV when 
all errors are summed.}

Finally, the total uncertainty is obtained simply by adding linearly
the scale and the  PDF uncertainties and, in the case of $gg\to H$, also
the EFT uncertainty on top of these. 
Nevertheless, for the dominant $gg\to H$ channel, as the results for the cross 
section at N$^3$LO are not yet publicly available at
energies above $\sqrt s=14$ TeV and the associated scale uncertainty
is not given in Ref.~\cite{Anastasiou:2015ema} for such higher
energies, we cannot  provide with a N$^3$LO description of
$\sigma(gg\to H)$. 
We will therefore use the following approximation 
for the observed Higgs boson with a mass $M_H=125$ GeV. At $\sqrt s=14$ TeV, 
the scale uncertainty  of $\sigma^{\rm NNLO} (gg\to H)$ is about $\pm 9\%$
and is thus of the same order of the sum of the scale uncertainty at  N$^3$LO
(which is about 3\%) and the EFT uncertainty (which is about 5--7\%). We will
therefore consider that this trend is valid at all energies and assume
that
\begin{align}
\Delta \sigma^{\rm NNLO}_{\mu} (gg\to H) =  \Delta \sigma^{\rm N^3LO}_\mu (gg\to H)
+ \Delta \sigma^{\rm NNLO}_{EFT} (gg\to H).
\end{align}

The SM input parameters used in the calculations throughout this paper
are as follows:
\begin{equation}
  \begin{matrix}
    M_W = 80.385\text{ GeV},\,\,\,\, M_Z = 91.1876\text{ GeV},\,\,\,\, M_t
    = 173.2\text{ GeV}, \nonumber\\
    M_H = 125\text{ GeV},\,\,\,\, m_b^{\rm pole} = 4.75\text{
      GeV},\,\,\,\, \bar{m}_b(\bar{m}_b) = 4.213\text{ GeV}, \nonumber\\
    \alpha_s^{\rm LO}(M_Z^2) = 0.13939,\,\,\,\, \alpha_s^{\rm
      NLO}(M_Z^2) = 0.12018,\,\,\,\, \alpha_s^{\rm  NNLO}(M_Z^2) =
    0.11707.
  \end{matrix}
\label{inputs}
\end{equation}

\begin{figure}[!ht]
\centering
\includegraphics[scale=.95]{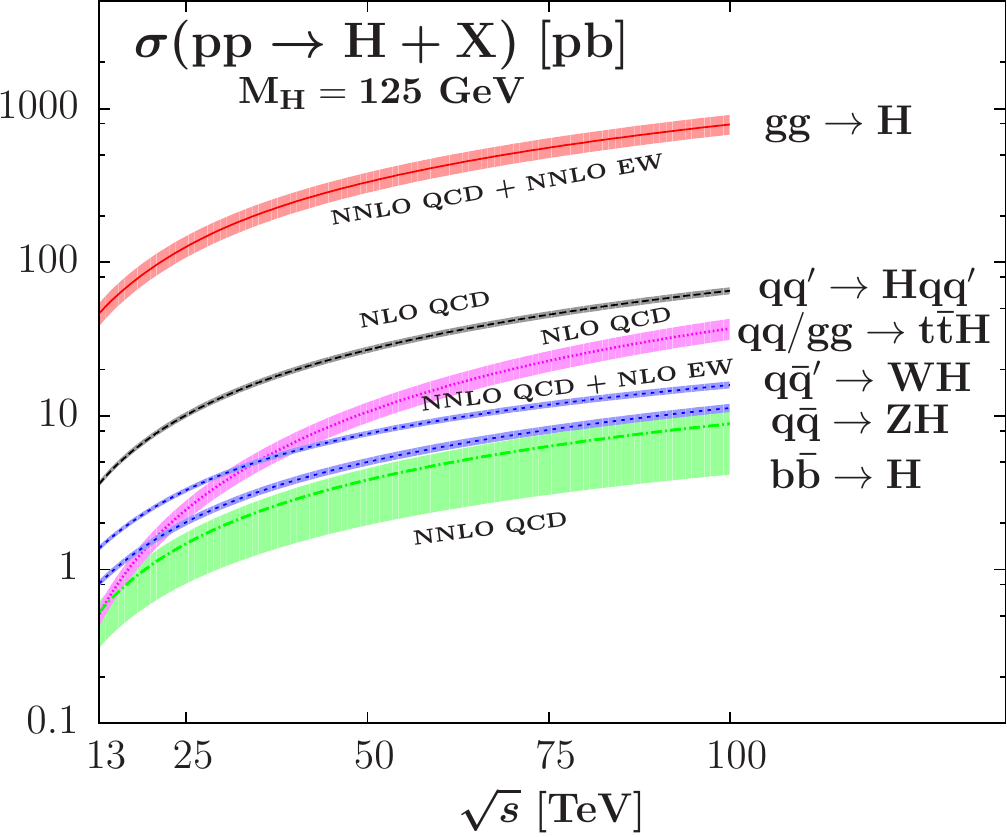}
\caption[]{The total cross sections for SM Higgs production in the main channels 
at a proton-proton collider as a function of the c.m. energy with $M_H = 125$ GeV: 
gluon fusion (red/full), VBF (grey/dashed), Higgs-strahlung  (blue/dotted), 
associated production with a top pair (violet/dotted with small dots) and 
bottom--quark fusion (green/dash-dotted). The MSTW2008
  PDF set has been used and theoretical uncertainties are included as
  corresponding bands around the central values.}
\label{fig:singleH}
\end{figure}

The results for the cross section in all channels for the single
production at a hadron machine of a Higgs boson with a mass $M_H=100$
GeV  are displayed in Fig.~\ref{fig:singleH} as a function of the
centre--of--mass energy,  starting from $\sqrt s=13$ TeV up to $\sqrt
s=100$ TeV. The total theoretical uncertainty bands are also
displayed. The gluon fusion 
channel remains the dominant production mechanism up to the FCC-hh
collider energies, with a cross section that ranges from $\sigma
\approx 50$ pb at $\sqrt s=13$--14 TeV to $\sigma \approx 800$ pb at
$\sqrt s=100$ TeV. It is followed by the VBF channel which has a cross
section  that is at least one order of magnitude  smaller in the
entire c.m. energy range. The Higgs--strahlung channels are the next
important ones for energies below $\sqrt s \approx 30$ TeV with
cross sections that are a factor 3--4 smaller than VBF. Above $\sqrt s
\approx 30$ TeV, it is in fact the $pp\to t\bar tH$ process which becomes
the third most important process. The $b\bar b \to H$ process closes
the list with a cross section that is slightly smaller than that of
the $HZ$ process for $\sqrt s \gsim 14$ TeV.  The numerical results are
summarised in Table~\ref{table:sigma_lhc} for all the considered
channels and the values of the  cross sections, together with the
scale, PDF and total uncertainties are displayed.

\begin{table}[!ht]
 \renewcommand{\arraystretch}{1.1}
  \begin{center}
   \small
\begin{tabular}{|c|c|cccc|}\hline
channel &      $\sqrt{s}$ [TeV] & $\sigma$ [pb] & Scale
      [\%] & PDF+$\alpha_s$ [\%] & Total [\%] \\\hline
&      $13$ & $46.68$ & ${+8.6}\;\;\;{-9.3}$ & ${+7.5}\;\;\;{-8.0}$ &
      ${+16.1}\;\;\;{-17.3}$ \\
$gg\to H$ &      $14$ & $52.43$ & ${+9.4}\;\;\;{-9.2}$ & ${+7.5}\;\;\;{-7.5}$ &
      ${+16.9}\;\;\;{-16.6}$ \\
&      $33$ & $189.5$ & ${+8.6}\;\;\;{-7.7}$ & ${+7.5}\;\;\;{-7.3}$ &
      ${+16.1}\;\;\;{-15.0}$ \\
&      $100$ & $788.6$ & ${+7.1}\;\;\;{-6.1}$ & ${+8.3}\;\;\;{-8.0}$ &
      ${+15.4}\;\;\;{-14.1}$ \\\hline
%
&      $13$ & $3.645$ & ${+0.6}\;\;\;{-0.6}$ & ${+3.8}\;\;\;{-3.4}$ &
      ${+4.5}\;\;\;{-4.0}$
      \\
VBF&      $14$ & $4.116$ & ${+0.7}\;\;\;{-0.6}$ & ${+3.8}\;\;\;{-3.3}$ &
      ${+4.5}\;\;\;{-3.9}$
      \\
&      $33$ & $15.12$ & ${+1.4}\;\;\;{-1.1}$ & ${+3.4}\;\;\;{-3.1}$ &
      ${+4.8}\;\;\;{-4.2}$
      \\
&      $100$ & $64.50$ & ${+2.2}\;\;\;{-2.1}$ & ${+3.1}\;\;\;{-3.2}$ &
      ${+5.3}\;\;\;{-5.2}$
      \\\hline
%
&      $13$ & $1.379$ & ${+0.3}\;\;\;{-0.2}$ & ${+3.9}\;\;\;{-3.5}$ &
      ${+4.2}\;\;\;{-3.6}$
      \\
$WH$ &     $14$ & $1.521$ & ${+0.3}\;\;\;{-0.3}$ & ${+3.8}\;\;\;{-3.4}$ &
      ${+4.0}\;\;\;{-3.6}$
      \\
&      $33$ & $4.705$ & ${+0.3}\;\;\;{-0.1}$ & ${+3.9}\;\;\;{-3.6}$ &
      ${+4.2}\;\;\;{-3.7}$
      \\
&      $100$ & $15.88$ & ${+0.7}\;\;\;{-0.1}$ & ${+5.0}\;\;\;{-4.7}$ &
      ${+5.7}\;\;\;{-4.8}$
      \\\hline
%
&      $13$ & $0.8137$ & ${+1.8}\;\;\;{-1.2}$ & ${+3.5}\;\;\;{-2.9}$ &
      ${+5.3}\;\;\;{-4.1}$
      \\
$ZH$ & $14$ & $0.9037$ & ${+1.8}\;\;\;{-1.3}$ & ${+3.3}\;\;\;{-2.9}$ &
      ${+5.1}\;\;\;{-4.3}$
      \\
&      $33$ & $2.969$ & ${+2.0}\;\;\;{-1.6}$ & ${+3.7}\;\;\;{-3.6}$ &
      ${+5.7}\;\;\;{-5.2}$
      \\
&      $100$ & $11.28$ & ${+1.8}\;\;\;{-1.7}$ & ${+4.5}\;\;\;{-4.3}$ &
      ${+6.3}\;\;\;{-6.0}$
      \\\hline
%
&      $13$ & $0.514$ & ${+6.7}\;\;\;{-9.7}$ & ${+6.9}\;\;\;{-7.1}$ &
      ${+13.6}\;\;\;{-16.8}$
      \\
$t \bar t H$ &      $14$ & $0.623$ & ${+6.8}\;\;\;{-9.7}$ & ${+7.0}\;\;\;{-6.9}$ &
      ${+13.8}\;\;\;{-16.6}$
      \\
&      $33$ & $4.51$ & ${+8.5}\;\;\;{-9.1}$ & ${+6.4}\;\;\;{-5.8}$ &
      ${+14.9}\;\;\;{-14.9}$
      \\
&      $100$ & $37.0$ & ${+9.5}\;\;\;{-10.2}$ & ${+6.3}\;\;\;{-5.4}$ &
      ${+15.8}\;\;\;{-15.6}$
      \\\hline
%
&      $13$ & $0.529$ & ${+12.6}\;\;\;{-35.3}$ & ${+4.9}\;\;\;{-6.0}$ &
      ${+17.6}\;\;\;{-41.3}$
      \\
$b\bar b \to H$ &      $14$ & $0.598$ & ${+12.7}\;\;\;{-35.8}$ & ${+5.1}\;\;\;{-5.9}$ &
      ${+17.7}\;\;\;{-41.6}$
      \\
&      $33$ & $2.20$ & ${+13.7}\;\;\;{-41.2}$ & ${+4.3}\;\;\;{-5.7}$ &
      ${+18.0}\;\;\;{-46.9}$
      \\
&      $100$ & $8.91$ & ${+15.9}\;\;\;{-47.4}$ & ${+5.2}\;\;\;{-5.9}$ &
      ${+21.1}\;\;\;{-53.4}$
      \\\hline
\end{tabular}
\caption[]{The total Higgs production cross sections (mostly at NNLO
  QCD + EW) in the main production processes for $M_H=125$ GeV at a
  proton-proton collider (in pb) for given c.m. energies (in TeV) at
  the central scales given in the text. The corresponding shifts due
  to the theoretical uncertainties from the scale variation and the
  90\%CL MSTW PDF+$\alpha_s$ uncertainties are shown, as well as the
  total uncertainty when all errors are added linearly. Note that in
  the case of gluon fusion, the NNLO scale uncertainty is an estimation
  of the combined the EFT+N$^3$LO uncertainty.}
\label{table:sigma_lhc}
\end{center}
\vspace*{-.5cm}
\end{table}

\begin{figure}[!ht]
\centering
\includegraphics[scale=.95]{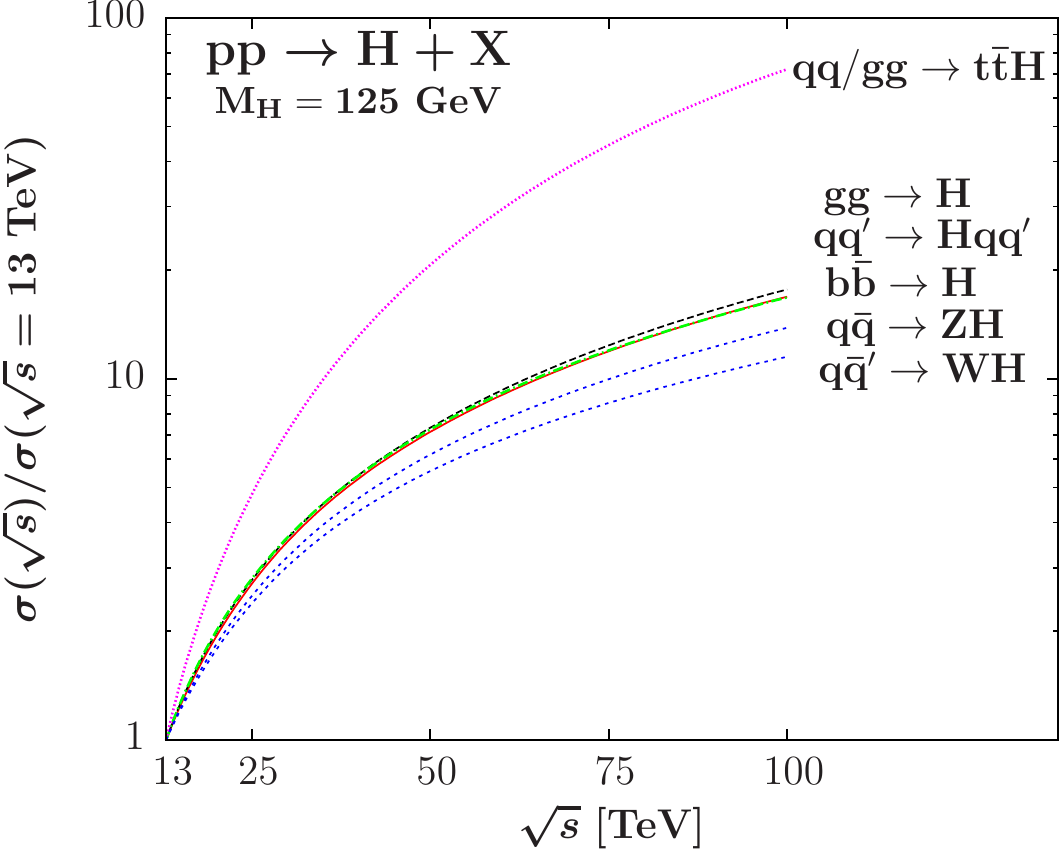}
\caption[]{The variation of the production cross sections for all channels 
with the c.m. energy relative to their values at $\sqrt s=13$ TeV. The
same graphical code as in Fig.~\ref{fig:singleH} is used.}
\label{fig:singleH-ratio}
\end{figure}

In Fig.~\ref{fig:singleH-ratio}, we show the relative rise of the
 cross sections as a function of $\sqrt s$ when they are normalised to
 their values at $\sqrt s=13$ TeV. One sees that compared to 13 TeV,
 there is a gain of a factor $\approx 17$--18 at $\sqrt s=100$ TeV in the
 case of gluon and vector boson fusion. Hence, with an integrated luminosity 
of a few ab$^{-1}$, the Higgs sample produced at a 100 TeV collider could allow for a 
measurement of some ratios of cross sections, such as $\sigma(pp\to H\to \gamma\gamma)/ 
\sigma(pp\to H\to ZZ^*)$ that would be made free of theoretical uncertainties 
\cite{Djouadi:2012rh,Djouadi:2015aba}, at the few per mille level. The  gain in cross section when
going from $\sqrt s=13$ TeV to 100 TeV is smaller for the Higgs--strahlung process 
($\approx 12$--14) but much larger ($\approx
 72$) for $t\bar tH$ production as a result of the opening of the
 phase--space. This clearly illustrates that the 100 TeV FCC-hh
 has a major potential in also improving the understanding of the  
 important bottom and top quark Yukawa couplings~\cite{Plehn:2015cta}.

\subsection{Single Higgs production in subdominant channels} 

We discuss briefly in this subsection the prospects for the associated
production of one Higgs boson with a pair of massive weak bosons,
namely $WW$, $WZ$ and $ZZ$ pairs as well as the associated production of 
a Higgs and a single top quark, $Wb \to Ht$ and considering the two 
channels $qb \to tHq'$ and $gq \to tHbq'$.  

\subsubsection{Associated production with vector boson pairs}

The channels $q\bar q \to VVH$ with $V=W,Z$ proceed through
$s$--channel gauge boson and/or $t$--channel quark exchanges, in which
the Higgs boson is radiated off the weak gauge bosons. The LO cross
sections are known for quite a while~\cite{Cheung:1993bm} and
calculations with modern PDFs for the LHC were presented in
Ref.~\cite{Djouadi:2005gi}. A few years ago the
NLO QCD corrections for $HWW$ and $HWZ$ channels were
released~\cite{Mao:2009jp,Liu:2013vfu} showing that 
as in the case of $HV$ production, they increase the rates
by approximately 50\% (with a few  percent contribution 
from the $gg$ fusion process). The cross sections are quite
small but nevertheless they may provide additional tests and
measurements at 100 TeV, for instance the $HWW$ coupling in the
process $pp\to H WW \to 4W$. {In addition the
associate production with vector boson pairs is a background process
for the production of a pair of Higgs bosons: For example the process
$p p\to H WW \to b\bar{b} WW$ is a background of the search channel $p
p \to H H\to b\bar{b} WW$.}

We will not perform an error analysis and postpone it to a future
publication~\cite{Baglio:2015eon} in which the NLO QCD corrections will
be given for all three processes matched to parton shower in the
POWHEG-BOX framework~\cite{Alioli:2010xd}. The numbers will then be
given in this section at LO only and they have been obtained with a
home-made computer program.

\begin{figure}[!ht]
\centering
\includegraphics[scale=0.85]{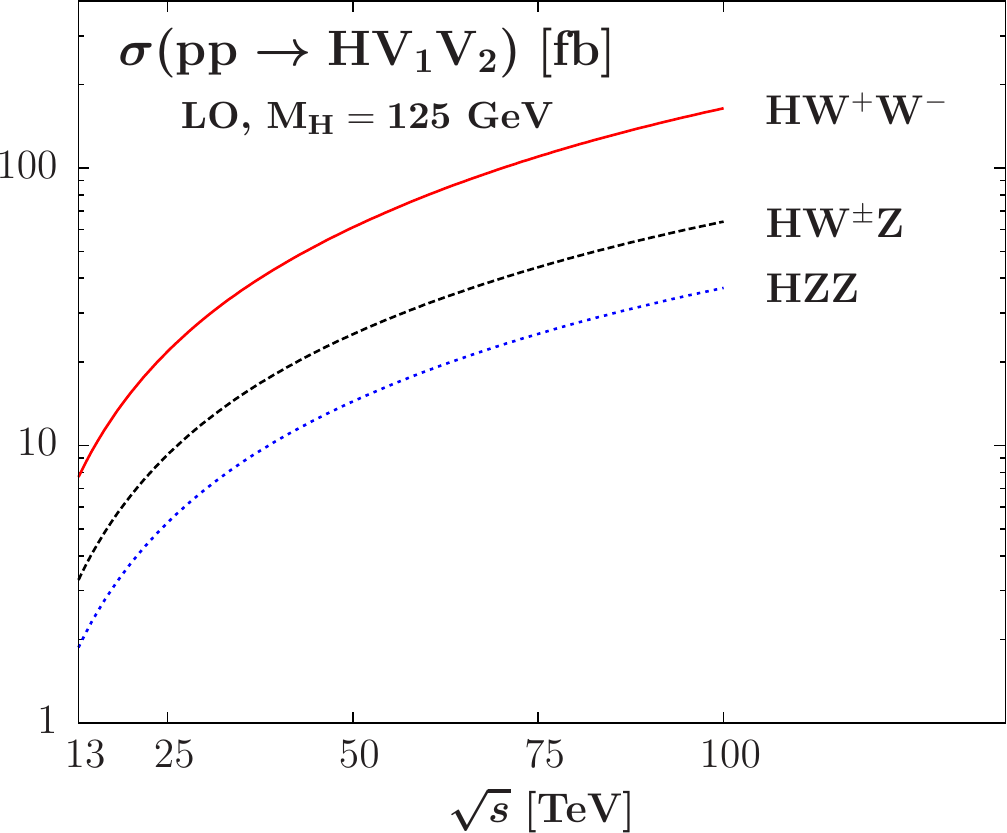}
\caption[]{The LO total cross sections for Higgs production in
  association with a weak boson pair $V_1 V_2$ at a
  proton-proton collider as a function of the c.m. energy with $M_H =
  125$ GeV. The MSTW2008  PDF set has been used.}
\label{fig:singleH_subdominant}
\end{figure}

The results are displayed in Fig.~\ref{fig:singleH_subdominant}. The
production of a Higgs boson in association with a $W$ pair is the
dominant, nearly three times the $HW^\pm Z$ production at 100 TeV. The
detailed numbers are summarised in Table.~\ref{table:hvv}. As a
central scale we have used the invariant mass of the three-particle
final state, $\mu_0 = M_{HVV}$.

\begin{table}[h!]
 \renewcommand{\arraystretch}{1.4}
  \begin{center}
   \small
\begin{tabular}{|c|ccc|}\hline
      $\sqrt{s}$ [TeV] & $\sigma^{\rm LO}_{pp\to H W^+ W^- }$ [fb] &
      $\sigma^{\rm LO}_{pp\to H W^\pm Z}$ [fb] &  $\sigma^{\rm
        LO}_{pp\to H Z Z}$ [fb] \\\hline
      $13$ & $7.70$ & $3.28$ & $1.87$ \\\hline
      $14$ & $8.72$ & $3.71$ & $2.12$ \\\hline
      $33$ & $33.18$ & $13.96$ & $7.96$ \\\hline
      $100$ & $164.1$ & $64.08$ & $36.98$ \\\hline
\end{tabular}
\caption[]{The total cross section (in fb) at LO QCD for the Higgs
  production in association with a weak boson pair at a proton-proton
  collider for given c.m. energies (in TeV) at the central scale
  $\mu_F=\mu_R= M_{HVV}$, for $M_H=125$ GeV. $VV$ means $W^+ W^-$,
  $W^\pm Z$ or $ZZ$.}
\label{table:hvv}
\end{center}
\vspace*{-6mm}
\end{table}

\subsubsection{Associated production with a single top quark}

At the partonic level, the process in which a Higgs boson is produced
in association with a single top quark is $Wb \to Ht$ and it proceeds
through the s--channel exchange of a $t$--quark or the $t$--channel
exchange of a $t$--quark or a $W$ boson. As in the case of the $t\bar
t H$ associated production process, it is directly proportional to the
top quark Yukawa coupling but it is in principle more favoured by 
phase--space at low energies. It has however a much smaller cross section
as it is formally of higher order in perturbation theory, since the
$W$ boson should come for a splitting parton. Nevertheless, it has
been shown in Ref.~\cite{Farina:2012xp} that the process is extremely
sensitive to the $ttH$ Yukawa coupling and, in particular, the $Wb \to
Ht$ cross section can be enhanced by more than one order of magnitude
compared to the SM case if the sign of the Yukawa coupling is reversed. 

At the hadronic level, two processes can generate the $tH$ final state
in a $t$--channel exchange. The first one is the $2\to 3$ process $qb
\to tH j$ with the $b$-quark treated as a parton taken directly from
the proton. The second one is the $2\to 4$ process  $qg \to tHjb$ in
which the $b$ quark is originating from gluon--splitting. In principle
both processes are equivalent (one is the NLO radiative correction of
the other) and, as in the $b\bar b\to H$ versus $gg \to b\bar bH$
processes discussed in the previous section, one has to have a
procedure to match the two processes calculated with four or five
active parton flavours in a four- or five- flavour scheme. However, as the
process is subleading, we will not enter into such sophisticated
details and calculate the two processes independently in a five-flavour
scheme, following closely the analysis performed in
Ref.~\cite{Farina:2012xp}. As also stated in an earlier
work~\cite{Maltoni:2001hu} as well as in a very recent
study~\cite{Demartin:2015uha} done at NLO in QCD, there is also a
$s$--channel process $q\bar{q}'\to t H \bar{b}$. The main advantage of
the five-flavour scheme is that the three production processes
presented here are totally independent at LO. The separation between
$s$--channel and $t$--channel productions still remains at NLO in the
five-flavour scheme~\cite{Demartin:2015uha}.
 
\begin{figure}[!ht]
\centering
\includegraphics[scale=0.85]{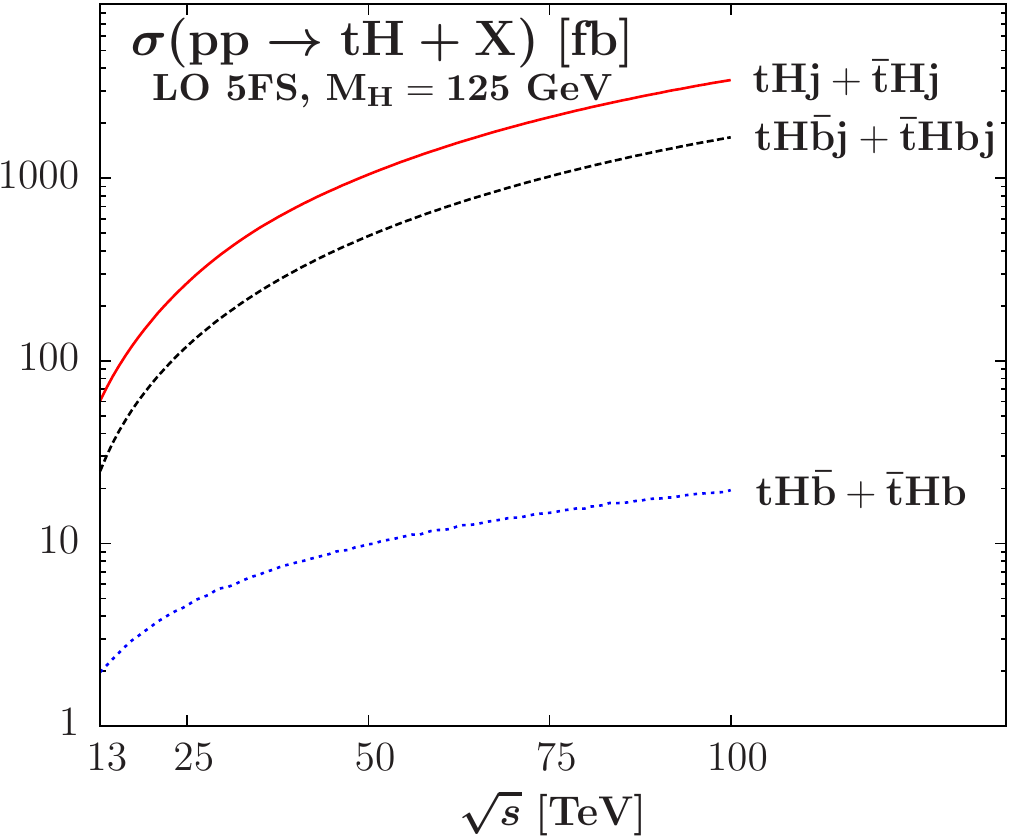}
\caption[]{The total cross section (in fb) at LO for the associated
  Higgs production with a single top quark in the five-flavour scheme
  at a proton-proton collider for given c.m. energies (in TeV) for
  $M_H=125$ GeV. The results for the two $t$--channels, $qb \to t H j$
  and  $qg \to t H j b$ as well as for the $s$-channel process
  $q\bar{q}'\to t H \bar{b}$ are shown. The MSTW2008  PDF set has been
  used and a sum over the two possible final states $ t H + X$ and
  $\bar{t} H + \bar{X}$ has been done.}
\label{fig:singletH}
\end{figure}

The cross sections for the three processes are calculated using the
program {\tt MadGraph 5} with the factorisation and renormalisation
scales set to $\mu_0 = \frac14 (M_H + M_t)$, a scale that minimises the
higher-order corrections and reduces the scale
uncertainty~\cite{Demartin:2015uha}; the MSTW2008 PDFs have been
adopted. While the inclusive rate is calculated in the $qb \to tHj $
and $q\bar{q}' \to t H {b}$ cases, we use the following cuts for
the detection of the $b$--quark (which allows an additional means to
suppress the QCD background) in the case of the $qg \to tHjb$: $p_T^b
>25$ GeV and $|\eta^b| <2.5$. This has allowed us to make a cross
check by comparing our results with those of Ref.~\cite{Farina:2012xp}
given at $\sqrt s=8$ and 14 TeV using their setup: a very good
agreement has been found. Our results for the two other processes have
been compared to Ref.~\cite{Demartin:2015uha} again using their set-up
and a perfect agreement has been found. The results for the cross
sections in the three processes are shown in Fig.~\ref{fig:singletH}
as a function of $\sqrt s$ and some numerical values are displayed in
Table.~\ref{table:ht} for some specific centre of mass energies. At
$\sqrt s=14$ TeV, one has $\sigma(qb \to tHj) \approx 73$ fb, i.e. it
is almost two order of magnitude smaller than $\sigma(pp  \to t\bar
tH)$, while $\sigma(qg \to tHjb)$ is approximately 3 times
smaller. The $s$--channel process $q\bar{q}' \to t H \bar{b}$ is totally
subdominant, of the order of 2 fb. At $\sqrt s=100$ TeV, the cross
section for $qb \to tHj$ has increased by a factor or $\approx 57$.

\begin{table}[h!]
 \renewcommand{\arraystretch}{1.4}
  \begin{center}
   \small
\begin{tabular}{|c|ccc|}\hline
      $\sqrt{s}$ [TeV] & $\sigma^{\rm LO}_{q \bar{q}'\to t H \bar{b}}$ [fb] &
      $\sigma^{\rm LO}_{q b\to t H q'}$ [fb] &  $\sigma^{\rm
        LO}_{q g\to t H \bar{b} q'}$ [fb] \\\hline
      $13$ & $1.97$ & $60.54$ & $24.90$ \\\hline
      $14$ & $2.17$ & $72.68$ & $30.90$ \\\hline
      $33$ & $6.34$ & $475.8$ & $216.0$ \\\hline
      $100$ & $19.61$ & $3453$ & $1676$ \\\hline
\end{tabular}
\caption[]{The total cross section (in fb) at LO in QCD for the two
  Higgs subleading production channels $qb \to tH j$ and  $qg \to
  tHjb$ as well as for the subleading production channel $q\bar{q}'
  \to t H \bar{b}$, at a few c.m. energies (in TeV) for $M_H=125$ GeV
  at the central scale $\mu_R=\mu_F=\frac14 (M_H + M_t)$. The
  five-flavour scheme has been used and a sum over the 
  final states $ t H + X$ and $\bar{t} H + \bar{X}$ is implicit.}
\label{table:ht}
\end{center}
\vspace*{-6mm}
\end{table}


\subsection{SM multi--Higgs production}

We will briefly review in this section the results for the pair production
of the SM Higgs boson, that allows for the determination of the
important trilinear couplings among the Higgs states. These were
already obtained up to 100 TeV in a previous
publication~\cite{Baglio:2012np} (which itself was an update of the
older analyses of Ref.~\cite{Djouadi:1999rca,Djouadi:1999gv} made for
the 14 TeV LHC and for an electron-positron collider) and we 
update in particular the gluon fusion channel in which new results
have appeared quite recently and the $t\bar{t}HH$ channels. We will
present both the cross sections 
at the various c.m. energies and the theoretical uncertainties, the
calculation of which follows the main lines already presented in the
previous subsection for single Higgs production in the dominant
channels. The case of triple Higgs production has already been studied
in the literature with hadronic energies up to 200 TeV and it has been
found that the cross section is so small for such c.m. energies that
it will be extremely difficult to observe
it~\cite{Plehn:2005nk,Binoth:2006ym,Maltoni:2014eza}. Nevertheless, a very
recent study \cite{Papaefstathiou:2015paa} which relies on very optimistic 
assumptions about the experimental setup,  considers this process to be 
interesting at 100 TeV despite the aforementioned limitations.
This calls  for further
investigations in the context of the FCC-hh (see also
Refs.~\cite{Chen:2015gva,Fuks:2015hna}) and with the theoretical
progresses as well as experimental improvements, one might hope for a
potential observation of this process at 100 TeV.
Therefore, in this section, although we will not include a
complete and detailed analysis, we will give the rates for the largely dominant 
 triple production  process $gg\to HHH$ for energies up to 
100 TeV.

\subsubsection{Status of the higher order corrections in Higgs pair
  production}


The gluon fusion channel, $gg\to HH$, is the dominant Higgs pair
production process, much like in the same way as it is the dominant
process for single Higgs production. It is also mediated by loops of heavy
quarks which are of two types in this case: triangular in the case of 
the diagram in which a virtual $H$ bosons is produced and splits to
real Higgs bosons (and which involves the trilinear Higgs coupling
that needs to be determined) and box--type in the case of the diagrams 
in which both Higgs bosons are emitted from the internal quark lines 
(and hence, do not involve the triple Higgs couplings and can be 
considered as an irreducible background in the measurement of the latter). 
Contrary to single Higgs production, bottom quark loops have a negligible 
contribution to the total cross section in this case, less than $1\%$ at
LO~\cite{Eboli:1987dy,Glover:1987nx,Dicus:1987ic,Plehn:1996wb}. 

The process was known for a long time at NLO QCD in the infinite top mass
approximation~\cite{Dawson:1998py}, and a  progress was made only 
quite recently, a progress that pushed this process to NNLO QCD in 
again the infinite top
mass approximation~\cite{deFlorian:2013uza,deFlorian:2013jea}. The
study of the NLO QCD structure has also made progress and top mass
expansion was performed in 2013~\cite{Grigo:2013rya} while the exact
real corrections were computed in the {\tt aMC@NLO}
framework~\cite{Frederix:2014hta,Maltoni:2014eza}. A full NLO
calculation including the exact quark mass dependence is still missing
though. Soft--gluon resummation was also performed up to NNLL
accuracy, first matched with the fixed order NLO
result~\cite{Shao:2013bz} and then matched with the NNLO
result~\cite{deFlorian:2015moa} where it has been found that the
resummation increases the cross section over the NNLO result by $\sim
+ 7\%$ at 14 TeV while reducing the scale uncertainty down to $\sim
\pm 5 \%$.


The VBF process, $qq \to V^* V^* qq \to HHqq$,   has also been known for 
quite a while only at
LO~\cite{Keung:1987nw,Eboli:1987dy,Dicus:1987ic,Dobrovolskaya:1990kx}. It
is very similar to the single Higgs production case, and the NLO QCD
corrections follow essentially the same trend and can be derived
simply by turning the tensor structure of the fusion of the weak
bosons from $V^* V \to H$ into $V^* V \to H H$ while using the exact
same QCD corrections to the quark lines as for single Higgs
production. These NLO QCD corrections were calculated quite
recently~\cite{Baglio:2012np,Frederix:2014hta} and are now available
in two different codes not only for the total rates but also for the
differential distributions~\cite{Baglio:2014uba,Frederix:2014hta}. The
NLO QCD corrections amount to a $\sim +7\%$ correction. The
approximate NNLO QCD corrections have also been computed in the
structure function approach~\cite{Liu-Sheng:2014gxa} and similarly to
the case of single Higgs production the increase of the rate is very
modest, half a percent at most, while it decreases the scale
uncertainty down to the percent level for the inclusive cross
section.


The production of a Higgs pair in association with a weak boson,
$q\bar q \to HHV$,  was calculated for the first time a while
ago~\cite{Barger:1988jk} and shares common aspects with the single
Higgs--strahlung process. In particular, the NLO and NNLO QCD
corrections can be implemented in complete analogy to the single Higgs
process, i.e. by adapting the corrections to the Drell--Yan
mechanism~\cite{Altarelli:1979ub,Hamberg:1990np} to this case. In
addition, one needs to consider again a new subchannel at NNLO for
$ZHH$ production, namely $gg\to ZHH$ that proceeds through triangle,
box and pentagons loops of heavy quarks. In sharp contrast to single
Higgs production, this subchannel give a  significantly more important
contribution which can amount to 30\% of the NNLO total
rate~\cite{Baglio:2012np}. The NLO QCD corrections were calculated in
Ref.~\cite{Baglio:2012np,Frederix:2014hta} and are available also for
the differential distributions while the NNLO QCD corrections are only
available for the total rates~\cite{Baglio:2012np}.


The last double Higgs production channel that we discuss is the 
associated production with a top--quark pair. This process has been
known only at LO and it is only recently that the NLO QCD
corrections have been made available~\cite{Frederix:2014hta}. This is
due to the complexity of the calculation that needs modern tools as
one has to calculate QCD corrections to a $2\to 4$ process involving
massive quarks in the final states. Depending on the chosen central
scale, they can reduce or enhance the total rate. In our case we have
chosen the central scale $\mu_0 = \mu_R = \mu_F = m_t +\frac12 M_{HH}$
and we see a very modest increase of the rates over the LO
results at 14 TeV, while the increase is larger at 100 TeV, of the
order of $+ 10\%$.

\subsubsection{Numerical results for pair production}

We present our numerical results, an update of
Ref.~\cite{Baglio:2012np}, using the following central scales:
\begin{align}
\mu_0^{gg\to HH} = M_{HH},\,\, \mu_0^{HHqq'} = Q_V^*,\,\,
\mu_0^{ V HH} = M_{V\! H H},\,\,  \mu_0^{t
  \bar{t}H H} = m_t + \frac12 M_{HH}.~~ 
\end{align}

We use the following program and procedures. For gluon fusion we use
the program {\tt HPAIR}~\cite{Michael-web,Spira:1997dg}  on top of
which we account for the NNLO QCD corrections in an approximate way by
multiplying our result by the following ratio $K^{\rm
  rescale}$~\cite{deFlorian:2013jea}:
\begin{align}
K^{\rm rescale}(\sqrt{s})  = 1.15 -0.33/ \sqrt{s} +
{0.33}/{s^{1/4}},
\end{align}
where $\sqrt{s}$ is the c.m. energy given in TeV. We also account for
the scale uncertainty using the formula (21) in
Ref.~\cite{deFlorian:2013jea}. The uncertainty related to the use of
the effective approach for the top loop is still kept this time
contrary to the single Higgs case as it has significant impact. For
the Higgs--strahlung and VBF processes, we simply collect our previous
results obtained in Ref.~\cite{Baglio:2012np} and replace the 8 TeV
LHC predictions by the case of the LHC at $\sqrt{s}=13$ TeV. The VBF
results are obtained using {\tt VBFNLO}~\cite{Baglio:2014uba} while
the Higgs--strahlung predictions are obtained with a computer program
of our own that was developed especially for this purpose. The results
for $t\bar{t}HH$ have been obtained \footnote{We thank Eleni
Vryonidou for having kindly provided the numbers.} with
MadGraph5/aMC@NLO~\cite{Alwall:2014hca}. The reduction of the scale
uncertainty is spectacular, from $\sim +30\% / -20\%$ down to $+2\%
/-7\%$ at 14 TeV.

\begin{figure}[!t]
\centering
\vspace*{-3mm}
\includegraphics[scale=0.9]{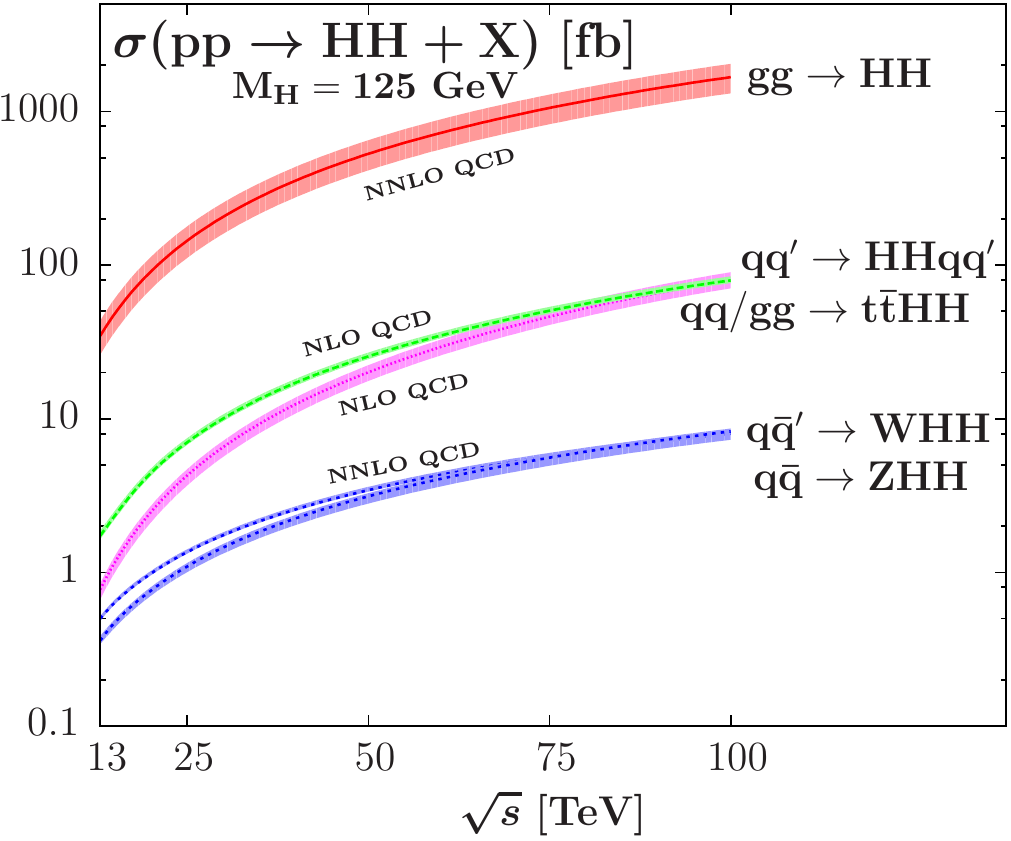}
\caption[]{The total cross sections for Higgs pair production at a
  proton-proton collider, including higher-order corrections discussed
  in the text, in the main production channels 
  as a function of the c.m. energy with $M_H = 125$ GeV. The MSTW2008
  PDF set has been used and theoretical uncertainties are included as
  corresponding bands around the central values.}
\label{fig:HH}
\vspace*{-1mm}
\end{figure}

The results are displayed in Fig.~\ref{fig:HH} as a function of the
c.m. energy starting from LHC 13 TeV up to the FCC--hh energy of 100
TeV. As stated above this clearly shows that the gluon fusion channel
is the dominant production on the full c.m. energy range as in single
Higgs production. This is followed by the VBF channel up to 100 TeV, but 
in this region it  is actually the associated production with a
top quark pair which dominates over VBF. Higgs--strahlung channels are
one order of magnitude smaller than VBF over the whole c.m. energy
range. The numerical results are summarised in Table~\ref{table:hh_lhc} 
for all channels. Hence, clearly, the gluon fusion channel will be the
most relevant channel at 100 TeV in much the same way as at 14
TeV. Nevertheless the VBF contribution can be important in the
analyses including two jets in the final state and is already included
in various 14 TeV analyses~\cite{Dolan:2013rja,Dolan:2015zja}, and
given the higher number of events at 100 TeV, this VBF contribution
could be even more interesting.

\begin{table}[t!]
 \renewcommand{\arraystretch}{1.1}
  \begin{center}
   \small
\begin{tabular}{|c|c|ccccc|}\hline
channel &      $\sqrt{s}$ [TeV] & $\sigma$ [fb] & scale
      [\%] & PDF [\%] & PDF+$\alpha_s$ [\%] &
       total [\%] \\\hline
$gg\to HH$ &      $13$ & $34.56$ & ${+8.5}\;\;\;{-8.5}$ & ${+4.0}\;\;\;{-4.0}$ &
      ${+7.1}\;\;\;{-6.2}$ & ${+25.6}\;\;\;{-24.7}$
      \\
&      $14$ & $41.11$ & ${+8.3}\;\;\;{-8.3}$ & ${+3.9}\;\;\;{-4.0}$ &
      ${+7.0}\;\;\;{-6.2}$ &  ${+25.3}\;\;\;{-24.5}$
      \\
&      $33$ & $247.93$ & ${+7.0}\;\;\;{-7.0}$ & ${+2.5}\;\;\;{-2.7}$ &
      ${+6.2}\;\;\;{-5.4}$ & ${+23.2}\;\;\;{-22.4}$
      \\
&      $100$ & $1670.83$ & ${+5.8}\;\;\;{-5.8}$ & ${+2.0}\;\;\;{-2.7}$ &
      ${+6.2}\;\;\;{-5.7}$ & ${+22.0}\;\;\;{-21.5}$
      \\\hline
VBF$\to HH$   &
      $13$ & $1.73$  & ${+1.7}\;\;\;{-1.1 }$ & ${+4.6}\;\;\;{-4.2}$&
      ${+5.9}\;\;\;{-4.2}$ & ${+7.6}\;\;\;{-5.2}$ \\
&      $14$ & $2.01$  & ${+1.7}\;\;\;{-1.1 }$ & ${+4.6}\;\;\;{-4.1}$&
      ${+5.9}\;\;\;{-4.1}$ & ${+7.6}\;\;\;{-5.1}$ \\
&      $33$ & $12.05$  & ${+0.9}\;\;\;{-0.5 }$ & ${+4.0}\;\;\;{-3.7}$&
      ${+5.2}\;\;\;{-3.7}$ & ${+6.1}\;\;\;{-4.2}$ \\
&      $100$ & $79.55$  & ${+1.0}\;\;\;{-0.9 }$ & ${+3.5}\;\;\;{-3.2}$&
      ${+5.2}\;\;\;{-3.2}$ & ${+6.2}\;\;\;{-4.1}$ \\ \hline
$WHH$ &
      $13$ & $0.50$  & ${+0.1}\;\;\;{-0.3 }$ & ${+3.6}\;\;\;{-2.9}$&
      ${+3.6}\;\;\;{-3.0}$ & ${+3.7}\;\;\;{-3.3}$ \\
&      $14$ & $0.57$  & ${+0.1}\;\;\;{-0.3 }$ & ${+3.6}\;\;\;{-2.9}$&
      ${+3.6}\;\;\;{-3.0}$ & ${+3.7}\;\;\;{-3.3}$ \\
&      $33$ & $1.99$  & ${+0.1}\;\;\;{-0.1 }$ & ${+2.9}\;\;\;{-2.5}$&
      ${+3.4}\;\;\;{-3.0}$ & ${+3.5}\;\;\;{-3.1}$ \\
&      $100$ & $8.00$  & ${+0.3}\;\;\;{-0.3 }$ & ${+2.7}\;\;\;{-2.7}$&
      ${+3.8}\;\;\;{-3.4}$ & ${+4.2}\;\;\;{-3.7}$ \\ \hline
$ZHH$ &
      $13$ & $0.36$  & ${+4.0}\;\;\;{-2.9 }$ & ${+2.8}\;\;\;{-2.3}$&
      ${+3.0}\;\;\;{-2.6}$ & ${+7.0}\;\;\;{-5.5}$ \\
&      $14$ & $0.42$  & ${+4.0}\;\;\;{-2.9 }$ & ${+2.8}\;\;\;{-2.3}$&
      ${+3.0}\;\;\;{-2.6}$ & ${+7.0}\;\;\;{-5.5}$ \\
&      $33$ & $1.68$  & ${+5.1}\;\;\;{-4.1 }$ & ${+1.9}\;\;\;{-1.5}$&
      ${+2.7}\;\;\;{-2.6}$ & ${+7.9}\;\;\;{-6.7}$ \\
&      $100$ & $8.27$  & ${+5.2}\;\;\;{-4.7 }$ & ${+1.9}\;\;\;{-2.1}$&
      ${+3.2}\;\;\;{-3.2}$ & ${+8.4}\;\;\;{-8.0}$ \\ \hline
$t\bar{t} HH$ &
      $13$ & $0.77$  & ${+2.4}\;\;\;{-7.3 }$ & ${+3.8}\;\;\;{-4.3}$&
      ${+6.7}\;\;\;{-6.6}$ & ${+9.2}\;\;\;{-13.9}$ \\
&      $14$ & $0.95$  & ${+3.1}\;\;\;{-7.5 }$ & ${+3.6}\;\;\;{-4.3}$&
      ${+6.5}\;\;\;{-6.6}$ & ${+9.6}\;\;\;{-14.1}$ \\
&      $33$ & $8.13$  & ${+6.9}\;\;\;{-8.2}$ & ${+2.5}\;\;\;{-3.0}$&
      ${+6.2}\;\;\;{-6.0}$ & ${+13.1}\;\;\;{-14.2}$ \\
&      $100$ & $79.86$  & ${+7.4}\;\;\;{-7.5 }$ & ${+1.6}\;\;\;{-2.0}$&
      ${+5.1}\;\;\;{-4.2}$ & ${+12.5}\;\;\;{-11.6}$ \\ \hline
\end{tabular}
\caption[]{The total Higgs pair production cross section (in fb) for
  given c.m. energies (in TeV) for $M_H=125$ GeV.  The corresponding
  shifts due to the theoretical uncertainties from the various sources
  are shown as well as the total uncertainty when all errors are added
  linearly. In the case of $gg\to HH$ an EFT uncertainty of order $\pm
  10\%$ has to be added. The central scales used for each process are
  defined in the text.}
\label{table:hh_lhc}
\end{center}
\vspace*{-6mm}
\end{table}


 \subsubsection{Expectations for triple Higgs production}

As stated already in the
literature~\cite{Plehn:2005nk,Binoth:2006ym,Maltoni:2014eza} the cross
section for triple Higgs production at at the LHC is very small (note
that the results in Ref.~\cite{Maltoni:2014eza} include approximate
NLO QCD corrections). We display in Table~\ref{table:hhh_lhc} 
the production rates in the dominant $gg\to HHH$ channel at a central 
scale $\mu_R=\mu_F = M_{HHH}$ for the parameter set of Eq.~(\ref{inputs}).
We have used our own implementation and checked our code against the
numbers given in Ref.~\cite{Binoth:2006ym} and found full agreement. The rates 
are indeed negligible at the LHC being three orders of magnitude smaller
than pair production. Nevertheless, at 100 TeV, the cross section reaches 
the level of 3 fb (maybe a factor 2 higher if the $K$--factors are the same 
as in single and pair production) and could thus lead to a few
thousand events with a  luminosity of 1 ab$^{-1}$. The extraction of
the signal, in particular in the $6b$ \cite{Papaefstathiou:2015paa},  
$4b2\gamma$ \cite{Chen:2015gva} and $4b2\tau$ \cite{Fuks:2015hna} 
detection channels,  requires very large integrated luminosities and 
is nevertheless extremely challenging in view of the formidable
backgrounds.

\begin{table}[h!]
\vspace*{-.2mm}
 \renewcommand{\arraystretch}{1.1}
  \begin{center}
   \small

\begin{tabular}{|ccccc|}\hline
$\sqrt{s}$ = & 13 TeV & 14 TeV & 33 TeV & 100 TeV\\\hline  
$\sigma^{\rm LO}_{gg\to HHH}$ [fb] &  $0.033$ & $0.040$  & $0.33$  & $3$ \\ \hline
\end{tabular}
\vspace*{-.1mm}
\caption[]{The total triple Higgs production cross section $gg\to HHH$
  (in fb) for given c.m. energies (in TeV) for $M_H=125$ GeV. The
  central scale is $\mu_R = \mu_F = M_{HHH}$.}
\label{table:hhh_lhc}
\end{center}
\vspace*{-5mm}
\end{table}


\section{Dark matter and the Higgs--portal}

\subsection{Higgs--portal dark matter models}

A very interesting scenario would be that the particles that form
partly or entirely the dark matter in the universe interact only with
the Higgs sector~\cite{Silveira:1985rk,McDonald:1993ex,Burgess:2000yq}
(see also Ref.~\cite{Mambrini:2011ik} for a review and more
references). The DM particles can be made stable by a $Z_2$ symmetry
and annihilate to SM states through the exchange of Higgs
bosons. These Higgs portal scenarios for DM can be of several kinds,
depending on whether the models contain additional Higgs and/or matter
particles or not, but the simplest one would be the scenario in which
the SM is extended to contain only the DM particle while its minimal
Higgs sector with one scalar doublet is kept unchanged. The DM
particles will then interact only with the observed $h$ state and
their annihilation into SM fermions and bosons, for instance, can
occur only through the $h$ exchange in the $s$--channel.

To describe the DM properties in this minimal Higgs-portal scenario,
it is convenient to work in a quite model--independent framework
(although the origin of the $Z_2$ parity that ensures the stability of
the DM particle is still model--dependent) in which the particle
consists of a real scalar $S$, a vector $V$ or a Majorana fermion $f$
that interacts with the SM fields only through the $h$
state~\cite{Kanemura:2010sh,Djouadi:2011aa,Djouadi:2012zc}. Hence, the
phenomenology of the model is described only by two additional
parameters in addition to those of the SM. These are, besides the
three spin assignment possibilities, the mass of the DM state and its
coupling to the $h$ boson. The relevant terms in the Lagrangians
describing the spin--0, spin--$\frac12$ and spin--1 cases have a
general form that can be simply written as
\begin{eqnarray}
\label{Lag:DM}
\!&&\Delta {\cal L}_S = -\frac12 m_S^2 S^2 -
  \frac14 \lambda_S S^4 - \frac14 \lambda_{hSS}  H^\dagger H
  S^2 \;, \nonumber \\
\!&&\Delta {\cal L}_V = \frac12 m_V^2 V_\mu V^\mu\! +\! \frac14
\lambda_{V}  (V_\mu V^\mu)^2\! +\! \frac14 \lambda_{hVV}  H^\dagger
H V_\mu V^\mu , \nonumber \\ \! 
&&\Delta {\cal L}_f = - \frac12 m_f \bar f f -  \frac14
{\lambda_{hff}\over \Lambda} H^\dagger H \bar f f \;.
\end{eqnarray}
where the self--interaction terms $S^4$ in the scalar and the $(V_\mu
V^\mu)^2$ term in the vector cases are not essential for our
discussion and can be ignored. In the fermionic case, the form that we
adopt here for the Higgs--DM coupling is not renormalisable, but as it
is a rather convenient parametrisation, we keep it in our
discussion\footnote{One could have a renormalisable Higgs coupling to
  spin--$\frac12$ Majorana DM particles similar to the one above and a
  good example would be the case of the MSSM where the $h\chi\chi$
  coupling (as well as the physical mass $M_\chi$) can be defined in
  terms of the elements of the matrix that diagonalizes the $4\times
  4$ mass matrix of the four neutralino states (the bino, wino and the
  two higgsinos) with the lightest one being identified with the DM
  particle. In the case where all the additional Higgs bosons except
  for $h$ and all the SUSY particles except for the stable lightest
  neutralino are very heavy, one would be in a
  situation~\cite{Gunion:1987yh,Djouadi:1992pu,Djouadi:1996mj,Djouadi:1996pj}
  that is quite similar to the $h$--portal fermionic DM scenario that
  we are discussing here.}. After electroweak symmetry breaking when
the neutral component of the doublet field $H$ is shifted to $(v + h)/
\sqrt{2}$ with $v = 246$ GeV, the physical masses of the DM particles
will be given by
\begin{eqnarray}
&& M_S^2 = m_S^2 + \frac{1}{4}\lambda_{hSS} v^2 \;, \nonumber \\ 
&&  M_V^2 = m_V^2 + \frac{1}{4}\lambda_{hVV} v^2 \;, \nonumber \\
&& M_f = m_f + \frac{1}{4}{\lambda_{hff}\over \Lambda} v^2 \;.
\end{eqnarray}

Since the cosmological relic density of the DM particles is obtained
by means  of the annihilation to SM particles through the exchange of
the $h$ boson, there is in principle a relation between the coupling 
$\lambda_{h\chi  \chi}$  and the mass $M_\chi$ of the $\chi$ DM particle 
if the Planck satellite constraints, $\Omega_{\rm DM} h^2 = 0.1186 \pm
0.020$~\cite{Ade:2015xua} with $h$ being the reduced Hubble constant,
is imposed. However, this is only true if the $\chi$ particle is 
absolutely stable and has to account for all the DM in the universe.
For a more general discussion in the context of collider physics that
we are interested in, we will ignore this constraint. Nevertheless,
there are also constraints from the rates for the direct and indirect
detection of the $\chi$ particles in astrophysical experiments. In
particular, the elastic DM interaction with nuclei occurs through the
$t,u$--channel exchange of the $h$ boson and the resulting nuclear
recoil or the spin--independent DM--nucleon cross section can be
interpreted in terms of the DM mass and coupling; see for instance
Ref.~\cite{Djouadi:2011aa} in which the present constraints from the
two most sensitive experiments XENON100  \cite{Aprile:2012nq} and LUX
\cite{Akerib:2013tjd} have been summarised. 

The issue that we will be concerned here is how to observe directly
the Higgs-portal DM particles at high energy proton colliders. There
are essentially two ways, depending on the $h$ versus $\chi$ particle
masses. If these particles are light enough, $M_\chi \lsim \frac12
M_h$, the invisible $h \to \chi \chi$ decay can occur and its partial
width will contribute to the total decay width of the observed Higgs
boson. It will then alter the branching ratios for the visible decays
and hence the signal rates in the various channels in which the $h$
boson has been detected at the LHC. In the previous LHC runs with
$\sqrt s=$7+8 TeV, the Higgs cross sections times branching ratios in
some channels like $pp\to h\to \gamma\gamma$ and $h\to ZZ\to
4\ell^\pm$ have been measured at the $\approx 20\%$ level and, as they
agree with the SM expectation, they set an upper bound ${\rm BR(h \to
  inv.)} \lsim 0.3$ for SM--like $h$ couplings; see
e.g. Ref.~\cite{Djouadi:2013qya}. The sensitivity will certainly
improve in the coming LHC run at $\sqrt s=13$ and 14 TeV, but the
observation of the invisible Higgs branching ratio would be extremely
difficult if ${\rm BR(h \to inv.)} \lsim 0.1$, in view of the large
QCD uncertainties that affect the Higgs production cross sections, in
particular in the main $gg\to h$ production channel, as discussed
earlier\footnote{The indirect observation of invisible Higgs decays
  would be much easier at a future $e^+e^-$ collider and it has been
  shown that at $\sqrt s \approx 500$ GeV with a sample 100 fb$^{-1}$
  data, the Higgs cross section times the visible branching  fractions
  can be determined with a percent level
  accuracy~\cite{Djouadi:1994mr,Accomando:1997wt,AguilarSaavedra:2001rg}.}.

On the other hand, DM particles could be directly detected by studying
the vector boson fusion and the Higgs--strahlung processes in which
the Higgs boson decays
invisibly~\cite{Choudhury:1993hv,Eboli:2000ze}. The ATLAS and CMS
searches at the 7+8 TeV run in these missing energy channels give the
constraint ${\rm BR(h \to inv.)} \lsim 30\%
$~\cite{ATLAS:2013pma,CMS:2013bfa} if the $h$ production and visible
decays rates are SM--like. From  parton level analyses, one does not
expect that this limit on the invisible branching ratio will be
significantly improved at the 14 TeV LHC upgrade even with a
sufficiently large amount of data\footnote{Here again, the situation
  is more favourable at an  $e^+e^-$ collider with a c.m. energy $\sqrt
  s \gsim 250$ GeV, as invisible decays  at the level of a few percent
  can be observed in the process $e^+e^- \to hZ$ by simply analysing
  the recoil of the leptonically decaying $Z$
  boson~\cite{Djouadi:1994mr,Accomando:1997wt,AguilarSaavedra:2001rg}.}~\cite{Godbole:2003it}.

In turn, if the mass of the DM particle is larger than half the $h$
mass, $M_\chi  \gsim \frac12 M_h$, there is no invisible Higgs decay
and the detection of the $\chi$ particles in collider experiments
becomes much more difficult. In fact, the only possible way to observe
the $\chi$ states would be through their pair production in the
continuum via the exchange of the Higgs
boson~\cite{Djouadi:2012zc,Quevillon:2014owa,Craig:2014lda}. The latter needs
to be produced in association with visible  particles and at hadron
colliders, three main processes are at hand:  $i)$ double production
in the Higgs--strahlung process $q\bar{q} \to  V^* \to V \chi \chi$
with $V$ being either a $W$ or a  $Z$ boson, $ii)$ the vector boson
fusion processes which lead to two jets and missing energy $qq \to V^*
V^* qq  \to \chi \chi qq$ in the final state and $iii)$ the gluon
fusion mechanism which is mainly mediated by loops of the heavy top
quark that couples strongly to the Higgs boson, $gg \to h^* \to \chi
\chi$, but in which additional jets are emitted in the final state in
order make the process visible.
   

\subsection{The cross sections for DM production through the Higgs
  portal}


We now present numerical results for the DM pair production cross
sections through Higgs splitting in the three possible processes
discussed before. Results are shown for a proton collider at $\sqrt
s\!=\! 100$ TeV and compared with those obtained at $\sqrt s\!=\! 14$
TeV. 

\begin{figure}[!ht]
\vspace*{-1mm}
\centering
\mbox{
\hspace{-1.0cm}
\includegraphics[scale=0.7]{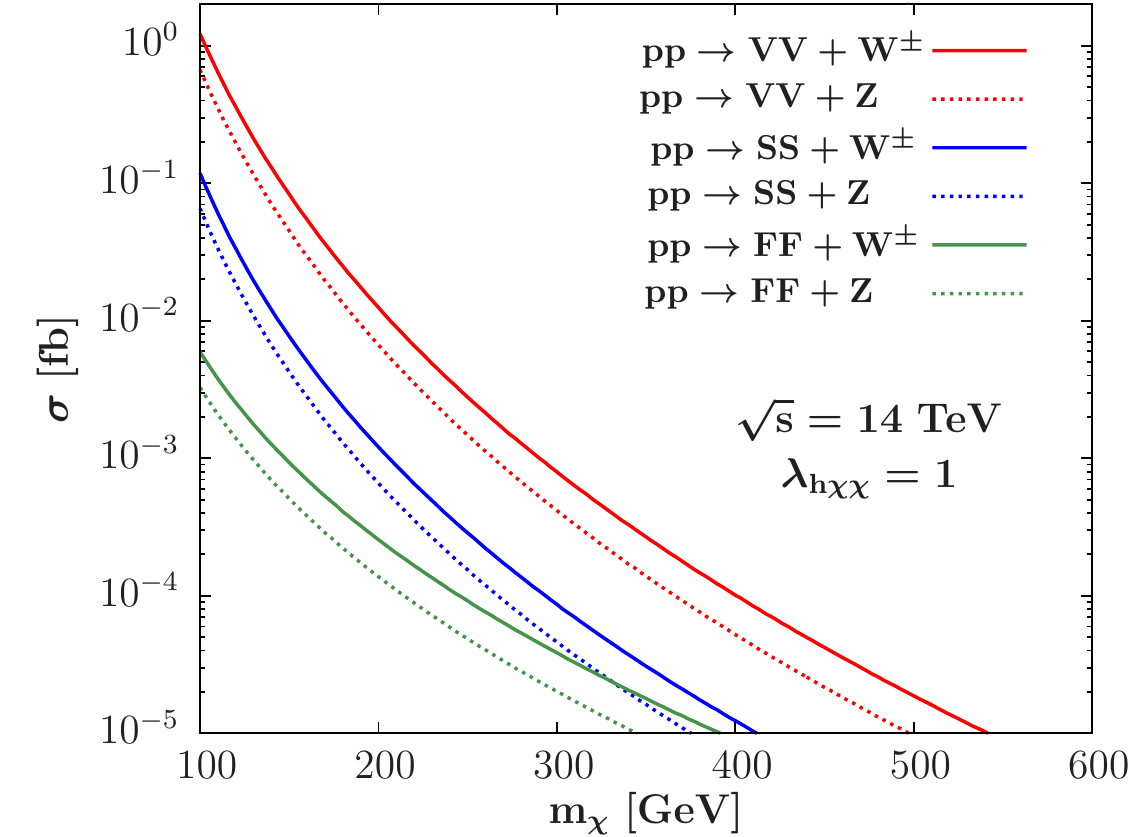}\hspace{-3mm}
\includegraphics[scale=0.7]{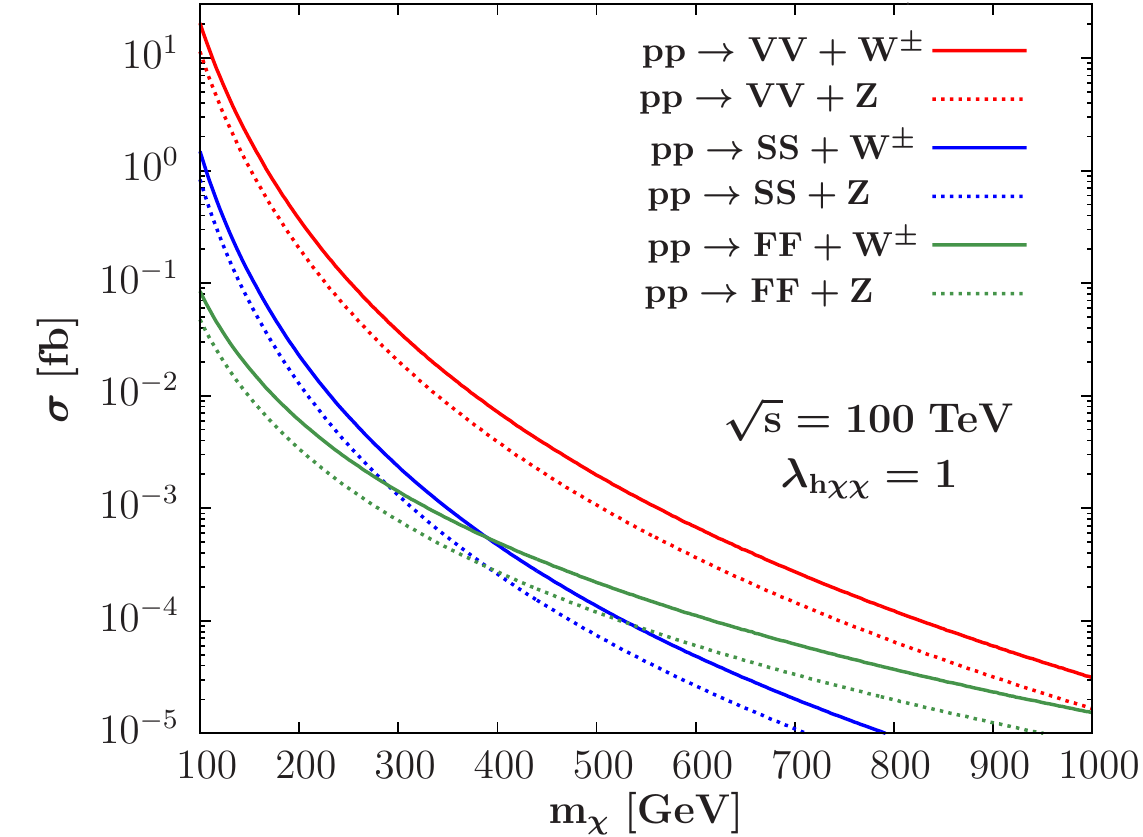}
}
\vspace*{-7mm}
\caption[]{DM pair production cross sections in the Higgs--strahlung process with $W$ and $Z$
boson final states and missing energy for c..m. energies of $\sqrt{s}=14$ TeV (left) and 
100 TeV (right) as a function of the DM mass  $M_\chi$ for a scalar, fermionic
and vectorial particles with $\lambda_{h\chi \chi}=1$.}
\label{prodDM-HV}
\end{figure}

For DM double production in the Higgs--strahlung process with either a
$W$ or a $Z$ boson, we have used the package
Feynrules~\cite{Alloul:2013bka} in which we implemented the Lagrangian
Eq.~(\ref{Lag:DM}) describing the DM interactions in the three spin
cases and exported the results into the MadGraph5/aMC@NLO
framework~\cite{Alwall:2014hca}. The cross sections include thus the
NLO QCD corrections and we adopted the MSTW set for the PDFs and the
MadGraph5 default cards (including kinematical cuts). The DM pair
production rates are shown in Fig.~\ref{prodDM-HV}. The left column
corresponds to a energy $\sqrt{s}=14$ and the one on the right
corresponds to 100 TeV. The cross sections are plotted as function of
the mass of the generic DM particle $\chi$ and we have set the DM
couplings to the $h$ portal to $\lambda_{h \chi \chi}=1$ (in the
fermionic case, we also set $\Lambda=1$ TeV)). For other couplings one
simply has to multiply the rates by  $\lambda_{h \chi \chi}^2$.

One see that for the three types of particles the cross sections are
extremely small at the LHC even for $M_\chi=100$ GeV.  For such a
$\chi$ mass, they do not exceed the fb level in the spin--1 case and
they are one and two order of magnitude smaller in, respectively, the
spin--$\frac12$ and spin--0 cases (with the rate for $W\chi \chi$
being twice as large as the one for $Z\chi \chi$ as it is usually the
case for such processes). At $\sqrt s=100$ TeV, there is an increase
of the cross section by almost two orders of magnitude at low
masses. The rates remain thus too low for the processes to be very
interesting, in particular at $M_\chi$ significantly above 100 GeV and
for the spin--0 and $\frac12$ cases.

\begin{figure}[!ht]
\vspace*{-1mm}
\centering
\mbox{
\hspace{-1.0cm}
\includegraphics[scale=0.7]{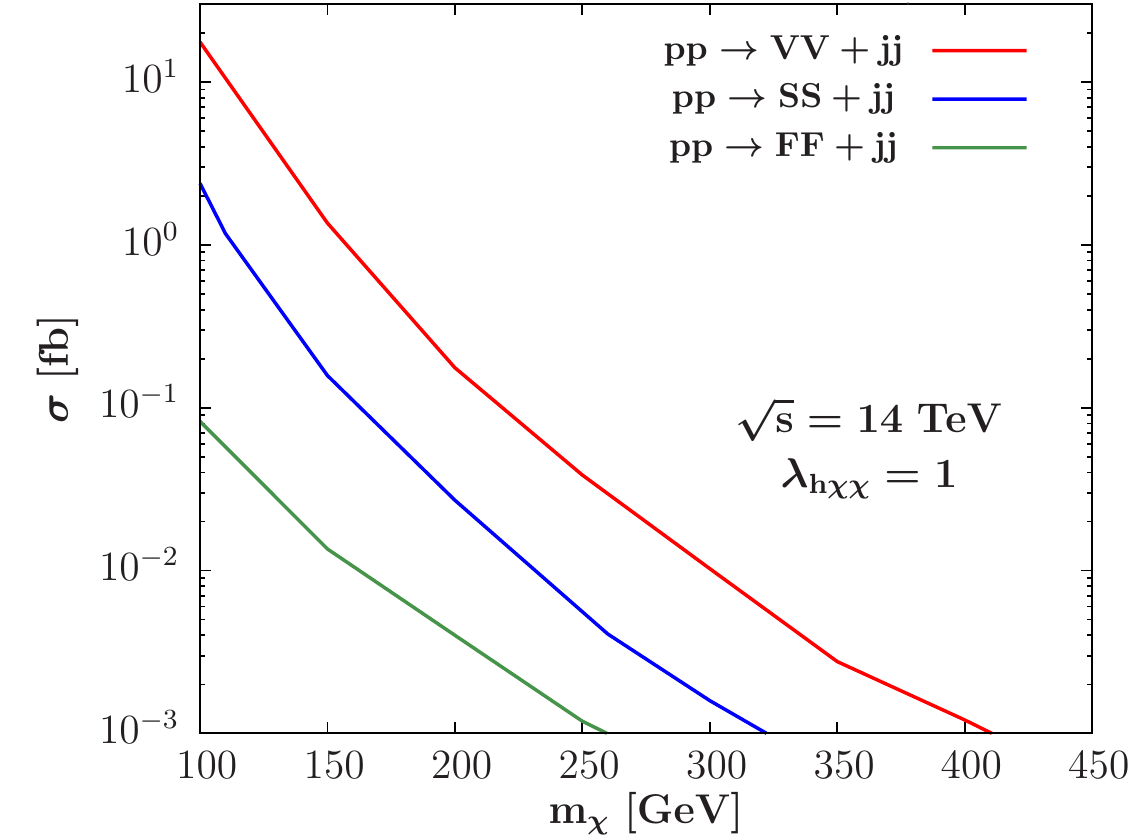}\hspace{-3mm}
\includegraphics[scale=0.7]{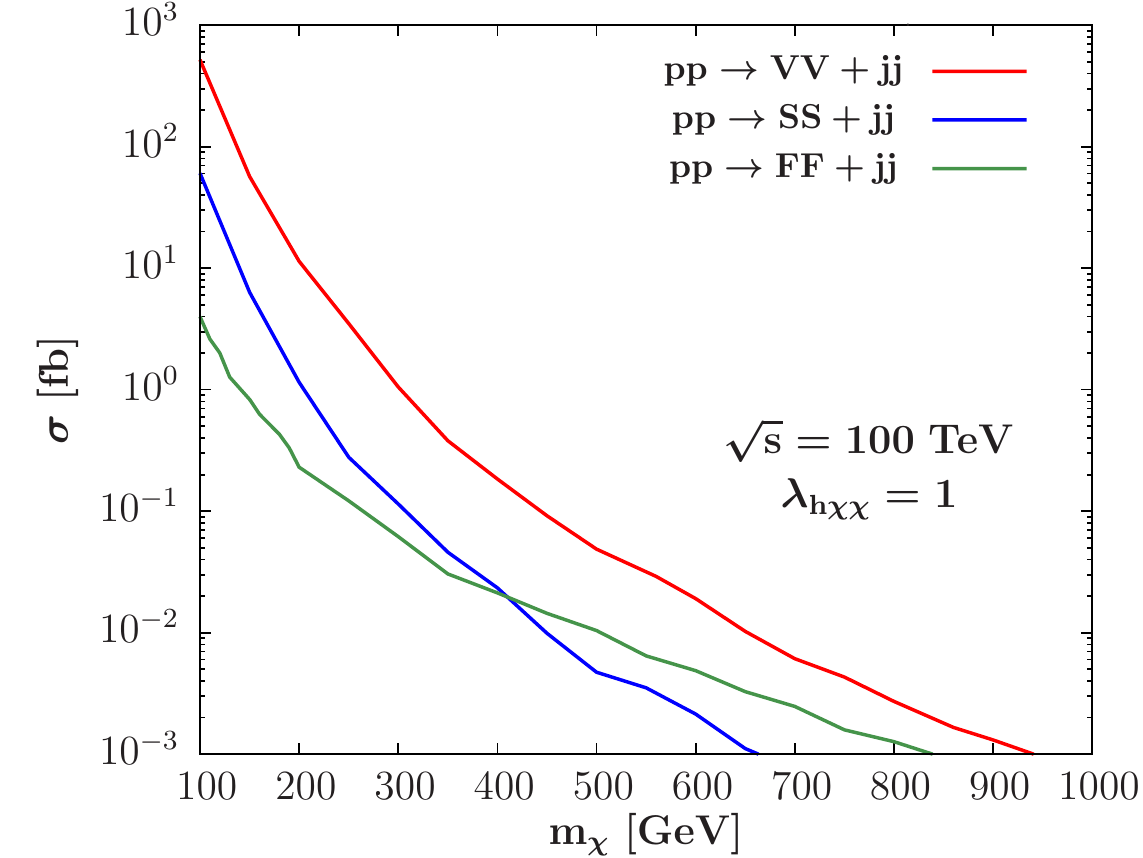}
}
\vspace*{-7mm}
\caption[]{DM pair production cross sections in the vector boson fusion process with 
2 jet final states and missing energy for c..m. energies of $\sqrt{s}=14$ TeV (left) and 
100 TeV (right) as a function of the DM particle mass  $M_\chi$ for a scalar, fermionic
and vectorial particles assuming $\lambda_{h\chi \chi}=1$.}
\label{prodDM-VBF}
\vspace*{-2mm}
\end{figure}

For the vector boson fusion case in which the pair of escaping DM
particle is produced in association with two jets, we also used the
same strategy as above for the calculation of the cross section and we
use Feynrules~\cite{Alloul:2013bka} to export our simple $h$--portal
extension of the SM into the MadGraph5
framework~\cite{Alwall:2014hca}. The cross sections are calculated at
LO only and we again adopt the MSTW PDFs and the MadGraph5 default
cards (but removing the cuts on the jet transverse momentum). The
production rates at $\sqrt s=14$ and 100 TeV are presented in
Fig.~\ref{prodDM-VBF} in the same configuration as Higgs-strahlung
that is, as functions of the mass $M_\chi$, setting the coupling to
$\lambda_{h \chi \chi}=1$ and considering the three spin
configurations.

The cross section are one order of magnitude larger than in the
Higgs--strahlung case for the three spin--configurations and the
hierarchy is the same: one order of magnitude larger for spin--1 than
for spin--0 and than for spin $\frac12$ if $\Lambda=1$ TeV.  Again,
the rates are 100 times larger at $\sqrt s=100$ TeV than at 14 TeV
and, for instance, in the vector case they almost reach the picobarn
level for $M_\chi =100$ GeV and hence, provide a chance to observe the
process. The cross sections fall steeply with the $\chi$ mass and even
in the optimistic spin--1 case, they drop to below the femtobarn level
for $M_\chi =300$ GeV.

\begin{figure}[!h]
\vspace*{-1mm}
\centering
\mbox{
\hspace{-1.0cm}
\includegraphics[scale=0.7]{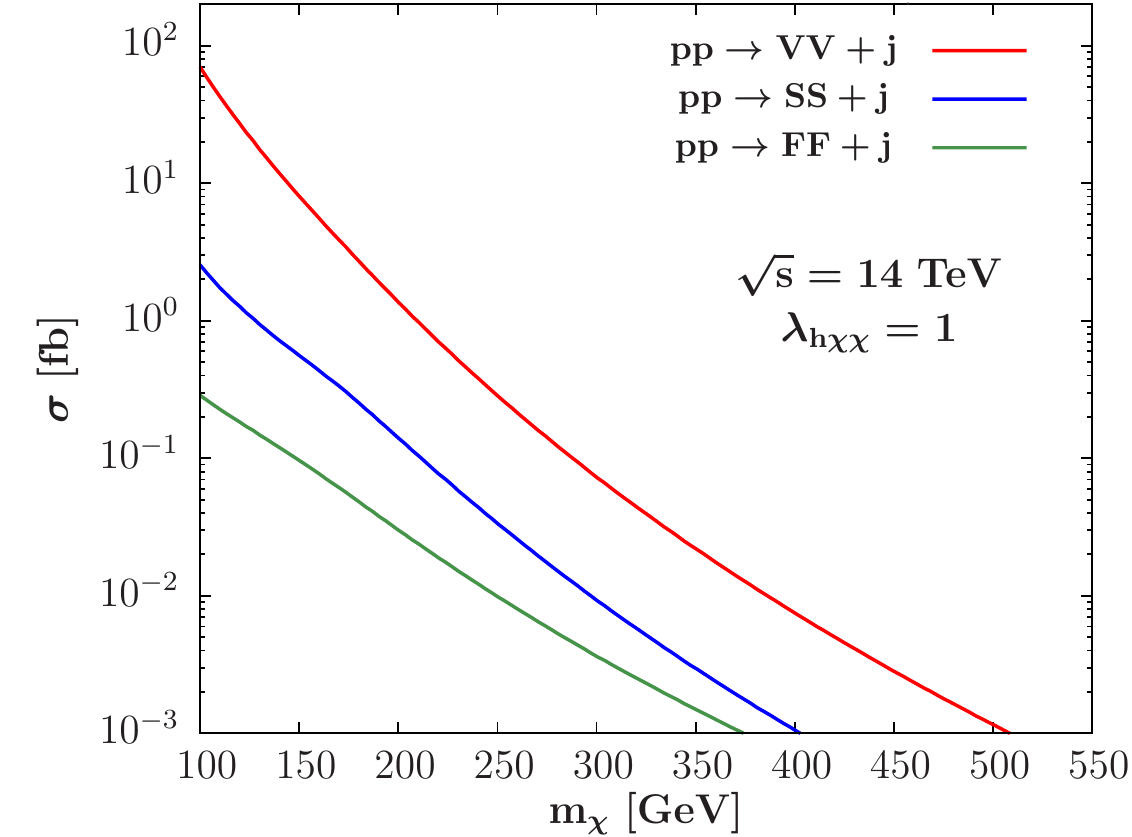}\hspace{-3mm}
\includegraphics[scale=0.7]{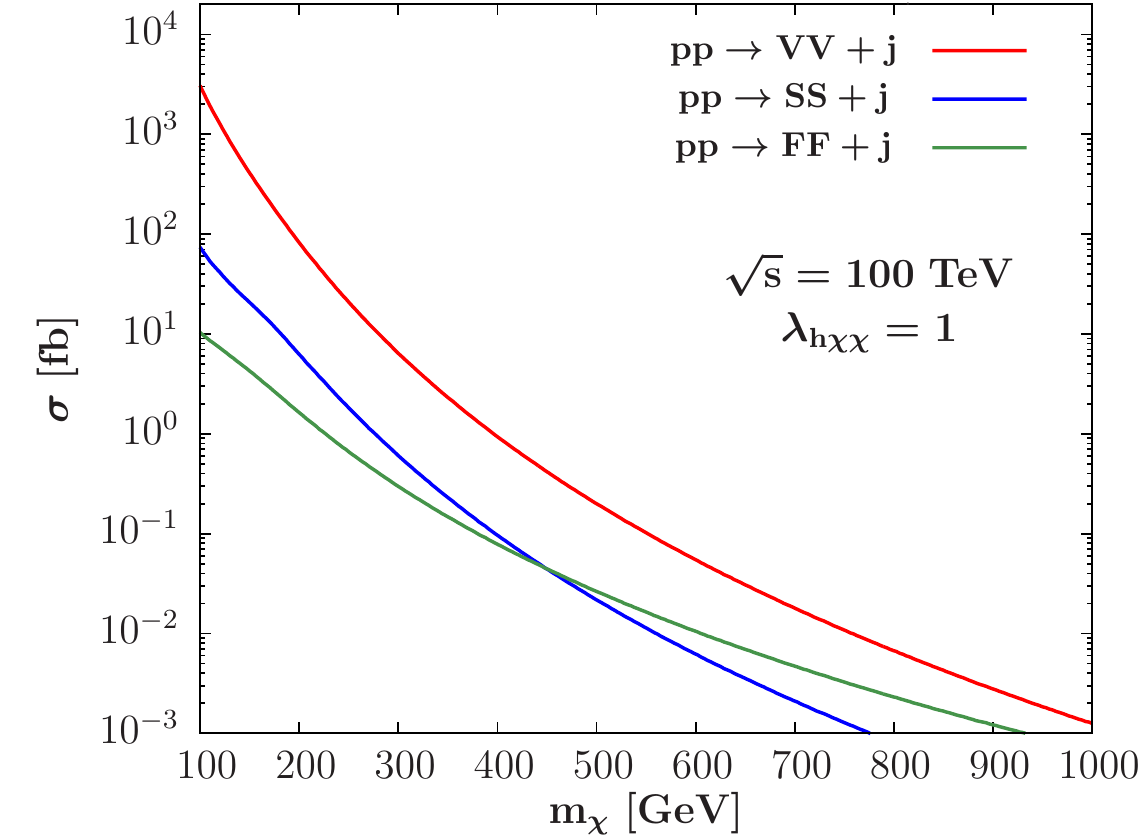}
}
\vspace*{-7mm}
\caption[]{DM pair production cross sections in association with one jet at NLO 
mainly from the gluon fusion process for c..m. energies of $\sqrt{s}=14$ TeV (left) and 
100 TeV (right) as a function of the DM particle mass  $M_\chi$ for a scalar, fermionic
and vectorial particles assuming $\lambda_{h\chi \chi}=1$.}
\label{prodDM-gg}
\vspace*{-1mm}
\end{figure}

Finally, concerning DM double production in association with one jet
either from gluon fusion $gg\to \chi\chi g$ or $qg$ annihilation $gq \to q \chi\chi$,  
we have used a modified version of the program {\tt
  HIGLU}~\cite{Spira:1995mt} that describes Higgs production and where
the couplings of Eq.~(\ref{Lag:DM}) are implemented. The cross
sections include the NLO QCD corrections (which are equivalent to the
NNLO ones in the SM Higgs case and are thus rather large) and we adopt
the MSTW set for the PDFs. The renormalisation and factorisation
scales have been set to $M_{\chi\chi}$. The production rates are shown
in Fig.~\ref{prodDM-gg} again at $\sqrt s=14$ TeV (left) and 100 TeV
(right) in the same configurations as in the two previous processes.

Here, the cross sections are an additional order of magnitude larger
than in the vector boson fusion case (except at very low $M_\chi$ for
the spin--0 case) and  approximately follow the same trend: about a
factor of 50 larger at 100 TeV than at 14 TeV and larger by a factor
of 10  and 100 in the vector case than in, respectively, the scalar
and the fermionic cases. At $\sqrt s=100$ TeV, one has large
production rates at reasonably low $\chi$ masses. This is particularly
the case for a vector DM particle where one obtains a few picobarns in
the low mass range, $M_\chi =100$ GeV. Here again, the cross sections
fall steeply with the $\chi$ mass. 


\section{The heavier Higgs bosons of 2HDMs and the MSSM}

\subsection{The physical set-up}

A very good benchmark for studying extended Higgs sectors are two Higgs doublet models
or 2HDMs for short~\cite{Gunion:1989we,Branco:2011iw}. Compared to the
SM with its unique Higgs particle, the Higgs sector of 2HDMs involves
five physical states after electroweak symmetry breaking and has a
phenomenology that is much richer. Indeed, the model has two CP--even
neutral states $h$ and $H$ that mix and share the properties of the SM
Higgs boson, a  CP--odd or pseudoscalar $A$ boson with properties that
are completely different from that of the SM Higgs and, above all, it
has two charged Higgs states $H^\pm$ that provide a unique signature
for new physics. While the pattern of the couplings of the two doublet
Higgs fields  to the gauge sector is somewhat fixed and relatively
simple, there are several possibilities for the  structure of the
couplings to standard fermions, leading to several types of
2HDMs~\cite{Gunion:1989we,Branco:2011iw}.

The Minimal Supersymmetric Standard Model or MSSM is a specific type
of 2HDM, the so--called type II in which one Higgs field doublet gives mass to up-type fermions and the other gives mass to down-type fermions~\cite{Gunion:1989we}. However, there are strong constraints
on the Higgs sector that reduce the number of free parameters to only
two inputs at tree--level. Nevertheless,  if the SUSY spectrum is
light, a number of complications and new features are introduced as
the superparticles can substantially affect the phenomenology of the
Higgs bosons, either indirectly through loop corrections or new
processes or directly by allowing for instance new production and
decay mechanisms. However, as the LHC data indicate that this scale is
rather high,  $M_S \gsim 1$ TeV, it is very likely that the SUSY
particles will not affect the MSSM Higgs sector making the
phenomenology in this model quite similar to that of 2HDMs.

In both the general 2HDM and the MSSM, the most important production
mechanisms of the neutral CP--even Higgs bosons are simply those of
the SM Higgs particle which have been discussed in detail in the
previous sections of this paper. However, major quantitative
differences compared to the SM case can occur since the cross sections
will depend on the specific Higgs mass and coupling patterns which can
be widely different; this is particularly the case for the
pseudoscalar Higgs boson. In the case of the charged Higgs particles,
new production mechanisms not discussed before occur. Another major
difference between the SM and these models is the Higgs decay pattern
which can be much more involved.

These are the issues that we will discuss in this section in which the
potential of a 100 TeV proton collider to probe this extended Higgs
sector is analysed in detail. But before that, let us summarise the
physical spectrum in these two models.

\subsubsection{The case of 2HDMs}

In our analysis, we will consider the CP--conserving two-Higgs-doublet-model with a 
softly broken $Z_2$ symmetry and here, we briefly highlight its main
features; for more details we refer the reader to
Ref.~\cite{Gunion:2002zf,Baglio:2014nea}, whose approach and notations
we adopt. The scalar potential of this model, in terms of the two
Higgs doublet fields $\Phi_1$ and $\Phi_2$, is described by three mass
parameters and five quartic couplings and is given by
\begin{eqnarray}
 V &=m_{11}^2\Phi_1^\dagger\Phi_1^{\phantom{\dagger}} +m_{22}^2\Phi_2^\dagger\Phi_2^{\phantom{\dagger}} -m_{12}^2 ( \Phi_1^\dagger\Phi_2^{\phantom{\dagger}} +\Phi_2^\dagger\Phi_1^{\phantom{\dagger}}) 
+\tfrac12 \lambda_1(\Phi_1^\dagger\Phi_1^{\phantom{\dagger}})^2
   +\tfrac12 \lambda_2(\Phi_2^\dagger\Phi_2^{\phantom{\dagger}})^2
 \nonumber \\  &\phantom{{}={}}
  +\lambda_3(\Phi_1^\dagger\Phi_1^{\phantom{\dagger}})
            (\Phi_2^\dagger\Phi_2^{\phantom{\dagger}})
  +\lambda_4(\Phi_1^\dagger\Phi_2^{\phantom{\dagger}})
            (\Phi_2^\dagger\Phi_1^{\phantom{\dagger}})
  +\tfrac12 \lambda_5[ (\Phi_1^\dagger\Phi_2^{\phantom{\dagger}})^2
                      +(\Phi_2^\dagger\Phi_1^{\phantom{\dagger}})^2].\label{eq:pot}
\end{eqnarray}
The masses of the two CP--even neutral states $h$ and $H$, as well as
that of the CP--odd neutral $A$ and two charged $H^\pm$ states that
are present in the model, $M_h, M_H, M_A$ and $M_{H^\pm}$,  are free
parameters and we will assume here that the lighter CP--even $h$ boson
is the observed Higgs resonance with a mass of $M_h=125$ GeV. Three
other parameters describe the model and among them there are two
mixing angles $\beta$ and $\alpha$: $\tb = v_2/v_1$ where
$v_1/\sqrt{2}$ and $v_2/\sqrt{2}$ are the vacuum expectation values of
the neutral components of the fields $\Phi_1$ and $\Phi_2$ (with
$v_1^2+v_2^2 = v^2 = {\rm (246~GeV)^2})$ and the angle $\alpha$ that
diagonalizes the mass matrix of the two CP--even $h$ and $H$
bosons. Finally; there is another mass parameter $m_{12}$ which enters
only in the quartic couplings among the Higgs bosons, 
\begin{eqnarray}
{\lambda}_{\phi_i \phi_j \phi_k}= g^\text{2HDM}_{\phi_i \phi_j \phi_k}/g^\text{SM}_{HHH}
 = f(\alpha, \beta, m_{12}) 
\end{eqnarray}
In this parametrisation, the neutral CP--even $h$ and $H$ bosons share
the coupling of the SM Higgs particle to the massive gauge bosons
$V=W,Z$ and one has at tree level, 
\begin{eqnarray}
g_{hVV}= g^\text{2HDM}_{hVV}/g^\text{SM}_{HVV}= \sin(\beta-\alpha) \ , ~~~
g_{HVV}= g^\text{2HDM}_{HVV}/g^\text{SM}_{HVV}= \cos(\beta-\alpha)
\end{eqnarray}
while, as a consequence of CP invariance,  there is no coupling of the
CP--odd $A$ to vector bosons, $g_{AVV}=0$. There are also couplings
between two Higgs and a vector boson which, up to a normalisation
factor~\footnote{The complete set of Feynman rules can be found in the Appendix A of Ref.~\cite{Gunion:1989we}}, are complementary to the ones above. For instance one has, using a normalisation that gives similar couplings as in the previous case,  
\begin{eqnarray}
g_{hAZ}= g_{h H^\pm W^\mp}=  \cos(\beta-\alpha) \ , ~~~ 
g_{HAZ}= g_{H H^\pm W^\mp} =  \sin(\beta-\alpha)
\end{eqnarray}
For completeness, additional couplings of the charged Higgs boson
which will be needed in our discussion.  They do not depend on any
SUSY parameter and are simply given by (here, we include the pre-factors in the 
couplings except for the momenta)
 \begin{eqnarray}
g_{A H^\pm W\mp}= e/2\sin\theta_W\ , ~~~ g_{H^+ H- \gamma} =  -e , ~~~ g_{H^+ H^- Z} = -e \cos2\theta_W / \sin 2 \theta_W 
\end{eqnarray}

In a general 2HDM, the interaction of the Higgs states with fermions
are model--dependent and there are generally two options which are
discussed in the literature~\cite{Branco:2011iw}. In Type II models,
the field $\Phi_1$ generates the masses of isospin down--type fermions
and $\Phi_2$ the masses of up--type quarks. In turn, in Type I models,
the field $\Phi_2$ couples to both isospin up-- and down--type
fermions. The couplings of the neutral Higgs bosons to gauge bosons
and fermions are given in Table~\ref{Tab:cpg-2HDM} in the two
models. The couplings of the charged Higgs boson to  fermions follow
that of the CP--odd Higgs boson.

\begin{table}[!h]
\begin{center}
\renewcommand{\arraystretch}{1.4}
\begin{tabular}{|c|c|c|c|c|c|c|} \hline
\ \ $\Phi$ \ \  &\multicolumn{2}{c|}{$g_{\Phi \bar{u}u}$}&  
                  \multicolumn{2}{c|}{$g_{\Phi \bar{d}d}$}&  
$g_{ \Phi VV} $ \\ \hline
& Type I & Type II & Type I & Type II & Type I/II \\   \hline
$h$  &  $\; \cos\alpha/\sin\beta       \; $ 
     &  $\; \cos\alpha/\sin\beta       \; $  
     &  $ \; \cos\alpha/\sin\beta \; $  
     &  $ \; -\sin\alpha/\cos\beta \; $  
     &  $ \; \sin(\beta-\alpha) \; $  \\
$H$  &  $\; \sin\alpha/\sin\beta \; $  
     &  $\; \sin\alpha/\sin\beta \; $  
     &  $ \; \sin\alpha/ \sin\beta \; $  
     &  $ \; \cos\alpha/ \cos\beta \; $  
     &  $ \; \cos(\beta-\alpha) \; $  \\
$A$  &  $\; \cot \beta \; $ 
     &  $\; \cot \beta \; $ 
     &  $ \; \cot \beta \; $    
     &  $\; \tan \beta \; $ 
     &  $ \; 0 \; $ \ \\ \hline
\end{tabular}
\end{center}
\caption[]{The couplings of the neutral Higgs bosons $h,H$ and $A$ to
  fermions and gauge bosons in 2HDMs of Type I and  II relative to the
  SM Higgs couplings; the $H^\pm$ couplings to fermions follow that of
  the $A$ boson. Other couplings are given in the text.}
\label{Tab:cpg-2HDM}
\vspace{-.2cm}
\end{table}

The Higgs couplings to fermions and gauge bosons depend only on the ratio $\tan\beta$ and 
the difference $\beta-\alpha$ between the two mixing angles. If one enforces the fact that 
the couplings of the observed $h$ boson should be SM--like, as the LHC Higgs data 
strongly indicate~\cite{HiggsCombo,Djouadi:2013lra}, one simply needs
to set  $\beta-\alpha=\pi/2$. This is called the alignment
limit~\cite{Pich:2009sp,Craig:2013hca,Carena:2013ooa,Dev:2014yca,Bernon:2014nxa}. In
this limit, the couplings $h$ couplings are SM--like $g_{hVV}=
g_{huu}=g_{hdd} \to 1$, while the couplings of the CP--even $H$ state
reduce to those of the pseudoscalar $A$ boson. In particular, there is
no $H$ coupling to vector bosons, $g_{HVV} \to g_{AVV} =0$ and the
couplings to up--type fermions are $g_{Huu} = \cot \beta$ while those
to down--type fermions are   $g_{Hdd} = \cot \beta$ and  $g_{Hdd} =
\tan \beta$ in, respectively, type I and II models.

In addition to $\tb$, at least the Higgs masses $M_H, M_A$ and
$M_{H^\pm}$ will enter the 2HDM phenomenology. These are in principle
free parameters and can have arbitrary values, except for the $H$
state that was assumed to be heavier than $h$. This makes any
phenomenological analysis rather complicated and to simplify our 
discussion here, we will make the
additional assumption that they are of the same order\footnote{More
  precisely, the discussion that we will have in the next subsections
  will hold if the differences of the masses of the $\phi_i,
  \phi_j=H,A,H^\pm$ states, $|M_{\phi_i} - M_{\phi_j}| \lsim M_W$,  is
  satisfied. This pattern will be more or less equivalent to the one
  in the MSSM close to the decoupling limit as will be seen later.}, 
\beq
M_H \approx M_A \approx M_{H^\pm}.
\label{2HDM-masses}
\eeq
Finally, in the alignment limit $\beta-\alpha=\frac\pi2$ the
expressions of two important  triple couplings among the CP--even
Higgs bosons simplify to
\begin{align}
\lambda_{hhh} & = 1 \, , \  \lambda_{Hhh} = 0 \, .
\label{eq:tripleSM}
\end{align}
This means that the triple $h$ coupling is SM like, while there is no
$Hhh$ coupling at the tree--level. The other couplings depend on the
additional parameter $m_{12}$ and they will not affect the discussion
that we will have in this paper, so we will ignore them.

\subsubsection{The SUSY case and the hMSSM approach}

The MSSM is essentially a two Higgs doublet model of type II, that is,
the field $\Phi_2$ generates the masses of isospin down--type fermions
and $\Phi_1$ the masses of up--type quarks. However, supersymmetry
imposes strong constraints on the Higgs sector and among the four
Higgs boson masses $M_h, M_H, M_A$ an $M_{H^\pm}$  (as well as the
parameter $m_{12}$) and the two mixing angles $\alpha$ and $\beta$,
only two of them are in fact independent at the tree level. These are
in general taken to be $M_A$ and $\tb$. Nevertheless, when the
radiative corrections are included in the Higgs sector, in particular
the dominant loop contributions from the top and stop quarks that have
strong couplings to the Higgs bosons, many supersymmetric parameters
will enter the
game~\cite{Okada:1990vk,Ellis:1990nz,Haber:1990aw,Carena:2002es,Allanach:2004rh,Heinemeyer:2004gx}. This
is for instance the case of the SUSY scale,  taken to be the geometric
average of the two stop masses $M_S= \sqrt {m_{\tilde t_1} m_{\tilde
    t_2} }$,  the stop/sbottom trilinear couplings $A_{t/b}$ or the
higgsino mass parameter $\mu$ (other corrections, that involve the
gaugino mass parameters $M_{1,2,3}$ for instance are rather small). 

In particular, the radiative corrections in the CP--even neutral Higgs
sector are extremely important and shift the value of the lightest $h$
boson mass from the tree--level value predicted to be $M_h \leq M_Z
|\cos2\beta| \leq M_Z$ to the value $M_h=125$ GeV that has been
measured experimentally.  In the basis $(\Phi_2,\Phi_1)$, the CP--even
Higgs mass matrix including the radiative corrections can be written
as:
\beq
M_{S}^2=M_{Z}^2 \left(
\begin{array}{cc}
  c^2_\beta & -s_\beta c_\beta \\ -s_\beta c_\beta & s^2_\beta \\
\end{array}
\right) +M_{A}^2
\left(
\begin{array}{cc}
 s^2_\beta & -s_\beta c_\beta \\
 -s_\beta c_\beta& c^2_\beta \\
\end{array}
\right)
+
\left(
\begin{array}{cc}
 \Delta {\cal M}_{11}^2 &  \Delta {\cal M}_{12}^2 \\
 \Delta {\cal M}_{12}^2 &\Delta {\cal M}_{22}^2 \\
\end{array}
\right)
\label{matrix-hMSSM}
\eeq
where we have used the short--hand notation $c_\beta \equiv \cos\beta$
etc$\dots$ and introduced the radiative corrections  by a general
$2\times 2$ matrix $\Delta {\cal M}_{ij}^2$. The neutral CP even Higgs
boson masses and the mixing angle $\alpha$ that diagonalizes the $h,H$
states, $H= \cos\alpha \Phi_2^0 + \sin\alpha \Phi_1^0$ and $h=
-\sin\alpha \Phi_2^0 + \cos\alpha \Phi_1^0$ can be then easily
derived. It is well known that in the $2\times 2$ matrix for the
radiative corrections, only the $\Delta{\cal M}^{2}_{22}$ entry which
involves the by far dominant stop--top sector
correction~\cite{Okada:1990vk,Ellis:1990nz,Haber:1990aw}, 
\beq 
\Delta{\cal M}^{2}_{22} \approx \Delta M_h^2|^{t/\tilde{t}}_{\rm 1loop} \sim 
\frac{3m_t^4}{2\pi^2v^2} \bigg[ \log \frac{M_{S}^2} {m_{t}^2} +\frac{X_{t}^{2}}{M_{S}^{2}}
- \frac{X_{t}^{4} }{12 M_{S}^{4}} \bigg]
\eeq 
where $M_S$ is the SUSY scale and $X_t=A_t-\mu/\tb$ the stop mixing
parameter, is relevant that is, $\Delta{\cal M}^{2}_{22} \gg
\Delta{\cal M}^{2}_{11}, \Delta{\cal M}^{2}_{12}$. It has been
recently
advocated~\cite{Maiani:2012ij,Maiani:2012qf,Djouadi:2013vqa,Djouadi:2013uqa,Quevillon:2014nka,Djouadi:2015jea}
that in this case, one can simply trade $\Delta {\cal M}^{2}_{22}$ for
the by now known $M_h$ value using
\beq
\Delta {\cal M}^{2}_{22}= \frac{M_{h}^2(M_{A}^2  + M_{Z}^2 -M_{h}^2) -
  M_{A}^2 M_{Z}^2 c^{2}_{2\beta} } { M_{Z}^2 c^{2}_{\beta}  +M_{A}^2
  s^{2}_{\beta} -M_{h}^2}
\label{m22-hMSSM}
\eeq
and write  the parameters $M_H$ and $\alpha$ in terms of
$M_A,\tb$ and $M_{h}$ in the simple form 
\begin{eqnarray}
h{\rm MSSM}:~~ 
\begin{array}{l} 
M_{H}^2 = \frac{(M_{A}^2+M_{Z}^2-M_{h}^2)(M_{Z}^2
  c^{2}_{\beta}+M_{A}^2
s^{2}_{\beta}) - M_{A}^2 M_{Z}^2 c^{2}_{2\beta} } {M_{Z}^2
c^{2}_{\beta}+M_{A}^2
s^{2}_{\beta} - M_{h}^2} \\
\ \ \  \alpha = -\arctan\left(\frac{ (M_{Z}^2+M_{A}^2) c_{\beta}
    s_{\beta}} {M_{Z}^2 
c^{2}_{\beta}+M_{A}^2 s^{2}_{\beta} - M_{h}^2}\right)
\end{array}
\label{wide} 
\end{eqnarray}
This is the so--called $h$MSSM
approach~\cite{Djouadi:2013uqa,Djouadi:2015jea} which has been shown
to provide a very good approximation of the MSSM Higgs
sector~\cite{Bagnaschi:2015hka}.

In the case of the $H^\pm$  masses, the radiative corrections are very
small at high enough $M_A$ and one has to a good
approximation~\cite{Frank:2013hba}
\begin{eqnarray}
M_{H^\pm} \simeq \sqrt { M_A^2 + M_W^2}
\end{eqnarray}
In this $h$MSSM approach, the MSSM Higgs  sector with  only the
largely dominant $\Delta {\cal M}_{22}^2$ correction, can be again
described with only the two parameters $\tb$  and $M_A$ as  the loop
corrections are fixed by the value of $M_h$. Another advantage of this
approach is that it allows to describe the low $\tb$ region of the
MSSM (see also Ref.~\cite{Bagnaschi:2015hka}) which was overlooked as
for SUSY scales of order 1 TeV, values $\tb \lsim 3$ were excluded
because they lead to $M_h<  125$S GeV. The price to pay is that for
such low $\tb$ values, one has to assume $M_S \gg 1$ TeV and, hence,
the model is excessively fine-tuned. 

In fact, the possibility that the SUSY scale is rather high was an
implicit assumption in this $h$MSSM approach. Indeed, one needs that
$M_S$ is much larger than the other SUSY parameters, and in particular
$M_S \gg |\mu|$, in such a way that $\Delta{\cal M}^{2}_{22} \gg
\Delta{\cal M}^{2}_{11,12}$ is indeed a good approximation. In this
case, the subleading corrections, e.g. $\propto |\mu|/M_S$, that enter
$\Delta{\cal M}^{2}_{11,12}$ are too small and can be ignored to write
Eq.~(\ref{m22-hMSSM}). Another implicit assumption is that the
CP--even Higgs sector can be indeed described by
Eq.~(\ref{matrix-hMSSM}) also at very high $M_S$, which is a
non--trivial statement but which has been verified in most
cases~\cite{Lee:2015uza}.

In the MSSM, the couplings of the CP--even $h$ and $H$ and the CP--odd $A$ states to
fermions and vector bosons are given by the type II entries of the
2HDM couplings shown in Table~\ref{Tab:cpg-2HDM} (again, the strength
of the $H^\pm$ couplings to fermions will be similar to that of
$A$). The only difference is that now, the angle $\alpha$ including
the radiative correction is fixed by the $h$MSSM relation
Eq.~(\ref{wide}). We note here that additional direct corrections
should in principle enter the Higgs couplings but because $M_S$ is
taken to be very large, they are assumed to have a small impact in the
$h$MSSM and will be ignored.

Another important set of couplings are the Higgs self--couplings and
in the $h$MSSM, they are again given in terms of $\beta$ and $\alpha$,
with the latter fixed by $\tb$ and $M_A$. For instance, the $hhh$ and
$Hhh$ self--couplings, up to a normalisation factor, read
\begin{eqnarray}
\lambda_{hhh} &=& 3\cos2\alpha \sin(\beta+\alpha) + 3 
\frac{\Delta {\cal M}^{2}_{22} }{M_Z^2} \frac{ \cos\alpha}{\sin\beta} \cos^2 \alpha
\nonumber \\[-3mm] 
\lambda_{Hhh} &=& 2\sin2\alpha \sin(\beta+\alpha) -\cos 2\alpha \cos(\beta+\alpha)  
+ 3 \frac{ \Delta {\cal M}^{2}_{22} }{M_Z^2} \frac{ \sin\alpha}{\sin\beta} \cos^2 \alpha
\label{eq:Hhh}
\end{eqnarray}
A last remark is that when  $M_A \! \gg \! M_Z$, one is in the so--called decoupling
regime~\cite{Haber:1995be} in which the $h$ state is light and as
$\alpha=\beta- \pi/2$, has almost exactly the SM--Higgs couplings,
$g_{hVV}= g_{hff}= 1$. The other CP--even $H$ and the charged $H^\pm$
bosons become heavy and degenerate in mass with the $A$ state, $M_H \!
\approx \! M_{H^\pm} \! \approx \! M_A$, and decouple from the massive
gauge bosons. The intensity of the couplings of the $H$ and $A$ states
are the same. In this regime, the MSSM Higgs sector thus looks almost
exactly as the one of the 2HDM of type II in the alignment limit,
especially if the additional assumption on the Higgs masses that we
made to further simplify the model, Eq.~(\ref{2HDM-masses}), is used.
The only exception will be the trilinear couplings which are different
because in the MSSM, there are additional loop corrections that make
for instance $\lambda_{Hhh}$  non--zero in the decoupling limit.

\subsection{Production of the Higgs bosons at hadron colliders}

\subsubsection{The gluon fusion process}

In 2HDMs, the neutral Higgs bosons can be produced in the gluon fusion mechanism,
$gg \to H,A$ as well as $h$,  via loops involving mainly the heavy bottom and top quarks. 
At the one--loop level, i.e. the Born approximation in this case, the cross 
sections will depend on the magnitude of the Higgs couplings to quarks but there is also 
a difference between the CP--even and CP--odd cases. The $\Phi gg$ amplitudes with
$\Phi\!=\!H,A$ follow the same trend except for the form factors which are different  
for the two CP Higgs cases. In the alignment limit of 2HDMs, the lightest $h$ boson 
has SM--like couplings and its cross section follows exactly that of
the SM Higgs particle which was also described in subsection 2.1.1. In
the case of the heavier $\Phi=H, A$ states,  the rates depend on
$M_\Phi$ and strongly on $\tb$.

For small values, $\tb \approx 1$, the dominant contribution to the
cross section comes from top quark loops as the $\Phi t \bar t$
couplings, $g_{\Phi t t} \propto 1/\tb$, are strong. For low $\Phi$
masses, $M_\Phi \lsim 2m_t$, one could use the EFT approach in which
the heavy top quark is integrated out and include not only the NLO QCD
corrections~\cite{Djouadi:1991tka,Dawson:1990zj,Spira:1995rr} but also
the NNLO corrections which are also known in this
case~\cite{Harlander:2002wh,Anastasiou:2002yz,Ravindran:2003um}. In
fact, it was shown that it is a good approximation to incorporate the
NLO corrections in this limit even for $M_\Phi \gsim 2m_t$, provided
that the Born term contains the full quark mass
dependence~\cite{Spira:1995rr}. Note that in the heavy top limit,
while the LO loop amplitudes  are different for the $Agg$ and $Hgg$
cases, as one has form factors for spin--$\frac12$ particles
$A^H_{1/2}= +\frac43$ and $A^A_{1/2}= +2$ for respectively the
CP--even and CP--odd cases, the QCD corrections at NLO and NNLO are
about the same~\cite{Spira:1993bb,Kauffman:1993nv,Harlander:2002vv}. 

At high values of $\tb$, $\tb \gsim 10$, the $\Phi$ couplings to top quarks are strongly
suppressed while those to the bottom quarks, $g_{\Phi b b} \propto \tb$, are enhanced. 
This makes that the  contribution of the $b$--quark loop to the $gg\to \Phi$ processes
(which was less than 10\% in the SM case) will become the dominant one. In fact, for
extremely large $\tb$ values, the cross section which grows as $\tan^2\beta$ 
and is enhanced by large logarithms $\log(m_b^2/M_\Phi^2)$, can be much larger than for 
a SM--like Higgs  of the same mass. In this case, as $M_\Phi \gg 2m_b$, one is in
the chiral limit  in which the rates are approximately the same in the CP--even and
CP--odd Higgs cases. In this limit, one cannot use anymore the EFT approach and 
integrate out the bottom quark to simply the calculation of the higher order terms. 
The QCD corrections can be thus included only to NLO where they have been calculated 
keeping the exact quark mass dependence \cite{Spira:1995rr}. At LHC
energies, the $K$--factors  are much smaller in this case, $K_{\rm
  NLO}^{\rm b\!-\!loop} \approx 1.2$, than in the case of the top
quark loop, $K_{\rm NNLO}^{\rm t\!-\!loop} \approx 2$~\cite{Dittmaier:2011ti}.  

For small to intermediate $\tb$ values, $\tb \approx 3$--10, the suppression of the $\Phi tt$
coupling is already effective while the $bb\Phi$ coupling is not yet
strongly enhanced, resulting in production cross sections that are
smaller than in the SM case. In fact, the minimum of the cross section
is obtained for the value $\tb \approx \sqrt{m_t / \bar m_b} \approx
7$ as one has $m_t\simeq 173$ GeV and $\bar m_b \simeq 3$ GeV for, 
respectively, the top--quark mass measured at colliders and the $b$--quark mass
in the $\overline{\rm MS}$ scheme evaluated at the scale of the Higgs mass.  Here
again, because the top and bottom loop contributions have a comparable weight, one
can include only the NLO QCD corrections which are known exactly (and of course, not the 
EW corrections which are known only for a SM--like  Higgs and not applicable in 
this case). 

In the case of the MSSM, the same discussion that we had above on a 2HDM of type II 
approximately holds, but with two differences. The first one is that for a light 
$A$ boson, say $M_A \lsim 300$ GeV for low values of $\tb$ and $M_A \lsim 150$ GeV
for $\tb \gg1$, we are not yet in the decoupling regime with  
$\alpha = \beta - \pi/2$ and the couplings of the $H$ state to fermions are not exactly 
the same as those of the pseudoscalar $A$ state. However, as discussed earlier, 
the difference between these two couplings should be small as the current Higgs data from 
the LHC indicate that we are close to this decoupling
regime~\cite{Djouadi:2013lra}. Nevertheless, even in this
decoupling case, there is a difference between 2HDMs and the MSSM that
is due to the kinematics: while in the former case the two masses
$M_H$ and $M_A$ were free parameters, they should be close to each
other in the MSSM, $M_A\approx M_H$, especially at high $\tb$.

Another difference between the two cases is that in SUSY theories, there are in principle 
additional contributions to the $\Phi gg$ couplings from squark
(mainly stop and sbottom) loops. These occur in the case of the
CP--even $H$ and not in the case of the CP--odd 
$A$ boson at leading order, as couplings  $A \tilde q_i \tilde q_i$ to squarks of the 
same flavours are forbidden by CP--invariance; squark loops can arise only at the 
two--loop level through the non--zero $A \tilde q_1 \tilde q_2$ non--diagonal coupling.  
These contributions can be particularly important in scenarios where large mixing 
effects occur in the sbottom and stop sectors making that the $\tilde t$ and/or $\tilde b$ 
are much lighter than the other squarks and their couplings to the
Higgs bosons strongly enhanced. Nevertheless, present direct limits
from SUSY searches at LHC and indirect limits
from the mass of the observed $h$ state indicate that these particles should be 
rather heavy and, hence, their impact on the $gg\to H$ production cross section 
should be limited. In any case, in the $h$MSSM approach that we adopt here, these SUSY 
effects are  ignored. 

\begin{figure}[!h]
\vspace*{-.1mm}
\centering

\mbox{
\includegraphics[scale=0.67]{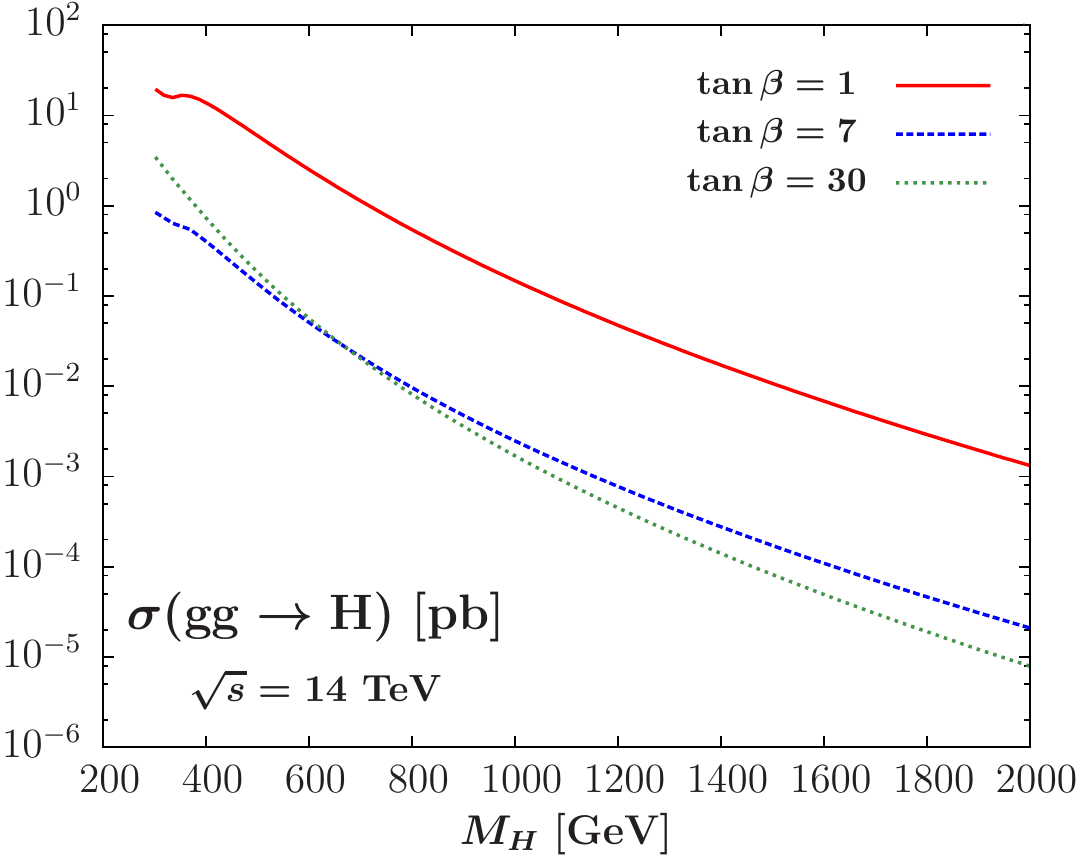}\hspace{2mm}
\includegraphics[scale=0.67]{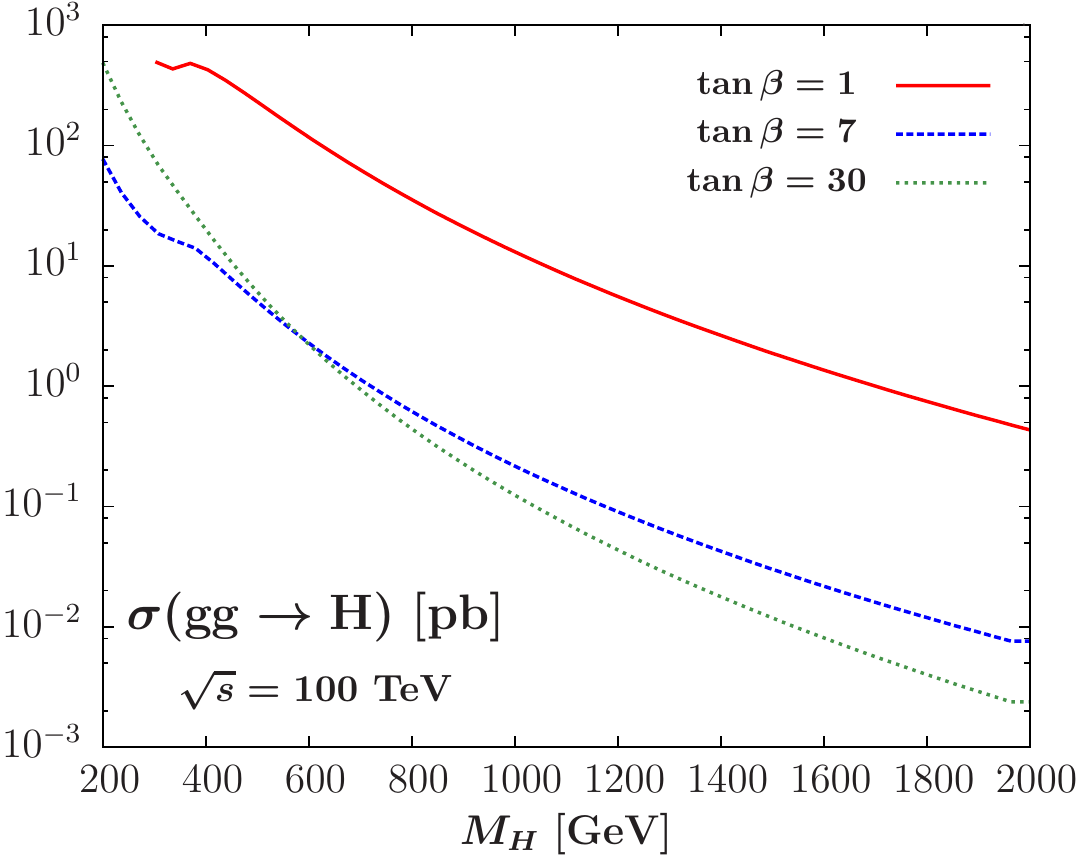}
}

\vspace*{2mm}

\mbox{
\includegraphics[scale=0.67]{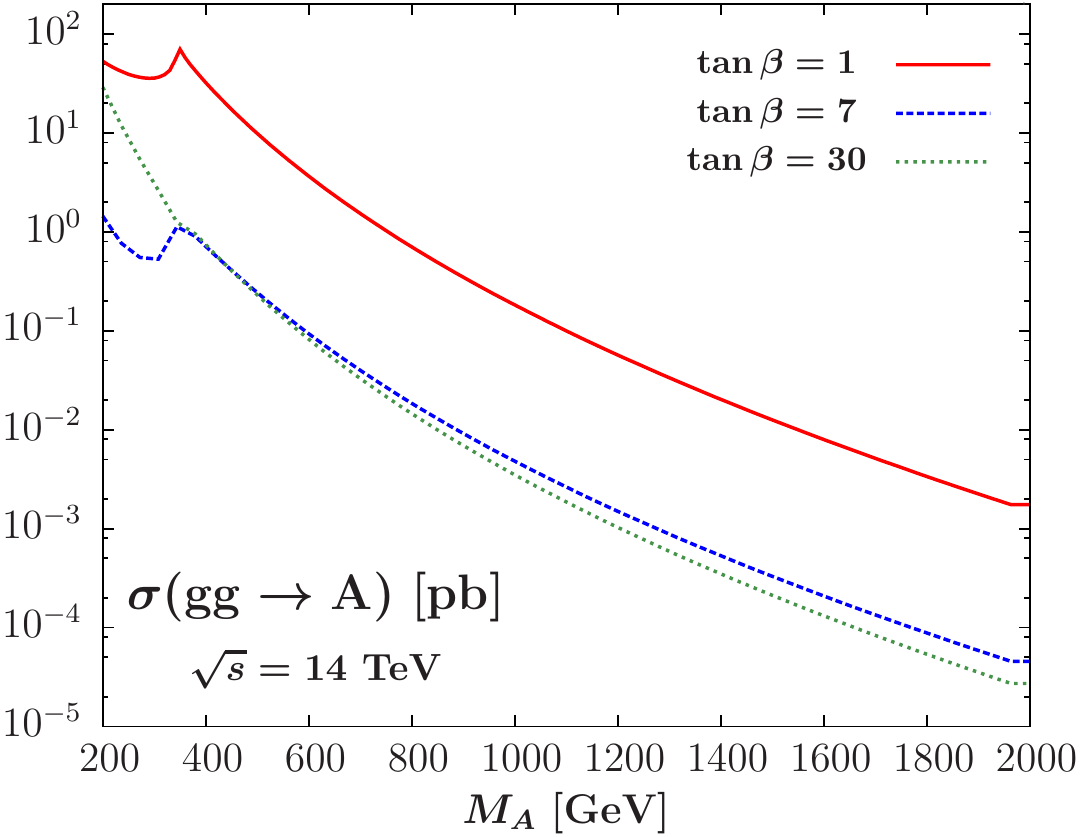}\hspace{2mm}
\includegraphics[scale=0.67]{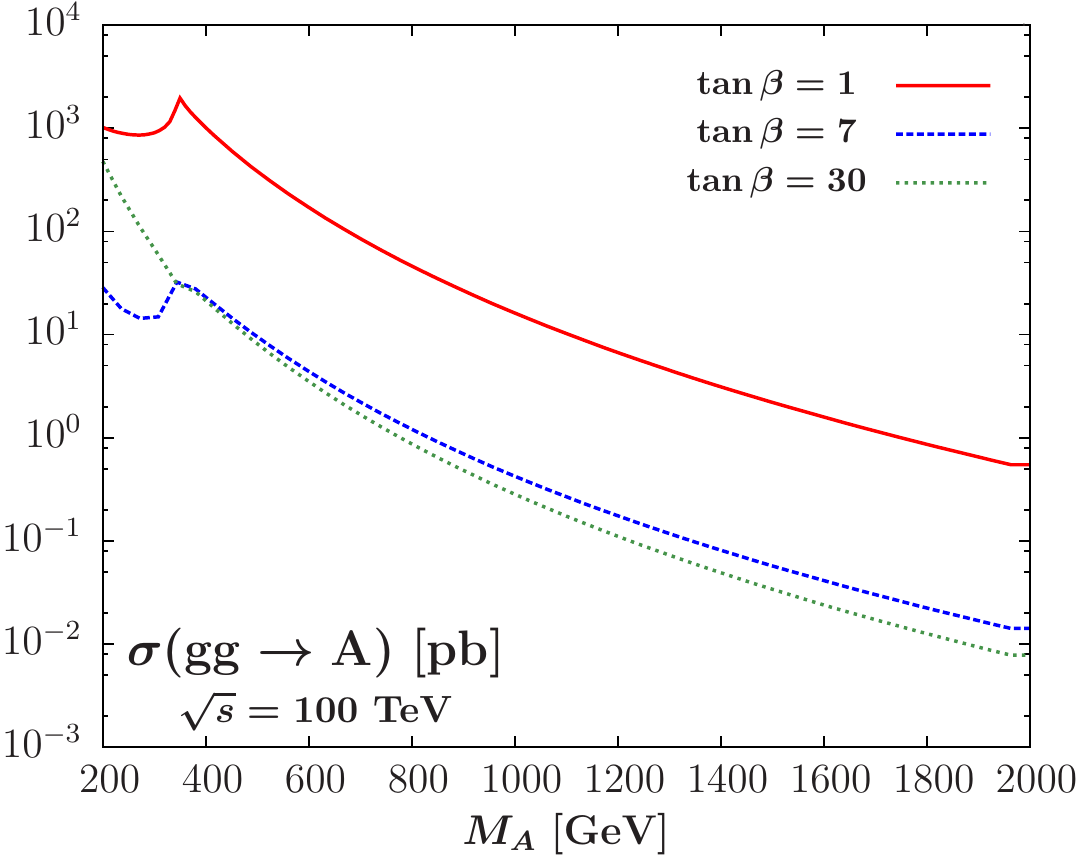}
}
\vspace*{-7mm}
\caption[]{The production cross sections of the CP--even $H$ (upper plots) 
and CP--odd $A$ (lower plots) bosons in the gluon fusion mechanism at 
the LHC with $\sqrt s=14$ TeV (left) and at $\sqrt s=100$ TeV (right) as a 
function of $M_\Phi$. We use the MSTW PDFs and the $h$MSSM with $\tb=1, 7, 30$.} 
\label{fig:prod-gg}
\vspace*{-2mm}
\end{figure}

We perform the numerical analysis of this process taking the example of the MSSM
and for this purpose, we use a the code {\tt SusHi}~\cite{Harlander:2012pb} in which the $h$MSSM approach is implemented.
The code implements the full top and bottom-loop contributions at NLO in QCD from Refs.~\cite{Spira:1995rr,Harlander:2005rq}, NNLO-QCD top contributions in the heavy-top limit from Refs.\cite{Harlander:2002wh,Harlander:2002vv}, and electroweak contributions by light quarks from Refs.~\cite{Aglietti:2004nj,Bonciani:2010ms}.
The cross sections for the production of the heavier CP--even $H$ (upper 
plots) and the CP--odd $A$ (lower plots) bosons are shown as functions of the 
respective Higgs mass in Fig.~\ref{fig:prod-gg} for the LHC at $\sqrt{s}=14$ TeV
(left plots) and at a proton collider with $\sqrt s=100$ TeV (right plots). 
The three values $\tb= 1,7$ and 30 are chosen to illustrate the low, intermediate
and high $\tb$ regimes. The MSTW PDF set has been adopted and the factorisation 
and renormalisation scales have been set to $\mu_F=\mu_R= \frac{1}{2} M_\Phi$.

As can be seen from the figure, and when comparing with the SM Higgs case discussed in 
section 2.1, the production rates for the MSSM CP--even $H$ state are smaller 
than for a SM--like state at low $\tb$ when the suppressed top quark
loop contribution is still dominant, and much larger at high $\tb$
values, when the $b$--quark loop contribution 
is strongly enhanced.  They are small for $\tb \sim 7$ when one has a 
maximal $g_{\Phi tt}$ suppression  and a minimal $g_{\Phi bb}$ enhancement. For
the value $\tb=30$ used for illustration, the $gg \to H$ cross sections is one to two orders 
of magnitude smaller than in the SM with a dominating top loop contribution. 
 At $\sqrt s=100$ TeV, the rates are 
enhanced with respect to LHC by one order of magnitude only for $M_H=200$ 
GeV but by three orders of magnitude for $M_H=2$ TeV due to the more favourable 
phase--space.  

The cross sections for $gg\to A$  are about the same as for $H$ production 
for $M_A\! \gsim\! 200\;$GeV, an approximation which improves at higher $\tb$ for which the 
decoupling regime is more quickly reached and the $b$--loop contributions more 
important resulting in almost equal $A gg$ and $Hgg$ amplitudes in the  chiral limit. 
A noticeable difference also occurs near the $2m_t$ threshold where the 
CP--odd amplitude $A_{1/2}^A$ develops a singularity (that is unphysical and due to 
the QCD corrections in the absence of a regulating finite width~\cite{Spira:1995rr}) while the 
CP--even one $A_{1/2}^H$ simply reaches a maximum. For low $\tb$ values, however, 
the amplitudes are slightly different for $H$ and $A$: first, because the couplings 
$g_{\Phi tt}$ are not the same at low $M_A$ values and second, because of the different 
one--loop $A_{1/2}^\Phi$ form factors.

In 2HDMs of type II in the alignment limit, the gluon fusion cross sections are the 
same as in Fig.~\ref{fig:prod-gg} for the CP--odd $A$ and also the CP--even 
$H$ boson in the decoupling limit, i.e. for $M_A \gsim 200$ GeV at sufficiently
high $\tb$ when $g_{Hff} \approx g_{Aff}$. In the case
of type I 2HDMs in the alignment limit, since one has $g_{H ff}\!= \!g_{Aff} \!= \!1/\tb$
to both top and bottom quarks, the rates are simply given 
by the ones for $\tb=1$ (when the top contribution dominates the $gg\Phi$ 
loop) in Fig.~\ref{fig:prod-gg} divided by $\tan^2\beta$. The rates are thus
approximately the same as in type II models for $\tb \! \lsim \! 3$ and much smaller at 
high $\tb$.

\subsubsection{Associated production with heavy quarks}

Many of the features discussed above for the gluon fusion process appear in 
associated production of the neutral Higgs particles $\Phi=H,A$ (and even $h$ which
is SM--like and thus has already been discussed before) with top and bottom 
quark pairs, $pp\to q\bar q,gg \to t\bar t \Phi$ and $b\bar b \Phi$. The two processes 
have been analysed in section 2 and the discussion holds in the case of both the CP--even 
and CP--odd Higgs bosons of the 2HDMs. The only differences are that
in 2HDMs one has to consider heavier particles and multiply the SM Higgs cross sections 
by the squares of the reduced Higgs Yukawa couplings to fermions
\beq 
\sigma (pp \to Q \bar Q \Phi )=g_{\Phi QQ}^2 \, \sigma_{\rm SM} (pp \to Q 
\bar Q \Phi) 
\eeq
Since $g_{\Phi Q\bar Q} \propto \tan^{-2I_Q^3}\beta$ with $I_Q^3=+\frac12 (-\frac12)$ 
for isospin up (down)---type quarks, the cross sections for the $t\bar t \Phi$ 
and $b\bar b \Phi$ processes are the same for type II 2HDMs in the alignment limit and 
for the MSSM in the decoupling limit\footnote{In fact, at high values
  of $\tb$, there are additional corrections that one should take into
  account in the MSSM: the so--called $\Delta_b$ corrections (see
  e.g. Refs.~\cite{Carena:1999py,Noth:2008tw}) that affect the $\Phi
  bb$ couplings and which grow as $\mu \tb$. These corrections are small for heavy
  sbottoms and gluinos and are ignored in the $h$MSSM approach that we
  adopt here; for a discussion on this issue, see
  Refs.~\cite{Djouadi:2013vqa,Djouadi:2015jea}.}.

In the case of $t \bar t \Phi$ however, there 
is an additional difference at low Higgs masses and moderate centre of mass energies 
where different but small mass effects, ${\cal O} (m_t^2/M_\Phi^2)$, in the matrix 
element squared appear between the CP--even and CP--odd cases as one is not close enough
to  the chiral limit, $M_\Phi \gg m_t$. Since one has $g_{\Phi tt} \propto 1/\tb$, the 
cross section for the $pp\to t\bar t  \Phi$ process is significant only for  
$\tb$ close to unity or lower, $\tb \lsim 3$. 

Note that because at the high collider energies that we are considering here,  the 
gluon luminosities are much larger compared to the quark luminosities, the cross 
sections for these channels are dominantly generated by the gluon fusion and not the  
$q\bar q$ annihilation subprocesses. The QCD corrections are known to NLO in the 
case of a CP--even Higgs state since a decade and lead to a $K$--factor that is of 
order unity, $K_{ttH} \approx 1.1$ for $M_H \approx 125$ GeV at $\sqrt
s=14$ TeV~\cite{Beenakker:2002nc,Dawson:2002tg}. In the case of a
CP--odd Higgs boson, the QCD corrections have been derived more
recently~\cite{Frederix:2011zi} and they lead to a $K$--factor that is
only slightly higher, $K_{ttA} \approx 1.18$ for the same Higgs mass
and c.m. energy. Considering the process at LO only, as will be done
here, is therefore a reasonable approximation in this case. 

\begin{figure}[!h]
\vspace*{1mm}
\centering

\mbox{
\includegraphics[scale=0.67]{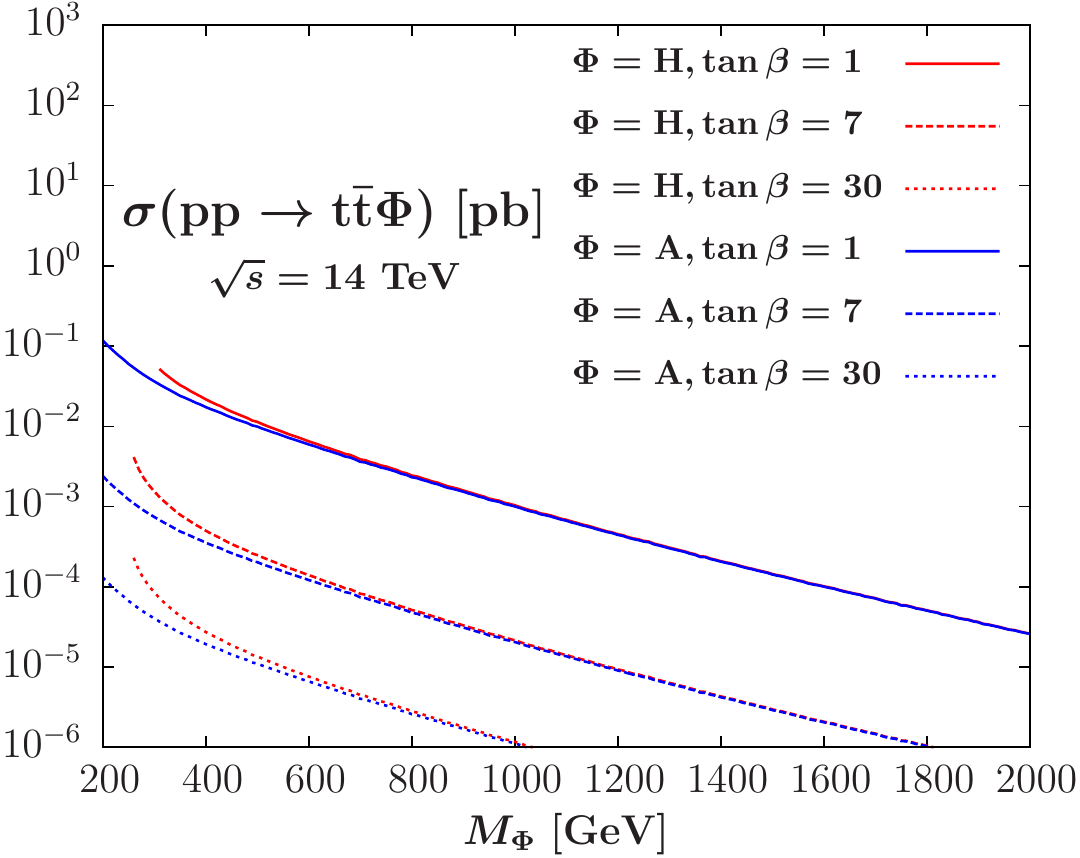}\hspace{2mm} 
\includegraphics[scale=0.67]{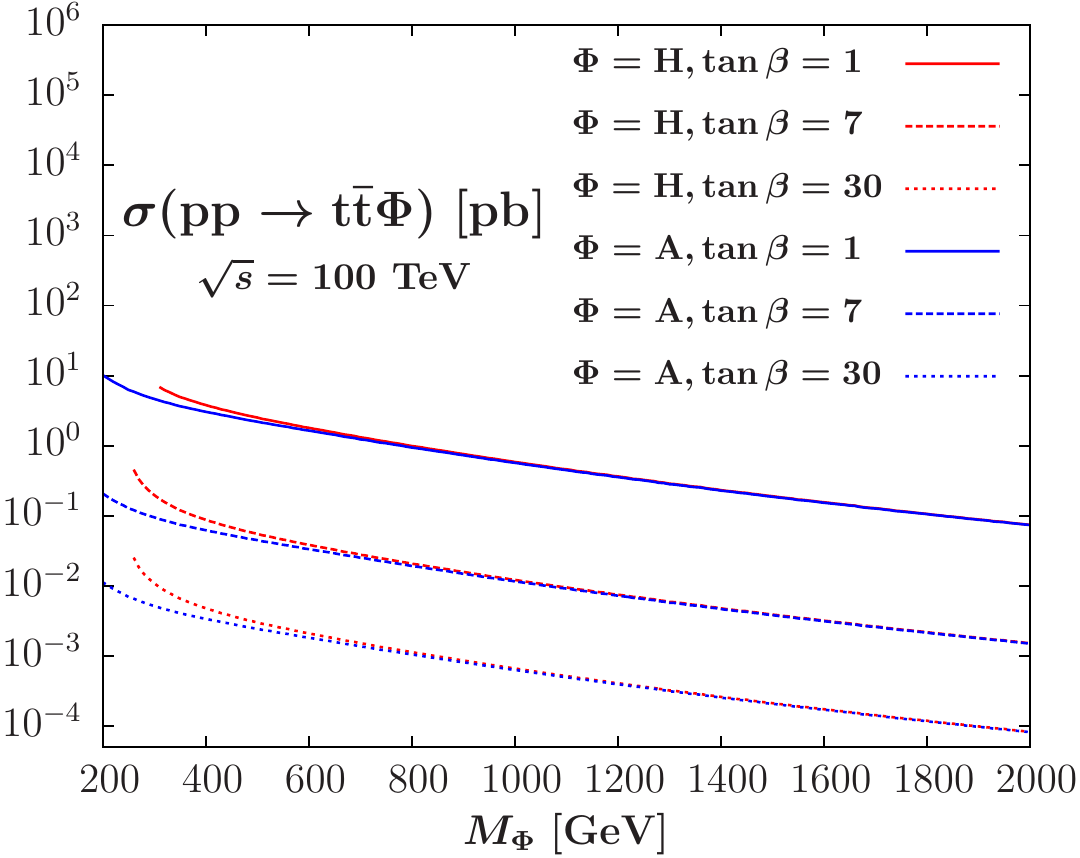}
}

\vspace*{4mm}

\mbox{
\includegraphics[scale=0.67]{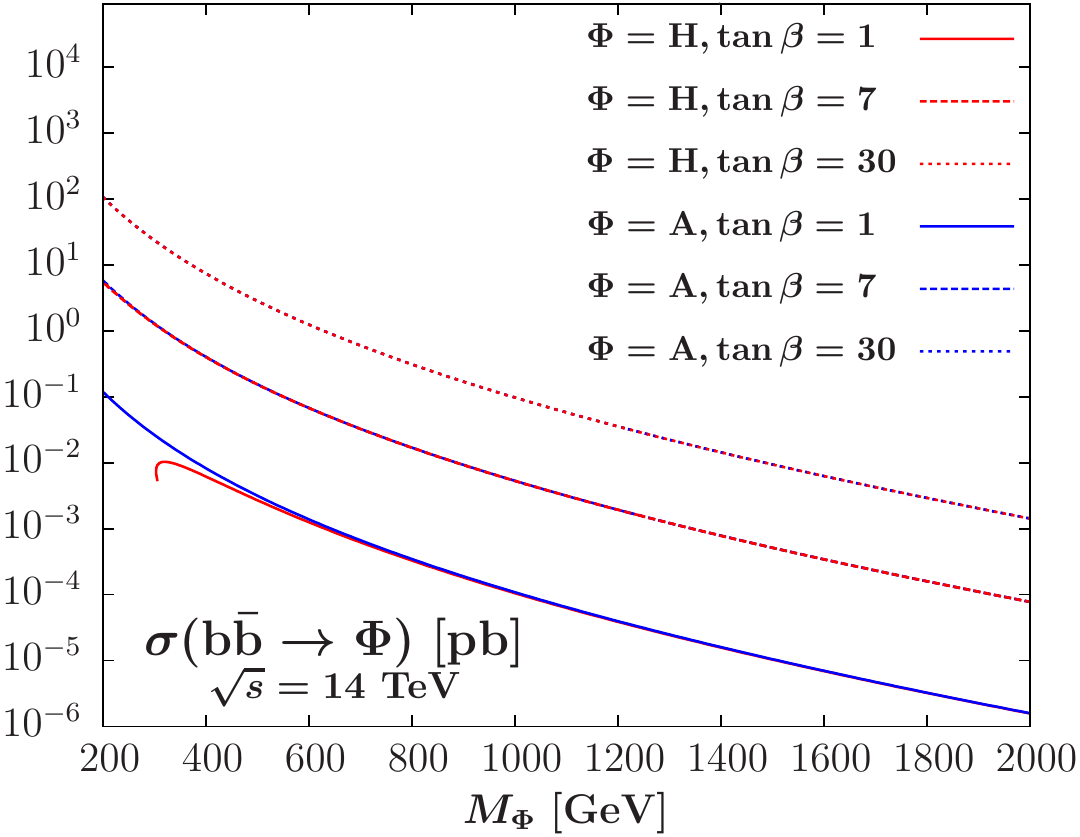}\hspace{2mm}
\includegraphics[scale=0.67]{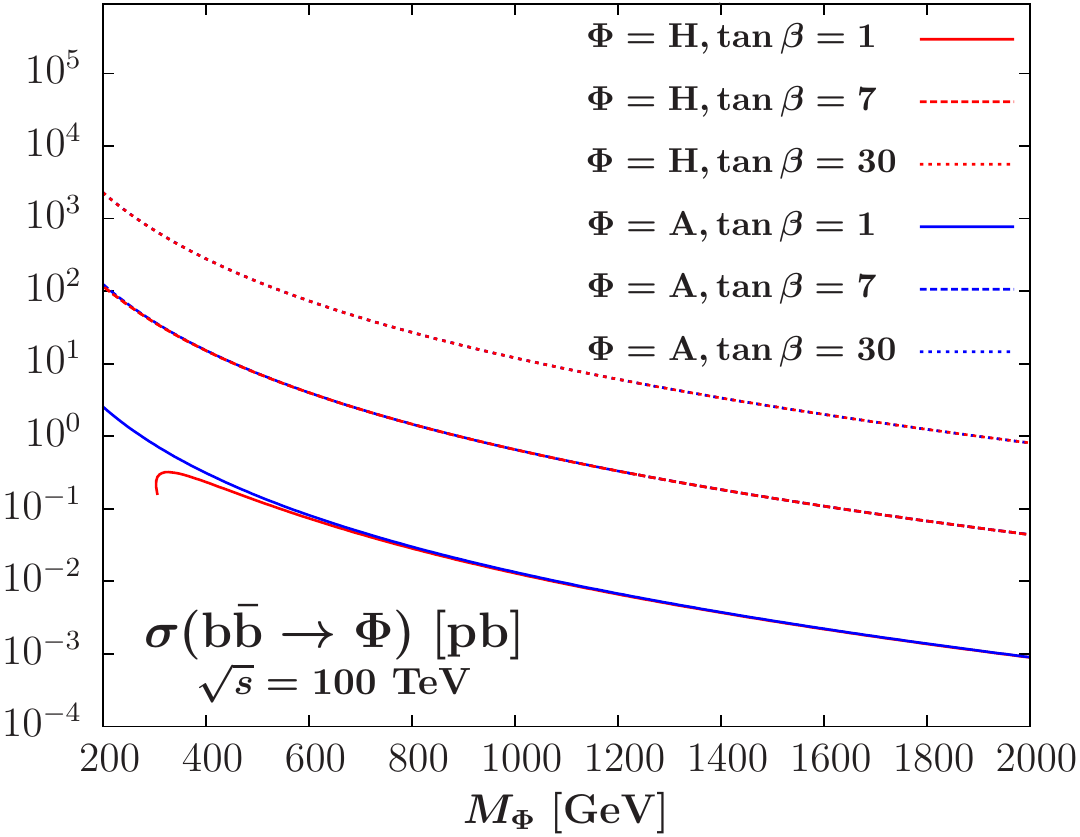}
}
\vspace*{-5mm}
\caption[]{The production cross sections in the   $pp\to t \bar t \Phi$ (upper plots) and 
$pp\to b \bar b \Phi$ (lower plots) for $\Phi=H$ or $A$ at the LHC
with $\sqrt s=14$ TeV (left) and at $\sqrt s=100$ TeV (right) as a
function of the Higgs masses. We use the MSTW PDFs and the $h$MSSM
approach with $\tb=1,7,30$.} 
\label{fig:prod-bb}
\vspace*{-2mm}
\end{figure}

The cross sections $\sigma(pp\to t\bar t \Phi)$ with $\Phi=H,A$ are shown in the 
upper part of Fig.~\ref{fig:prod-bb} as a function of $M_\Phi$ at both $\sqrt s=14$ TeV 
(left) and $\sqrt s=100$ TeV (right) in the $h$MSSM with $\tb=1,7$ and 30. They have been 
obtained with the help of a modified version of the LO program {\tt HQQ}~\cite{Michael-web,Spira:1997dg} with 
the renormalisation and factorisation scales fixed to $\mu_0= m_t +\frac12 M_\Phi$ 
and again using the MSTW set of structure functions. As can be seen, the cross sections 
are sizeable only for $\tb\!=\!1$ and not too high values of $M_\Phi$. For $M_\Phi=1$ 
TeV for instance, they are at the fb level at LHC but reach the pb level 
at $\sqrt s=100$  TeV. Note the difference between the rates of $A$ and $H$ at low 
$M_A$ and $\tb$ as a result of the different masses and couplings in
this area which is outside the decoupling regime. One has
approximately the same rates in 2HDMs in the alignment limit.

In contrast, the production rates for the $pp \to b\bar b \Phi$ process with $\Phi=H,A$ 
are strongly enhanced at high $\tb$ values in type II 2HDMs like the MSSM since one has 
$ g_{\Phi bb} \propto \tan\beta$. The cross sections are almost identical for the production 
of the CP--even $H$ and CP--odd $A$ bosons as  chiral symmetry holds since $M_\Phi \!
\gg \! m_b$ and as one has $g_{Hbb}= g_{Abb}$ in the alignment limit of 2HDMs and the MSSM 
at $\tb \! \gsim \! 5$ where  the decoupling limit is quickly reached. In type I models, 
$ g_{\Phi bb} \propto \cot \beta$ and the cross section follows that of $t\bar t 
\Phi$; it is thus negligible for $\tan\beta >1$ as  the $bb\Phi$ Yukawa 
coupling, $\propto \bar m_b$, is small.

The NLO QCD corrections to the $pp \to b\bar b \Phi$ processes~\cite{Dittmaier:2003ej} are  
the same as those discussed in the SM case. Since $m_b$ is very small compared to $M_\Phi$ 
and chiral symmetry holds, the corrections are now the same for the 
CP--even and CP--odd states.  This small $m_b$ value leads to another major difference 
between the $\Phi b\bar b$ and $\Phi t\bar t$ cases: the cross sections $\sigma (gg \to 
b\bar b \Phi)$ develop (from the splitting of gluons to $b\bar b$ pairs)  large logarithms, 
$\log(Q^2/m_b^2)$, with the scale $Q$ being typically of the order of the factorisation 
scale  and the $b$--quark transverse momentum, $Q \approx \mu_F \approx p_{T}^b$.  
Hence, while one has reliable results for $\sigma( gg\to b\bar b\Phi)$ at high $p_{T}^b$, 
the  convergence of the perturbative series is poor in the opposite
case unless the large logarithms are resummed. As discussed in section
2, this resummation is performed by
treating the $b$--quark as a massless parton in the proton, using the bottom PDF in a 
five-flavour scheme (5FS) and considering instead the fusion process $b\bar{b}\to\Phi$~\cite{Dicus:1988cx}. 
The requirement of one or two additional high--$p_T$ final state $b$ quarks
is fulfilled by considering the NLO~\cite{Dicus:1998hs,Balazs:1998sb,Campbell:2002zm,Maltoni:2003pn}
or NNLO~\cite{Harlander:2003ai} QCD corrections to this process.   

The cross sections for the fusion process $b\bar b\to \Phi$ with $\Phi=H,A$ are displayed 
as a function of $M_\Phi$ in the lower part of Fig.~\ref{fig:prod-bb} at the usual 
c.m. energies, $\sqrt s=14$ TeV (left) and $\sqrt s=100$ TeV (right) and
adopting again the MSTW parton densities. We have used 
a modified version of the public code {\tt SusHi}~\cite{Harlander:2012pb} in  which 
the $h$MSSM approach was implemented and chosen again $\tb=1,7$ and 30 for illustration. 
The NNLO QCD corrections, which are known in this case~\cite{Harlander:2003ai}, 
 are included with renormalisation and factorisation scales set at $\mu_R = 
\mu_F = \frac12 M_H$ and we use $\bar m_b(M_\Phi)$ in the $b$--quark Yukawa coupling 
which significantly improves the convergence of the perturbative series.
  As expected, the production cross sections  are 
extremely large at the value $\tb=30$: at the LHC, the production rates are approximately 
the same as in the $gg \to \Phi$ fusion process at low $M_\Phi$ but decrease less steeply 
with increasing Higgs mass. At $\sqrt s=100$ TeV, the cross sections increase by one or 
two  orders of magnitude depending whether we are at low or high Higgs masses, respectively. 
In type II 2HDMs like the MSSM and  at high $\tb$, the $pp \to b\bar b \Phi$ processes 
are the dominant ones for neutral Higgs production at hadron colliders.  

\subsubsection{Other processes for single production}

In the SM case, there were two additional processes for single Higgs production: 
Higgs--strahlung and vector boson fusion. Since these two processes directly
involve the Higgs couplings to the massive gauge bosons $V=W$ or $Z$, they do not occur 
in the case of the pseudoscalar Higgs particle as the  $AVV$ coupling is forbidden 
by CP--invariance, $g_{AVV}=0$. In fact, even in CP--violating MSSMs and more generally 
2HDMs, these couplings are absent at tree--level and can be generated only at higher 
orders; they are thus also very small and lead to negligible cross
sections for these processes at the LHC and beyond\footnote{Note that
  in 2HDMs, one can have a resonant production of the $H$ boson which,
  if phase--space allowed, could decay through $H\to AZ$, leading to
  $AZ$ final states that are not Higgs--strahlung though. In the MSSM,
  one has $M_H \approx M_A$ and these processes occur only at the
  suppressed three--body level~\cite{Djouadi:1995gv}.}.

The Higgs--strahlung and vector boson fusion processes are in principle allowed only for 
the CP--even $H$ state, besides of course for the SM--like $h$ boson which was discussed 
in section 2.1.  Nevertheless, because in the alignment limit of
2HDMs and in the decoupling limit of the MSSM one has $g_{HVV}=
\cos(\beta-\alpha)$, the $HVV$ coupling is also very 
small or (nearly) vanishing so that the cross sections $\sigma( q\bar q \to HV)$ and 
$\sigma(qq \to  V^* V^* qq \to Hqq)$ are also tiny. The observation of such
processes thus signals a departure from decoupling or alignment in
two--Higgs doublet extensions and would allow for a direct measurement
of the $HVV$ coupling as both cross sections scale as $g_{HVV}^2$ (in
addition to the indirect measurement that can be made by studying the
rates of the lighter $h$ boson, whose coupling is $g_{hVV}^2 =1-
g_{HVV}^2= \sin^2(\beta-\alpha)$, in SM--like processes).

\begin{figure}[!h]
\vspace*{-2mm}
\centering

\mbox{
\includegraphics[scale=0.67]{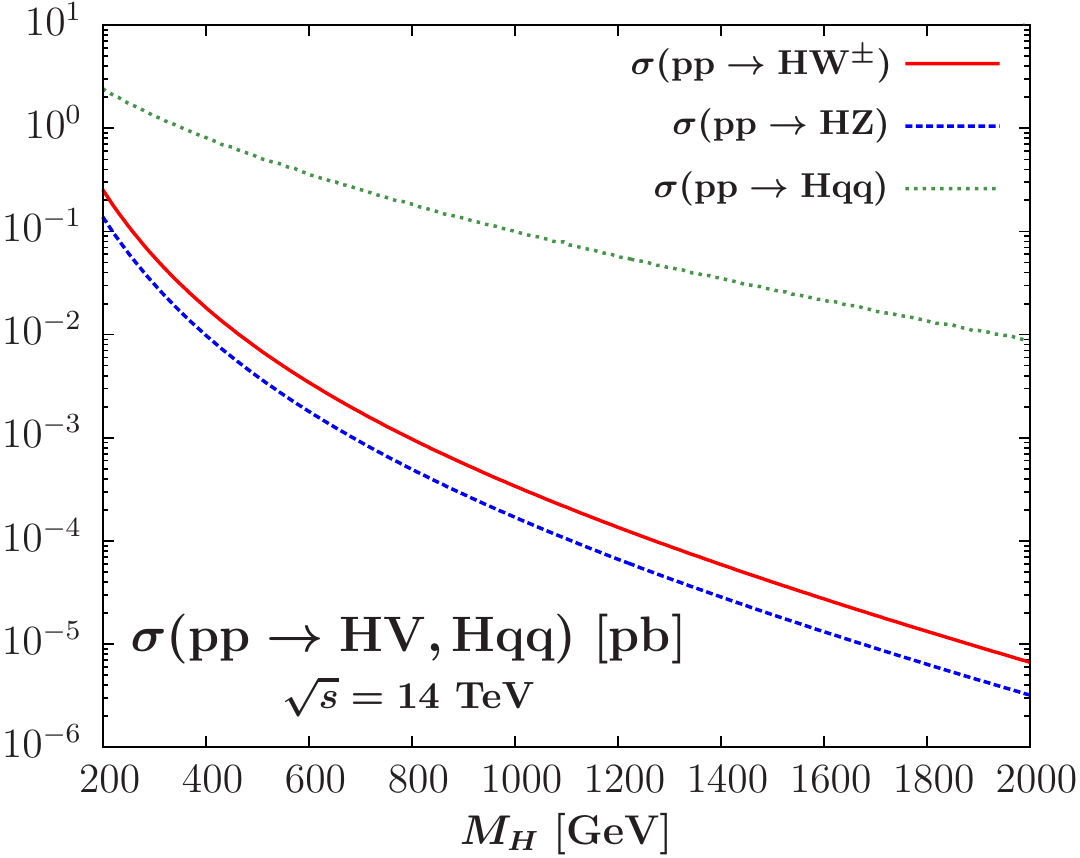}\hspace{1mm}
\includegraphics[scale=0.67]{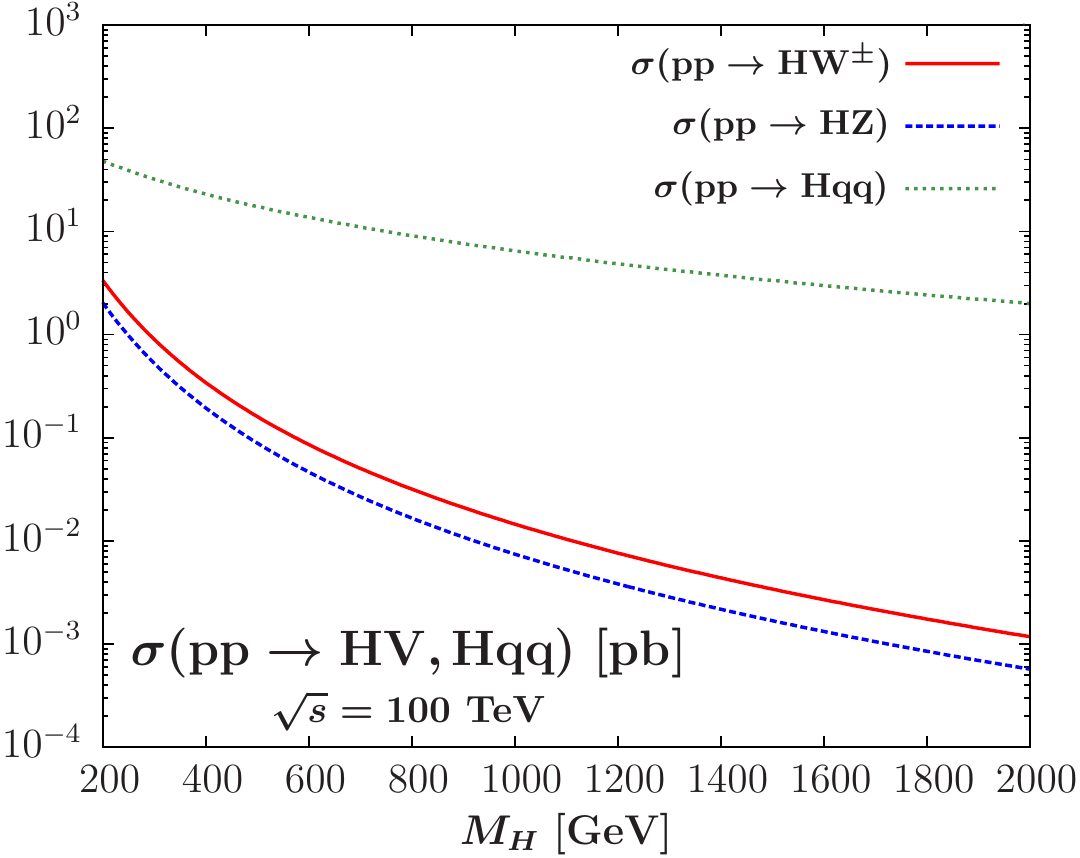}
}
\vspace*{-7mm}
\caption[]{The production cross sections for the CP--even $H$ state in the  $pp \to HV$ 
and $pp \to Hqq$ channels as a function of $M_H$ at $\sqrt s=14$ TeV (left) and $\sqrt 
s=100$ TeV (right). We use the MSTW PDFs and set the $HVV$ coupling to its SM value, 
$g_{HVV}=1$.} 
\label{fig:prod-V}
\vspace*{-1mm}
\end{figure}

In Fig.~\ref{fig:prod-V} we assume a full strength $HVV$ coupling $g_{HVV}=1$ for
illustration and plot the cross sections for the production of the CP--even $H$
boson in the Higgs--strahlung and vector boson fusion processes as a function of $M_H$
at $\sqrt s=14$ TeV and 100 TeV. The rates follow the same trend as
for the SM--Higgs boson and are much larger in the VBF than in the
Higgs--strahlung case and drop less steeply with $M_H$. In the VBF
case and at $\sqrt s=100$ TeV, the cross section stays at the picobarn
level for masses up to  $M_H \approx $ 2 TeV and even for a departure
from the alignment or decoupling limits as small as
$g_{HVV}^2=10^{-3}$,  one still obtains reasonable production cross
sections if the luminosity is high enough, few fbs.

\subsubsection{Neutral Higgs boson pair production}

In a general two Higgs doublet model, the production of pairs of neutral Higgs particles 
in the continuum can be achieved in two main processes: $q \bar q$ annihilation, leading 
to $hA$ and $HA$ final states through the exchange of a virtual $Z$ boson, 
\beq
q\bar{q} \ra Z^* \to hA \, , \ HA \nonumber
\eeq
or $gg$ fusion induced by heavy 
quark box and triangle diagrams, the latter being sensitive to the triple Higgs 
couplings, leading to various Higgs final states,  
\beq
gg \to hh \, , \ HH \, , \ hH \, , \ AA \ \ {\rm and} \ hA \, , \ HA \nonumber
\eeq

The partonic cross sections for neutral Higgs pair production in $q\bar q$ annihilation, 
$q\bar q \to \phi A $  with $\phi=h$ or $H$ are, up to couplings factors,  those of the
associated $\phi$ production with a $Z$ boson with another change in the phase--space factor 
to account for the production of two spin--zero bosons instead. Hence,
as in Higgs--strahlung, the QCD corrections at NLO are simply those of
the Drell--Yan process with an off--shell 
$Z$ boson to be evaluated at the optimal scales $\mu_R\!=\!\mu_F\!=\!M_{A\phi}$. 
The cross sections are proportional to the square of the reduced $\phi AZ$ coupling which,
in 2HDMs, are simply given by $g_{\phi AZ}^2= 1- g_{\phi VV}^2$. Thus, in the alignment
or the decoupling limit of 2HDMs, one would have $g_{hAZ}=\cos(\beta -\alpha)\to 0$ and 
$g_{HAZ}= \sin(\beta -\alpha)\to 1$. The cross section for $pp\to hA$ is thus expected
to be very small (and would be another way to measure the departure from the decoupling 
limit) while that of $pp\to HA$ would be suppressed only by the
phase--space. An analysis of gluon-fusion Higgs pair production in
2HDM has been presented in Ref.~\cite{Hespel:2014sla}.

Note that $A+h/H$ production, as well as the production of all possible
combinations of  pairs of Higgs bosons, are also accessible in bottom quark
fusion, $b\bar b \to \Phi_1 \Phi_2$ with $\Phi_i=h,H,A$ (which are equivalent 
to $gg \to b\bar b  \Phi_1 \Phi_2$ since in the former processes, $b$--quarks 
also come from gluon splitting). The lower $b$--quark luminosities compared to
those of light quarks may be compensated for by large values of $\tb$ which in 
principle strongly enhance the cross sections. Nevertheless, the rates stay at a rather 
modest level even for very high energies and very large $\tb$ values. 

\begin{figure}[!ht]
\vspace*{-3mm}
\centering
\mbox{
\hspace{-1.0cm}
\includegraphics[scale=0.65]{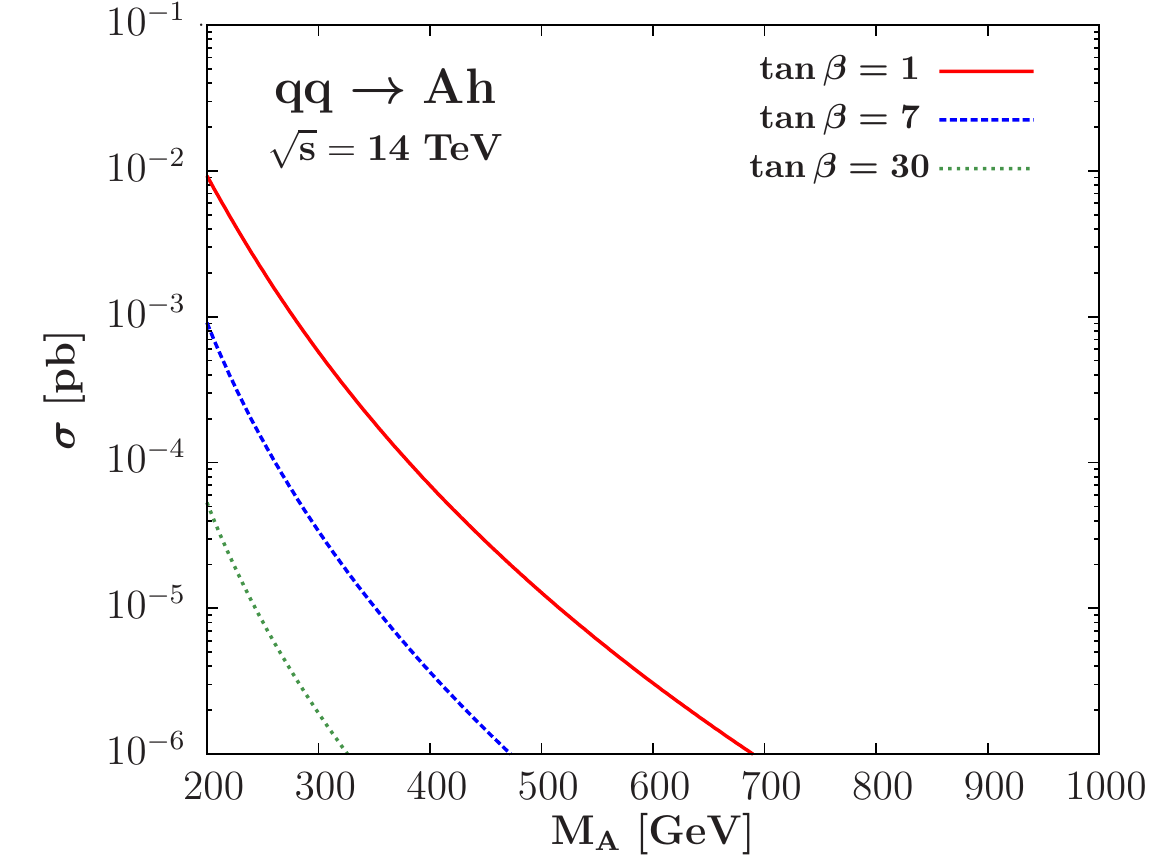}
\includegraphics[scale=0.65]{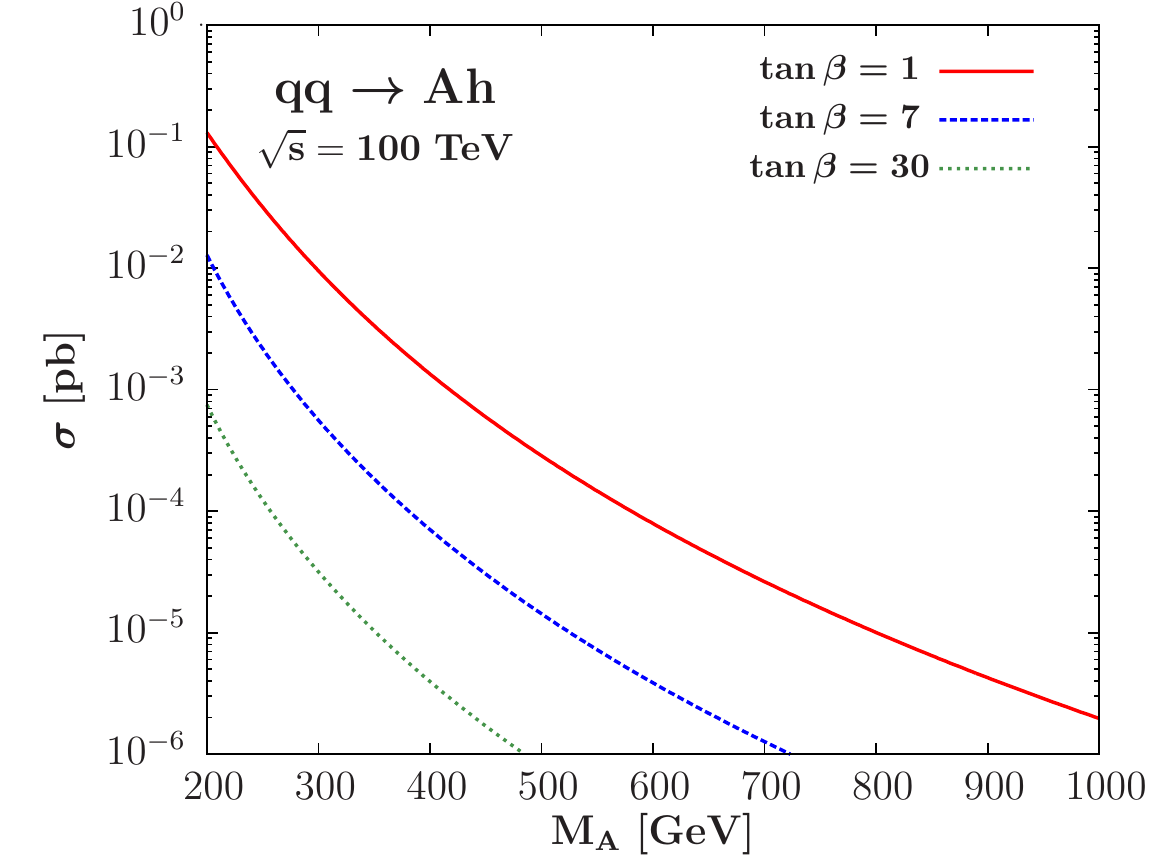}}

\vspace*{1mm}

\mbox{
\hspace{-1.0cm}
\includegraphics[scale=0.65]{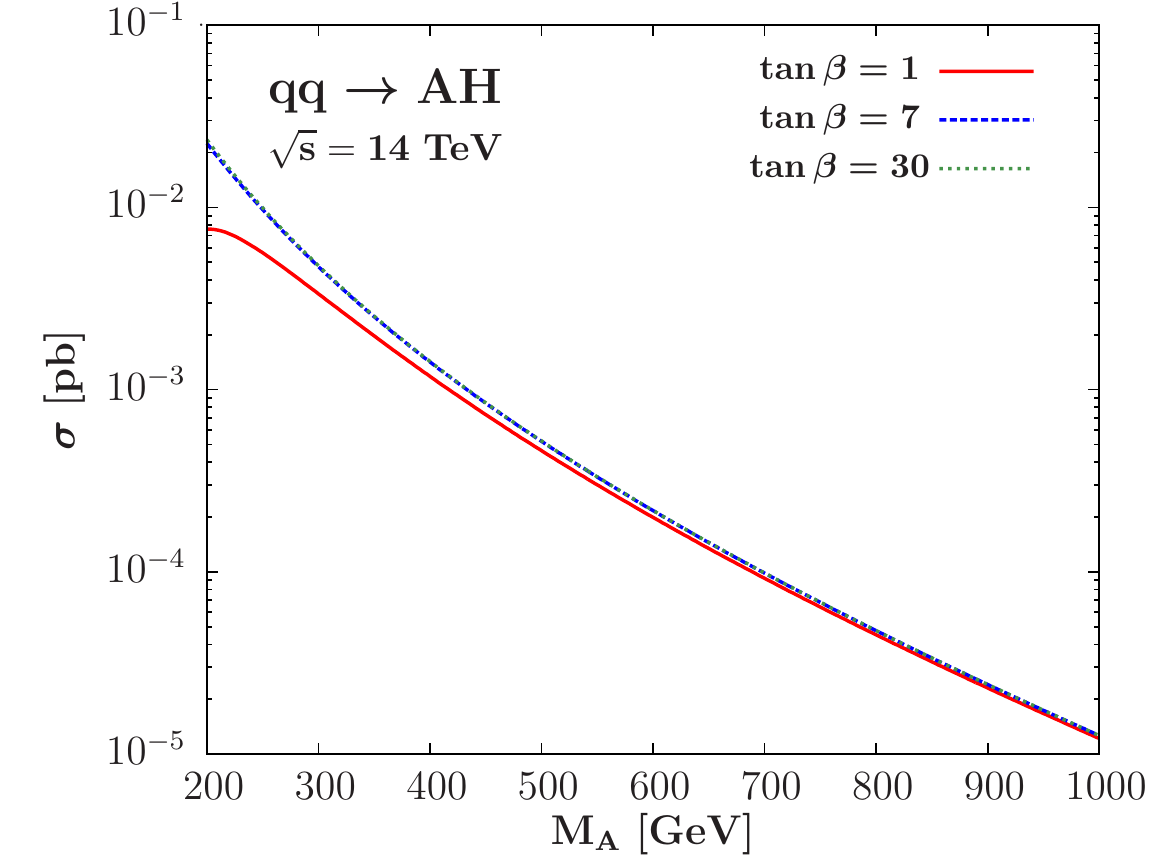}
\includegraphics[scale=0.65]{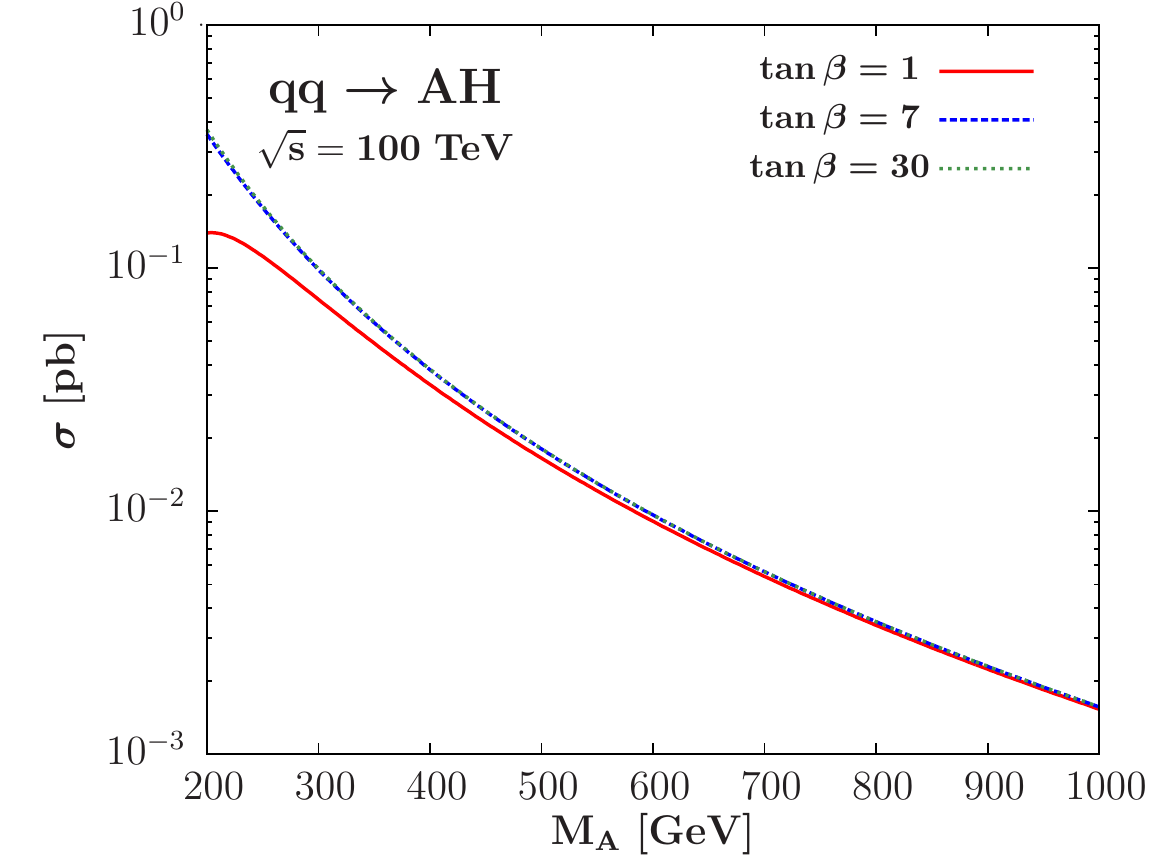}}
\vspace*{-3mm}
\caption[]{The cross sections for associated neutral Higgs pair production in the  
$q\bar q$ annihilation channels, $\sigma( q\bar q \to hA)$ (upper curves) and $\sigma( 
q\bar q \to HA)$ (lower curves). The rates are as functions of $M_A$ in the 
$h$MSSM approach with $M_h=125$ GeV and $\tb=1,7$ and 30. The c.m. energies of 
$\sqrt{s}=14$ TeV (left) and 100 TeV (right) have been assumed and the MSTW
PDFs have been used.}
\label{fig:prodAHqq}
\vspace*{-3mm}
\end{figure}

The cross sections for the $q\bar q \to hA$ and $HA$ processes are shown in 
Fig.~\ref{fig:prodAHqq} again as a function of $M_A$, for $\tb=1, 7, 30$ assuming 
the $h$MSSM; the c.m. energies are also fixed at 14 TeV (left) and 100 TeV (right). 
In the case of $Ah$ production, the cross section are sizeable only far from the 
decoupling limit, i.e. low $\tb \sim 1$ and not too high $M_A$. The rates for $AH$ 
doe not depend on $\tb$ in general and are significant for $M_A \lsim 0.5$ TeV, 
especially at $\sqrt s=100$ TeV where they still can be at the few 10 fb level. Note 
that in 2HDMs, one can relax the mass equality $M_H \approx M_A$ that holds in the MSSM
so that phase--space effects can be more (or less) favourable than in the $h$MSSM. 

In the $gg$ fusion mechanism, a large number of processes for Higgs pair production is 
accessible. The corresponding Feynman diagrams involve top and  bottom quark loops 
(and in SUSY theories,  possibly their scalar partners when these particles are 
relatively light)  that appear in  box and triangular loops. In the latter channels,  
 the pair production proceeds with the virtual exchange of the neutral CP--even $h$ and 
$H$ states (for $AA, Hh,HH$ and $hh$ production) or the CP--odd state
$A$ (for $Ah$ and $AH$ production).  The continuum production can be
supplemented by resonant production with  
Higgs boson decaying into pairs of lighter ones when phase--space allowed (in addition 
to channels in which the resonant Higgs state decays into a lighter Higgs and a $Z$ 
boson as mentioned previously). 
In the MSSM, the only possibility would be a the resonant production of the $H$ state 
$gg \rightarrow H$,  which then decays into two lighter CP--even Higgs bosons, $H 
\rightarrow hh$. However, in the decoupling limit corresponding to $M_A \gsim 350$ GeV
for $\tb \approx 1$, both the $Hhh$ coupling  and the branching fraction for the $hh$ 
decay ($H$ decays are dominated by final states that are either $b\bar b$ at high $\tb$ 
or  $t\bar t $ at low $\tb$)  are very small. In 2HDMs in the alignment
limit, the coupling $g_{Hhh}$  vanishes and there are no $H\to hh$ decays. Instead, 
because  in this case the hierarchy of Higgs masses are different from the MSSM, one 
could have additional resonant channels such as  $gg\to H\to AA$ (or $H\to H^+ H^-$) for instance.  

 \begin{figure}[!ht]
\vspace*{-3mm}
\centering
\mbox{
\hspace{-1.0cm}
\includegraphics[scale=0.65]{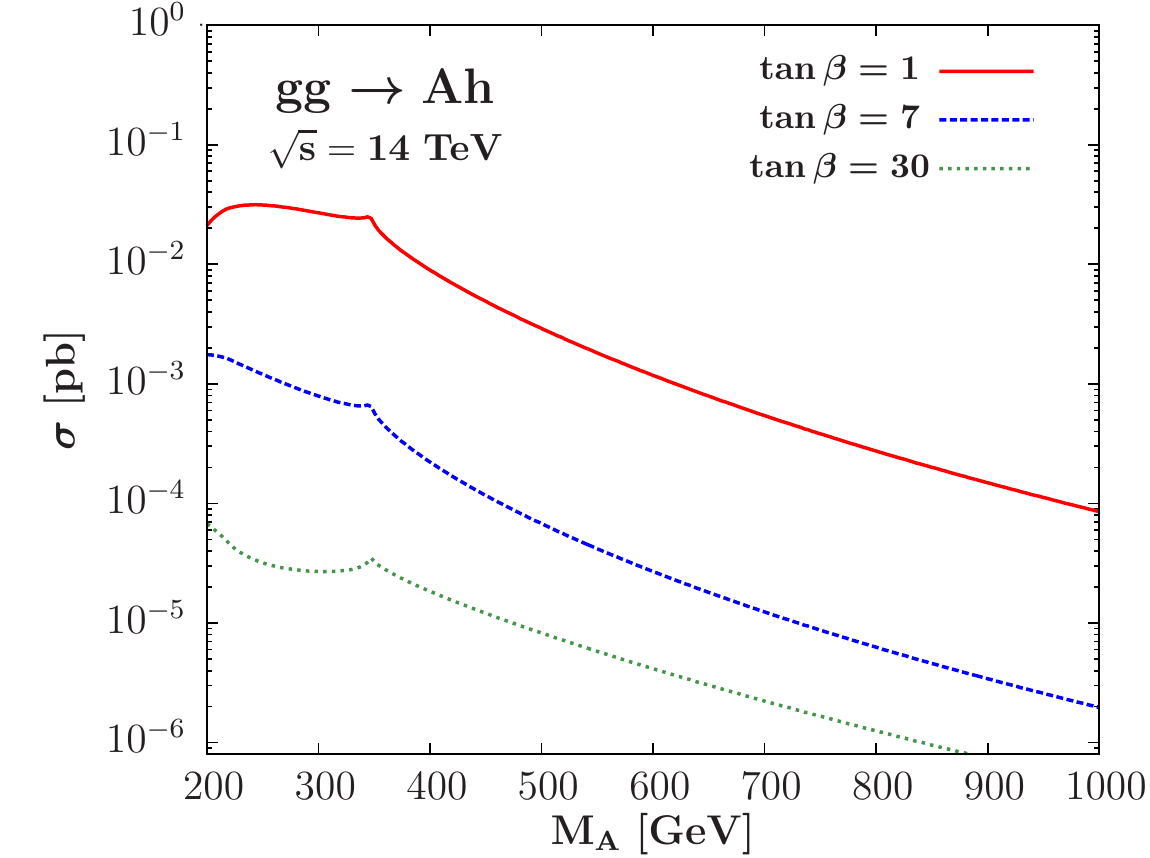}
\includegraphics[scale=0.65]{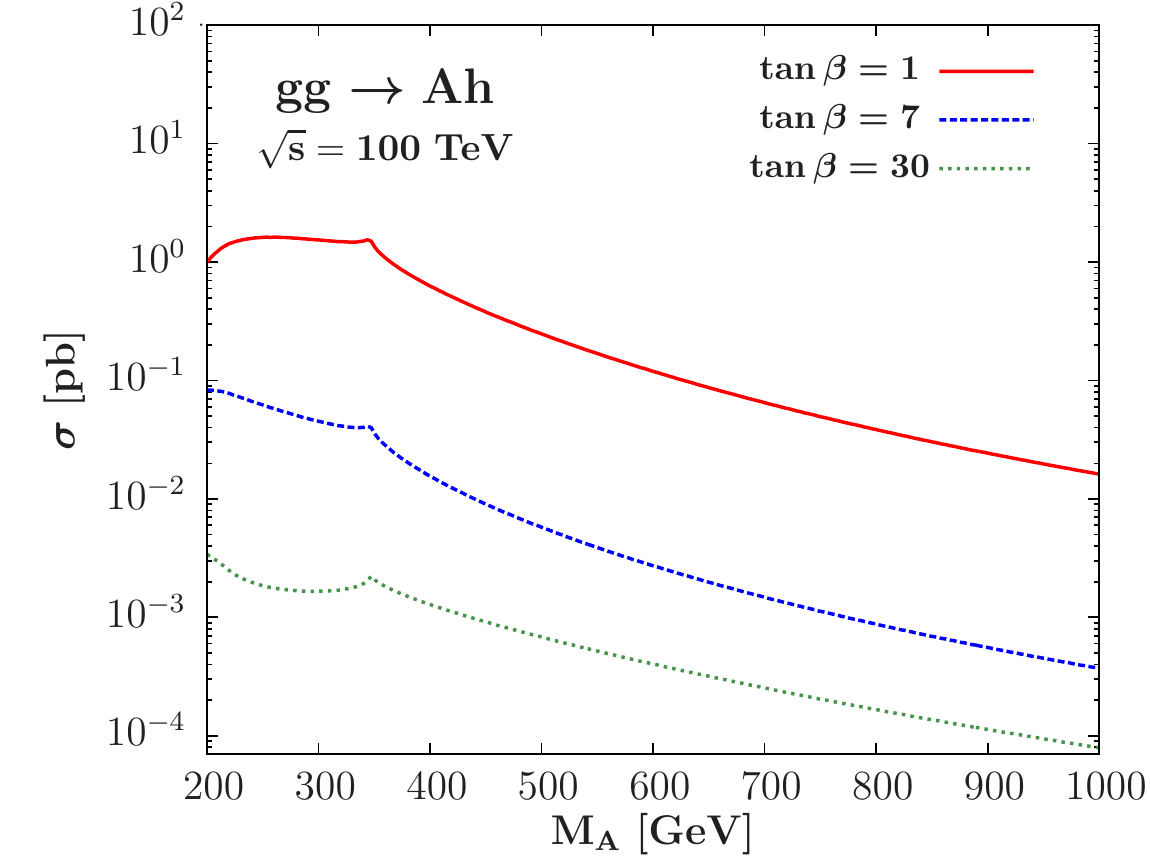}}\bigskip

\mbox{
\hspace{-1.0cm}
\includegraphics[scale=0.66]{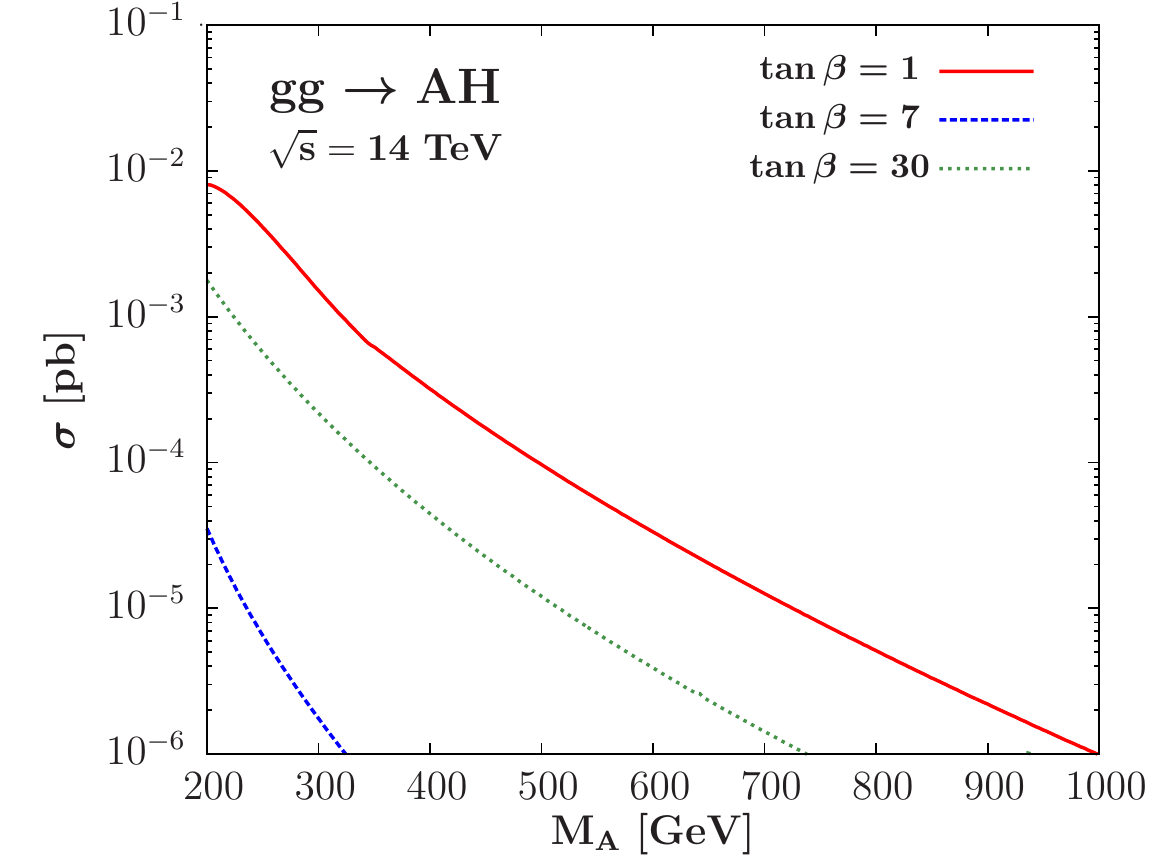}
\includegraphics[scale=0.66]{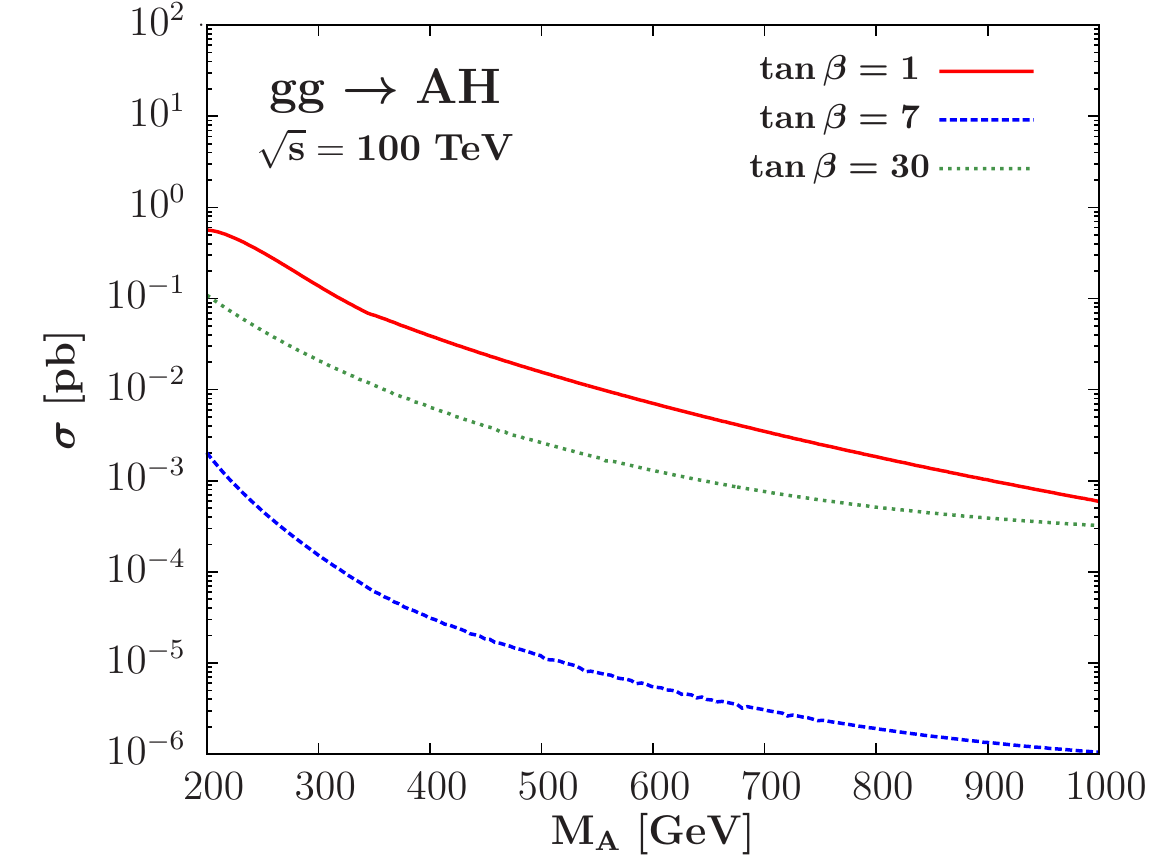}}
\vspace*{-3mm}
\caption[]{The cross sections for neutral MSSM Higgs pair production 
in $gg$ fusion, $gg \rightarrow hA$ (top) and $HA$ (bottom) as a function of $M_A$ for 
$\tb=1,7$ and 30. Shown are the rates for $\sqrt{s}=14$ TeV (left) and 100 TeV (right),
with the MSTW PDFs.}
\label{fig:prodAHgg}
\vspace*{-3mm}
\end{figure}

For the numerical analysis, we will however stick to the $hA$ and$HA$ processes
discussed above for $q\bar q$ annihilation, since the results for $hH$ and $HH$ 
production are similar to the former and latter cases respectively and the SM--like $hh$
case has been discussed in section 2. In both cases, there is no resonant process as
the particle that is exchanged in the $s$--channel is the $A$ boson. The rates 
are evaluated at LO  using the program {\tt HPAIR}~\cite{Michael-web}
(the NLO corrections are not known for the dominant $b$--loop
contributions at high $\tb$) with the scales fixed to the invariant
mass of the two final Higgses.

The cross sections $\sigma(gg \to hA)$
and $\sigma(gg \to HA)$ in the $h$MSSM are shown in
Fig.~\ref{fig:prodAHgg} again  as a function of $M_A$ for $\tb=1,7$
and 30. The rates for $HA$ are most significant either 
at low or at high $\tb$ when one of the $\Phi tt$ or $\Phi bb$ couplings  is 
sufficiently large, but one is limited by the phase--space (and by the fact that the 
process is of high order) and  for $M_\Phi=500$ GeV, the cross section is a the few ten  
fb level even at $\sqrt s=100$ TeV for $\tb \approx 1$ or 30. Again, in 2HDMs with  
$M_A \neq M_H$, the rates can be larger if $H$ is much lighter than $A$. For the $gg\to 
Ah$ process, the phase--space is more favourable but the cross sections
are  significant only for $\tb \approx 1$ since $h$ has SM--like
couplings and they are mainly generated by the box diagram with top
quark exchange (there is no enhancement of the $hbb$ coupling at high
$\tb$).

\subsubsection{The production of the charged Higgs bosons}

The  by far dominant process for charged Higgs production at hadron colliders 
is certainly the top and antitop quark decays $t \rightarrow H^+ b$ and $\bar{t} 
\rightarrow H^- \bar{b}$ for sufficiently light $H^\pm$ states,  $M_{H^\pm} \lsim 
m_t-m_b \sim 170$ GeV.  The branching fractions of the decay $t \to H^+b$ (which has
to compete only with the dominant $t \to bW$ decay) are large at low or 
high $\tb$ values when either the top or bottom quark component of the $H^\pm tb$ 
coupling, $g_{H^\pm tb} \propto m_t/\tb+ m_b \tb$, is significant, but even at intermediate 
values $\tb \approx 7$, it is above the percent level.  As the production cross sections 
for top quark pairs $q \bar q, gg \to  t\bar t$ (with gluon fusion largely dominating at 
LHC energies and beyond) are extremely large at hadron colliders, they allow to probe 
all $\tb$  values provided that the decay is not phase--space suppressed. Through the
dominant $H^\pm$ decay for $\tb \gsim 1$, $H\to \tau \nu$, searches have been 
conducted at the previous LHC run and masses $M_{H^\pm} \lsim 140$ GeV have been 
excluded for any $\tb$ value by the ATLAS and CMS collaborations. Masses up to the
kinematical limit of $M_{H^\pm} \sim 170$ GeV could be probed at the LHC with $\sqrt 
s=14$ TeV and high--luminosity as will be discussed in the  next subsection, 
leaving little space for a 100 TeV collider.      

If the charged Higgs boson is heavier than the top quark, one has to resort to direct
production processes,  the most relevant one at high energies being associated production with  
top and bottom quarks in $q\bar q$ annihilation and mainly $gg$ fusion, $ pp \rightarrow gg, 
q\bar q \rightarrow tH^- \bar{b} + \bar{t}H^+ b$. Here again, the production cross section 
are most significant at low or high $\tb$ when the $H^+ tb$ coupling is large, and are minimal 
at intermediate values $\tb \approx \sqrt{m_t/ \bar m_b}$ when the top component of the 
$H^\pm tb$ coupling is suppressed and the bottom component  not sufficiently enhanced. 
The NLO QCD corrections to this process are
significant~\cite{Flechl:2014wfa,Dittmaier:2009np,Berger:2003sm} and
exhibit logarithms that involve the ratio of the $b$--quark mass and
the factorisation scale, $\log (m_b/\mu_F)$. 
As in the case of associated $H/A$ production 
with $b\bar b$ pairs, these large logarithms can be resumed  by treating the bottom quark 
as a parton and considering the process $gb \to H^+ t$ in a five-flavour scheme. The  
NLO QCD corrections in this case are also known and part of the corrections corresponds 
in fact to the original process in the four-flavour scheme, $gg \to
tbH^\pm$. The optimal choices for the renormalisation and
factorisation scales have been shown to be $\mu_F= \frac12 \mu_R =
\frac12 (M_{H^{\pm}}+m_{t})$. 

\begin{figure}[!ht]
\vspace*{1mm}
\centering

\mbox{
\hspace{-1.0cm}
\includegraphics[scale=0.65]{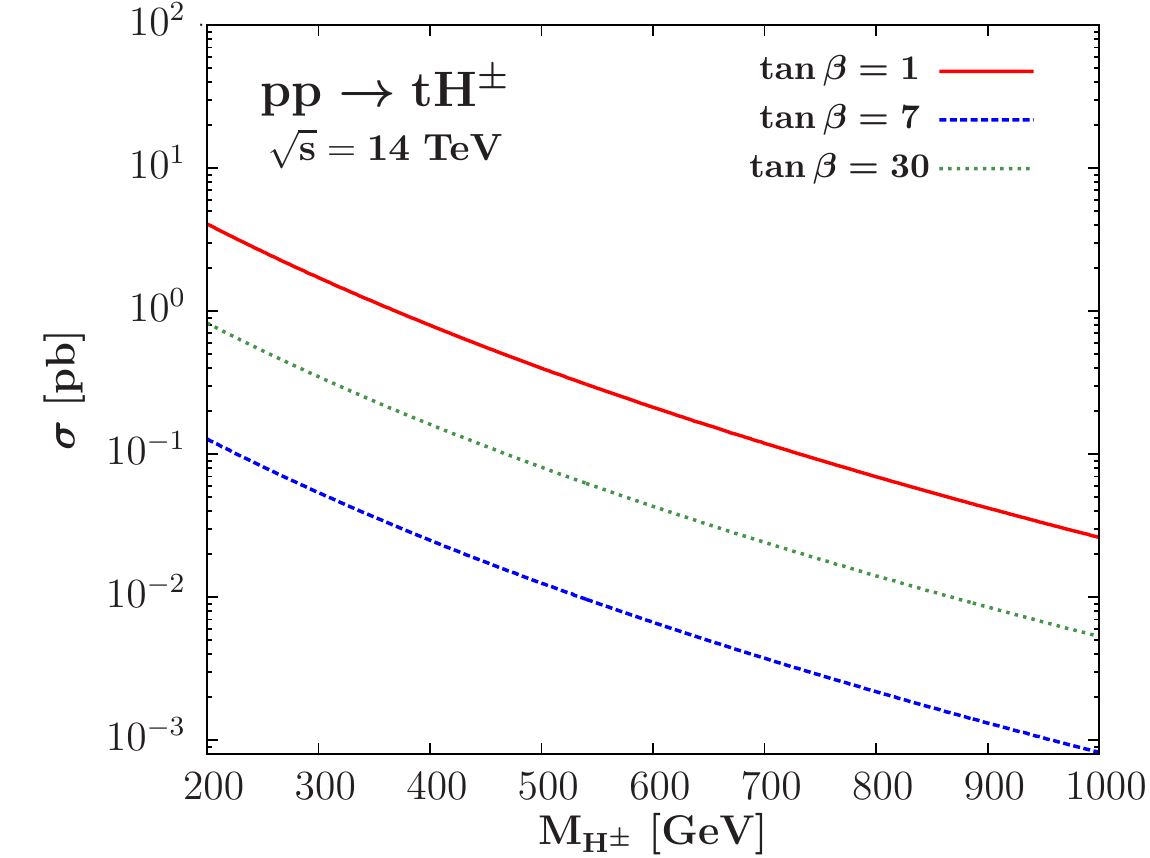}
\includegraphics[scale=0.65]{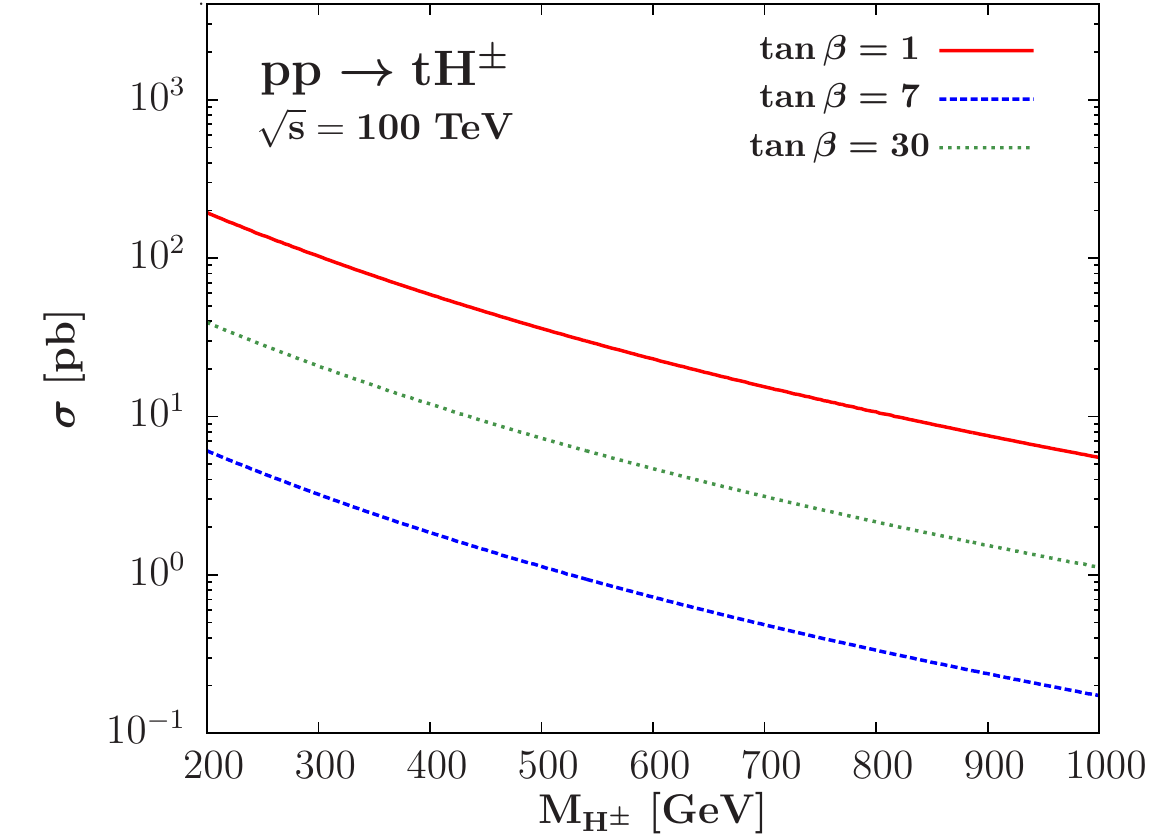}}

\vspace*{5mm}

\mbox{
\hspace{-1.0cm}
\includegraphics[scale=0.65]{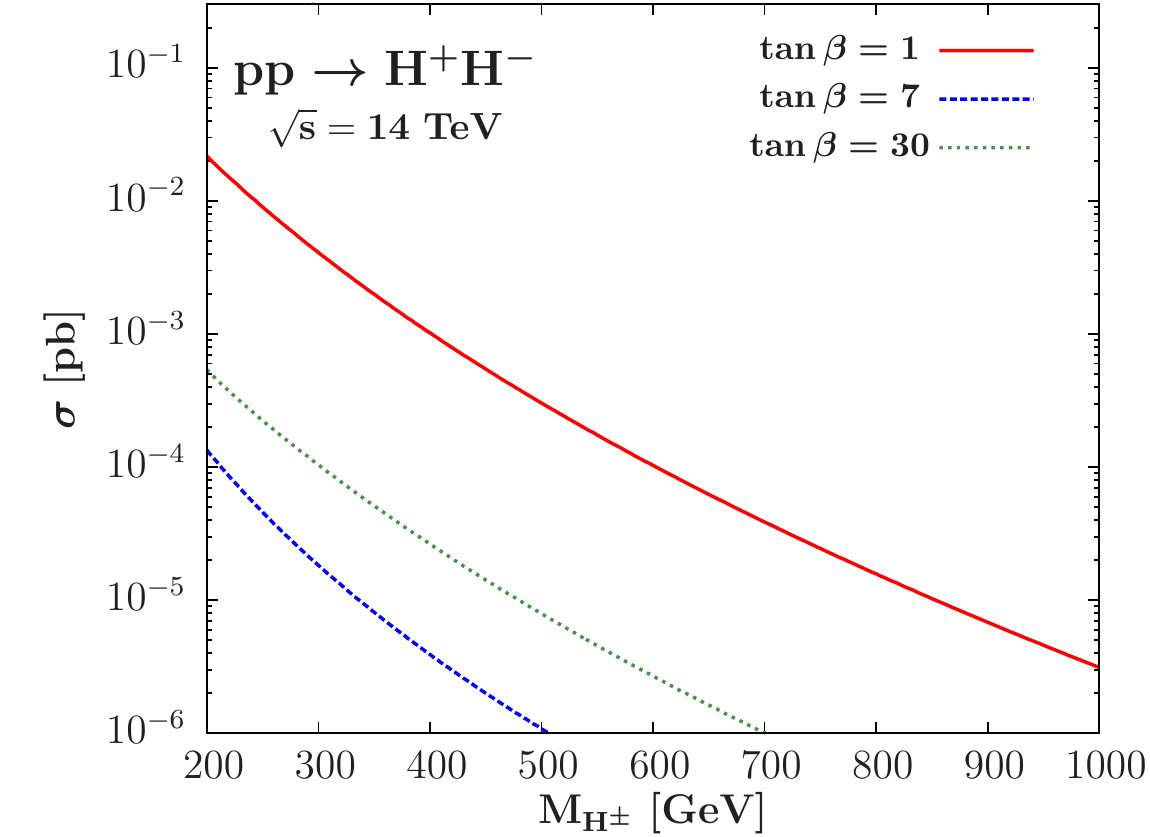}
\includegraphics[scale=0.65]{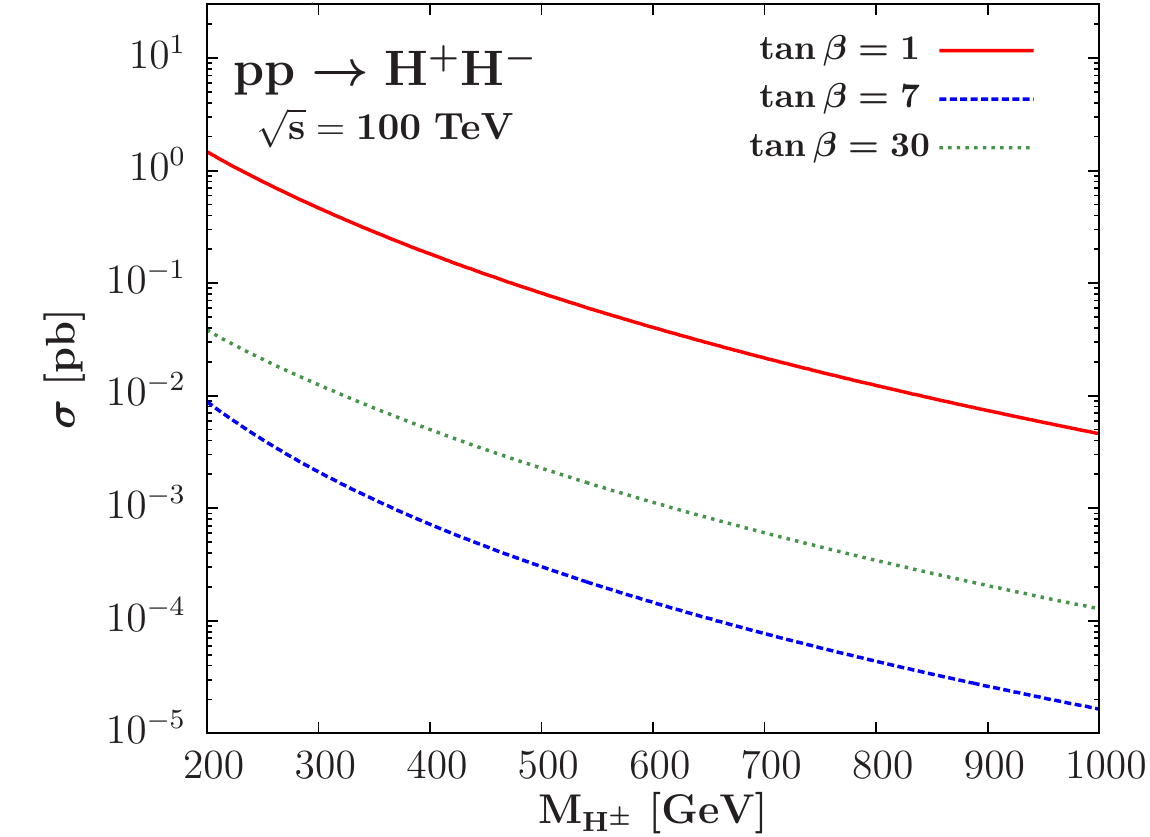}}

\vspace*{5mm}


\mbox{
\hspace{-1.0cm}
\includegraphics[scale=0.65]{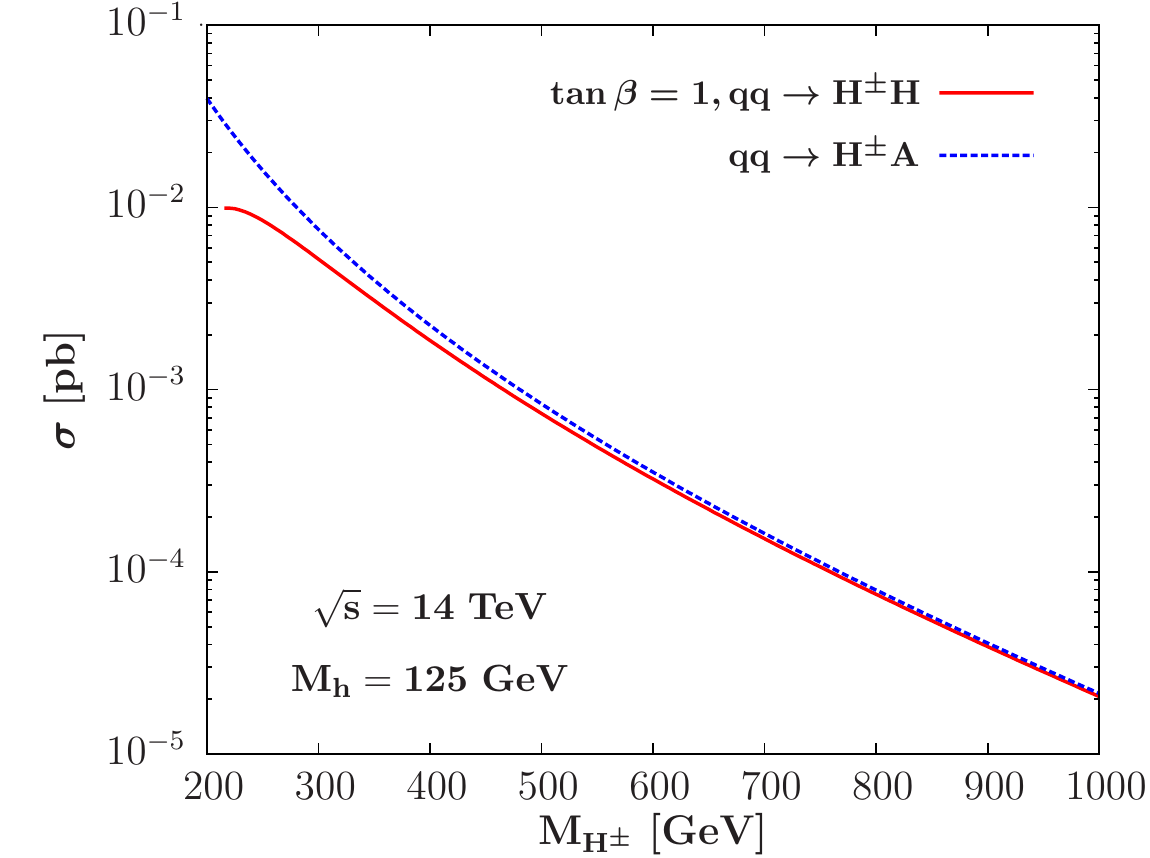}
\includegraphics[scale=0.65]{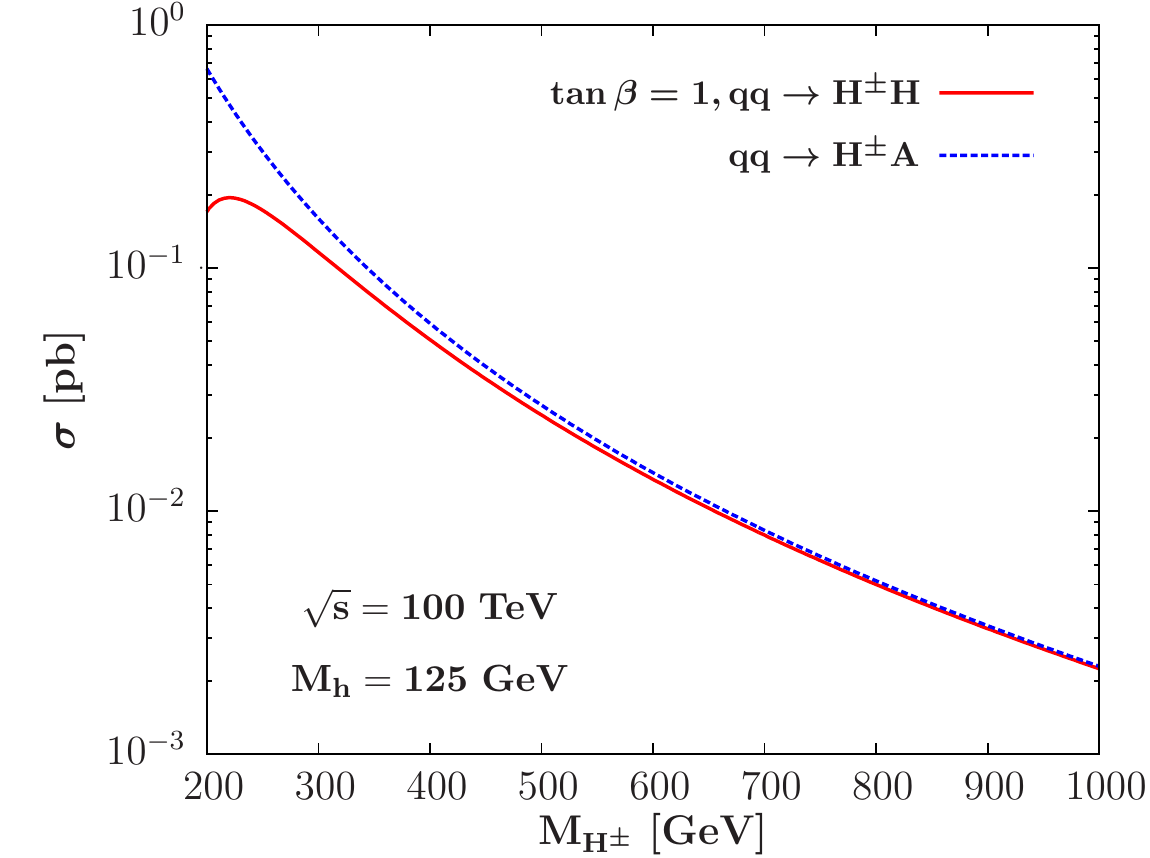}}

\vspace*{-2mm}
\caption[]{The production rates of the charged Higgs bosons as  functions of $M_{H^{\pm}}$
for $\sqrt{s}=14$ TeV (left) and $\sqrt s=100$ TeV (right) for $\tb=1,7$ and 30. Three
processes are considered:  associated production with a top quark $gb \to H^- t$ (upper 
plots), pair production in  $q\bar q$ annihilation and loop induced $gg$ fusion  
$pp\to q\bar q + gg \to H^+H^-$  (middle plots) and associated production with $h,H,A$ states 
$q\bar q \to H^\pm +h,H,A$ for $\tb=1$ only (lower plots).}
\label{fig:prodH+}
\vspace*{-5mm}
\end{figure}

The numerical results for the cross sections in this process are displayed in  the 
upper part of Fig.~\ref{fig:prodH+} as a function of $M_{H^\pm}$ again at $\sqrt s=
14$ TeV (left) and 100 TeV (right) for the usual three values $\tb=1,7$ and $30$.  
They have been evaluated using the program {\tt Prospino} of
Ref.~\cite{Beenakker:1996ed,Plehn:2002vy} and include the NLO QCD
corrections in the five-flavour 
scheme with the scales set a the values given above. Here we used the
CTEQ PDF set~\cite{Nadolsky:2008zw} instead.
For the low and high $\tb$ values, as they scale as $m_t^2 \cot^2\beta$ or $\bar 
m_b^2\tan^2\beta$ respectively, the cross sections exceed the 10 pb level only for 
$H^\pm$ masses up to $M_{H^\pm} \sim 200$ GeV  at $\sqrt s =14$ TeV but up to 
$M_{H^\pm} \sim 700$ GeV for $\sqrt{s}=100$ TeV and they drop
quickly with increasing charged Higgs mass.

Note that since the $H^\pm$ properties depend only on $\tb$ and $M_{H^\pm}$, 
the discussion above holds in both the MSSM and type II 2HDMs. However, in the 
MSSM there are SUSY--QCD corrections that are proportional to $m_b\tb$ and can be large for not
too heavy SUSY particles: the notorious $\Delta_b$
corrections~\cite{Carena:1999py,Noth:2008tw} that appear also in
$gg\to b\bar b+ H/A$ which were discussed previously. We also ignore
these corrections here as  in the  $h$MSSM approach, we assume a high
SUSY scale. In addition, in 2HDMs of type I, the cross sections simply
scale as $(m_t^2 +m_b^2) \cot^2\beta$ and hence are significant only
at small $\tb$ values, with QCD corrections that do not involve large
logarithms and are hence moderate.

Charged Higgs bosons can also be produced in pairs. At LO, the process proceeds via $q\bar q$
annihilation with the exchange of a virtual photon and $Z$ boson and the cross section 
depends only on $M_{H^\pm}$ and no other MSSM parameter. The QCD corrections are again 
those of Drell-Yan~\cite{Altarelli:1979ub} which at NLO lead to an increase of the rate by $\approx 30\%$
at scales $\mu_F= \mu_R =2 M_{H^{\pm}}$. 
Another important pair production process at high energies is the $gg$ fusion mechanism, 
$gg\to H^+ H^-$, which proceeds through loops involving
top and bottom quarks, including contribution from the channels $gg\to h,H$ with  
$h, H \to H^+H^-$. In the MSSM both $s$--channel  particles are off shell 
but in more general 2HDMs, one could have 
$M_{H}\gsim 2M_{H^\pm}$ so that the $H \to H^+H^-$ decay is resonant, a situation similar
to $gg\to AA$. The process is known at LO and we take 
the scales to be $\mu_F= \frac12 \mu_R =\frac12 M_{H^{\pm}}$. 

The combined cross sections at NLO for $q\bar q$ and LO for $gg$, evaluated with the 
program Prospino of Ref.~\cite{Beenakker:1996ed,Plehn:2002vy}, are shown at 
$\sqrt{s}=14$ TeV and 100 TeV  
in the middle frames of Fig.~\ref{fig:prodH+} as a function of $M_{H^\pm}$ for 
the three values $\tb=1,7$ and 30. Here again, we use he CTEQ PDF set~\cite{Nadolsky:2008zw} and  take
the renormalisation and factorisation scales equal to the values mentioned above. 
The rates are dominated by $gg$ fusion and are about two order of magnitude 
higher at 100 TeV than at 14 TeV. For either very low or very high $\tb$, the rates 
can be at the level of 10 fb for $M_{H^\pm}=1$ TeV at the highest energy. 

Finally, associated $H^\pm$ production with a neutral Higgs boson, $q\bar q' \rightarrow  
\Phi H^\pm$ with $\Phi=h,H$ and $A$, is mediated by virtual $W$ exchange and the cross 
section is again simply the one in SM Higgs--strahlung  $q\bar q \rightarrow H_{\rm SM} 
W$, with the proper change of the coupling and phase--space factors. For associated 
production with the CP--even $h,H$ states $q\bar q'\rightarrow hH^\pm, H H^\pm$, the 
cross sections follow exactly the same trend as the ones for $hA,HA$ production 
except for the overall normalisation factor: once the two $H^\pm$ charges are summed, 
the rates are larger than for $A$ by about a factor of two for the same 
mass  $M_{H^\pm} \sim M_A$. The rates for $AH^\pm$ production follow those of $HH^\pm$ 
in the decoupling limit of the MSSM or the alignment limit of 2HDMs. The rates
are shown at NLO in the lower part of Fig.~\ref{fig:prodH+} at $\sqrt s=14$ and 100 TeV for 
$\tb = 1$ only. For $\tb=7$ and 30, they are almost the same for  $AH^\pm$ and 
$HH^\pm$ production while they becomes negligible for $hH^\pm$ production. 
  
\subsection{The sensitivity on the extended Higgs sectors}

In this subsection, we will attempt to quantify the increase in sensitivity on 
the heavier Higgs states that one can obtain by moving from a c.m. energy
of  14 TeV to 100 TeV. We will assume that that 3000 fb$^{-1}$  of data will 
be collected both at FCC-hh and the high luminosity (HL)  
LHC option. We will illustrate this gain in sensitivity in the $h$MSSM, 
following exactly the results obtained in Ref.~\cite{Djouadi:2015jea}
where the HL--LHC  case was discussed in detail (see also
Ref.~\cite{Hajer:2015gka} for an another recent analysis at 14 and 100
TeV). We will also discuss the 
case of 2HDMs of type II in the alignment limit and in the simple case 
where the masses of the three heavy Higgs states $H,H,H^\pm$ are comparable. 
We will concentrate on the direct searches of the heavy Higgs particles since 
in both the alignment and decoupling limits of the two models, the lighter
Higgs particle will behave as the SM Higgs boson discussed in section 2
and we ignore the (complementary) indirect constraints that can be obtained from 
measuring its couplings to fermions and gauge bosons in the various production
and decay channels.

To analyse the discovery reach of the heavy states at the LHC and beyond, 
it is necessary to know their various decay modes. In the case of the MSSM, 
the decays modes have been discussed in great detail and we refer the reader 
to the recent account~\cite{Djouadi:2015jea} that was made in the context of the $h$MSSM 
and which we will closely follow here. In the case of 2HDMs, the situation
will more difficult to describe than in the $h$MSSM since the masses of 
the four Higgs states, as well as the two mixing angles $\alpha$ and $\beta$, 
are free parameters making any attempt to perform a full analysis a daunting 
task. Even when including the results on the observed $h$ state, i.e. by fixing 
its mass to $M_h\!=\! 125$ GeV and forcing its couplings to be SM--like by adopting 
the alignment limit that leads to $\alpha=\beta-\frac12 \pi$, one still has four 
input parameters to deal with, instead of two in the $h$MSSM for instance. 

In the present analysis, we will further simplify our discussion in type II 2HDMs
by assuming that similarly to the $h$MSSM case, the heavier $H,A,$ and
$H^\pm$ states  have a comparable mass, $M_{H^\pm} \approx M_H \approx
M_A$  in such a
way that decays of one Higgs particle to another one and a gauge boson is kinematically 
not allowed. This makes that complicated decays such as $H\to AZ, H^\pm  W^\mp$,
$A\to HZ, H^\pm  W^\mp$ and $H^\pm \to  AW^\mp, HW^\mp$ are kinematically not
allowed at the two--body level. Our justification is that in fact, even if
these modes are allowed, they cannot compete with the fermionic decays  of the 
$H^\pm, H,A$ states involving top or bottom quarks (and tau--leptons) at either low 
or high $\tb$. Adding the fact that, in the alignment limit, decays such
as $H\to WW,ZZ$ or $A \to hZ$ which involve the reduced couplings $g_{HVV} = 
g_{AhZ}=\cos(\beta-\alpha) =0$ as well as the decay $H\to hh$, are absent or have 
small branching ratios, one will have a simple decay pattern. 

Indeed, in this alignment limit with $M_{H^\pm} \approx M_H \approx M_A$, the only
important decays of the neutral $H,A$ states will be into $t\bar t, b\bar b$ and 
$\tau^+\tau^-$ final states, while those of the charged Higgs boson will be into 
$tb$ and $\tau \nu$. In fact, at low $\tb$ only the decays 
$A, H \to t\bar t$ for masses $M_H \approx M_A \gsim 2m_t$ and $H ^+ \to t b$ for 
$M_{H^\pm} \gsim 180$ GeV are relevant with branching ratios of order unity, 
while at high $\tb$, one would have ${\rm BR}(A/H \to b\bar b) \approx 
{\rm BR}(H^+ \to t  \bar b) \approx 90\%$ and ${\rm BR}(A/H \to \tau^+\tau^-) 
\approx {\rm BR}(H^+ \to \tau\nu) \approx 10\%$, a simple reflection of 
$3 \bar m_b^2/m_\tau^2 \approx 10$ (with 3 being the colour factor). At intermediate
values of $\tb$ and above the top threshold the two sets involving
top and bottom+tau decays would have comparable branching rates.

In the $h$MSSM outside the decoupling regime, i.e for low $\tb$ values with
$M_A \lsim 350$ GeV, the decay pattern can be rather involved  as discussed
in Ref.~\cite{Djouadi:2015jea} for instance and a summary is as follows: $i)$ above
the $2M_V$ threshold, the $H\! \to \! WW$ and $ZZ$ decay rates are still significant 
as $g_{HVV}$ is not completely suppressed;  $ii)$ for $2M_h \lsim M_H \lsim 2m_t$, 
$H\to hh$ is the dominant $H$ decay mode at low $\tb$ as the $Hhh$ 
self--coupling is large in this case; $iii)$ for $M_h+M_Z \lsim M_A\! \lsim 2m_t$, 
the decays $A \to hZ$ decays would occur with large rates at low $\tb$; $iv)$ this is also
the case for the channel  $H^+\! \to \! Wh$ which is important for $M_{H^\pm}\! \lsim 
\! 250$ GeV, but at intermediate $\tb$ values this time. 

In the context of the $h$MSSM, the impact of the various searches that have been 
performed by the ATLAS and CMS collaborations at $\sqrt s =7$+8 TeV with up to $\approx 
25$ fb$^{-1}$ data have been used to constrain the $[M_A, \tb]$ parameter space of the 
model~\cite{Djouadi:2015jea}. 
The searches that have been considered are essentially the fermionic Higgs decays 
$H/A \rightarrow \tau\tau$ and $H^\pm \rightarrow \tau\nu$ and the bosonic ones, 
$H\rightarrow WW, ZZ,hh$ and $A\rightarrow Zh$. The channels $H^+ \to tb$ as well as the 
$H,A \to t\bar t$ in some approximation have also been included. As experimental input, 
the following Higgs searches and measurements, published by the ATLAS and CMS 
collaborations, have been used.

\begin{center}
\vspace*{-3mm}
\begin{tabular}{l|ll}
search  channel & \ \ \ ATLAS & \ CMS \\ \hline
$A/H \to \tau\bar{\tau}$ & Ref.~\cite{Aad:2014vgg}          & Ref.~\cite{CMS:2013hja} \\
$A \to Zh$        & - & Ref.~\cite{CMS:2014yra} \\
$H \to hh$             & Ref.~\cite{Aad:2014yja}           & Ref.~\cite{CMS:2014ipa} \\
$H \to WW$             & -           & Ref.~\cite{Chatrchyan:2013iaa} \\
$H \to ZZ$             & -           & Ref.~\cite{CMS:2013pea,Chatrchyan:2013mxa} \\
$H^{+} \to \tau\nu/tb$             & Ref.~\cite{ATLAS:2013wia,ATLAS-CONF-2014-050}~~~           & Ref.~\cite{CMS:2014cdp,CMS:2014pea} \\
\hline
\end{tabular}
\end{center}

Concerning the $H/A \rightarrow t\bar t$ analysis at low $\tb$ with the $A/H$
states dominantly  produced in the $gg$ fusion mechanism, there was no
dedicated analysis from the ATLAS and CMS collaborations. As a first approximation,
we thus estimated the sensitivity in this channel by considering searches that have been 
performed for high mass spin--1 electroweak resonances (in particular new $Z'$ and 
Kaluza--Klein electroweak gauge bosons) that decay into top quarks and
adapting them to our case. One should note however that this
adaptation is quite naive as for spin--1 particles, the main
production channel is $q\bar q$ annihilation and there is no
interference with the (coloured) QCD $q\bar q \to t\bar t$ background
and the resonances show up as peaks in the $t\bar t$ invariant mass
distribution. In the Higgs case, the main process is $gg\to H/A$ and
would thus interfere with the $gg\to t\bar t $
background~\cite{Dicus:1994bm,Barger:2006hm,Frederix:2007gi,Bernreuther:2015fts}. This
interference depends on the mass and total width of the Higgs and on
their CP--nature, making it either constructive or destructive.  This
leads to a rather complex signature with a peak--dip structure of the
$t\bar t$ mass distribution that is not experimentally addressed yet.

The constrains obtained in Ref.~\cite{Djouadi:2015jea} from the ATLAS and CMS searches with
the 25 fb$^{-1}$ data collected at 7+8 TeV in the channels above 
 are quite impressive and a large part of the parameter space (in particular 
at high $\tb$ for $M_A \lsim 300$ GeV) has been already excluded. If no new signal is 
again observed, they can be still vastly improved at the next LHC phase with a 
c.m. energy of $\sqrt s=14$ TeV and with one or two orders of 
magnitude accumulated data. The projections for this 
case, and the procedure to obtain them have been discussed in 
Ref.~\cite{Djouadi:2015jea} to which we refer for the details. Here, we simply summarise
how the projections are obtained at a given c.m. energy and luminosity.

For a specific search channel, on starts with the expected median 95\%CL exclusion limits 
that have been given by the ATLAS and CMS collaborations in the searches performed 
at 7+8 TeV with $\approx 5$+20 fb$^{-1}$ data. One then assumes that the 
sensitivity will approximately scale with the square root of the number of
expected events and does not include any additional systematical effect.
In addition, having only the information on the signal cross sections for
a given Higgs mass, and  not the corresponding background rates for the same mass bin, 
one needs to make the additional naive assumption that the background also 
simply scale as the signal cross sections (which is true for many channels).
  
\begin{figure}[!h]
\vspace*{-.3cm}
\centering
\includegraphics[scale=0.8]{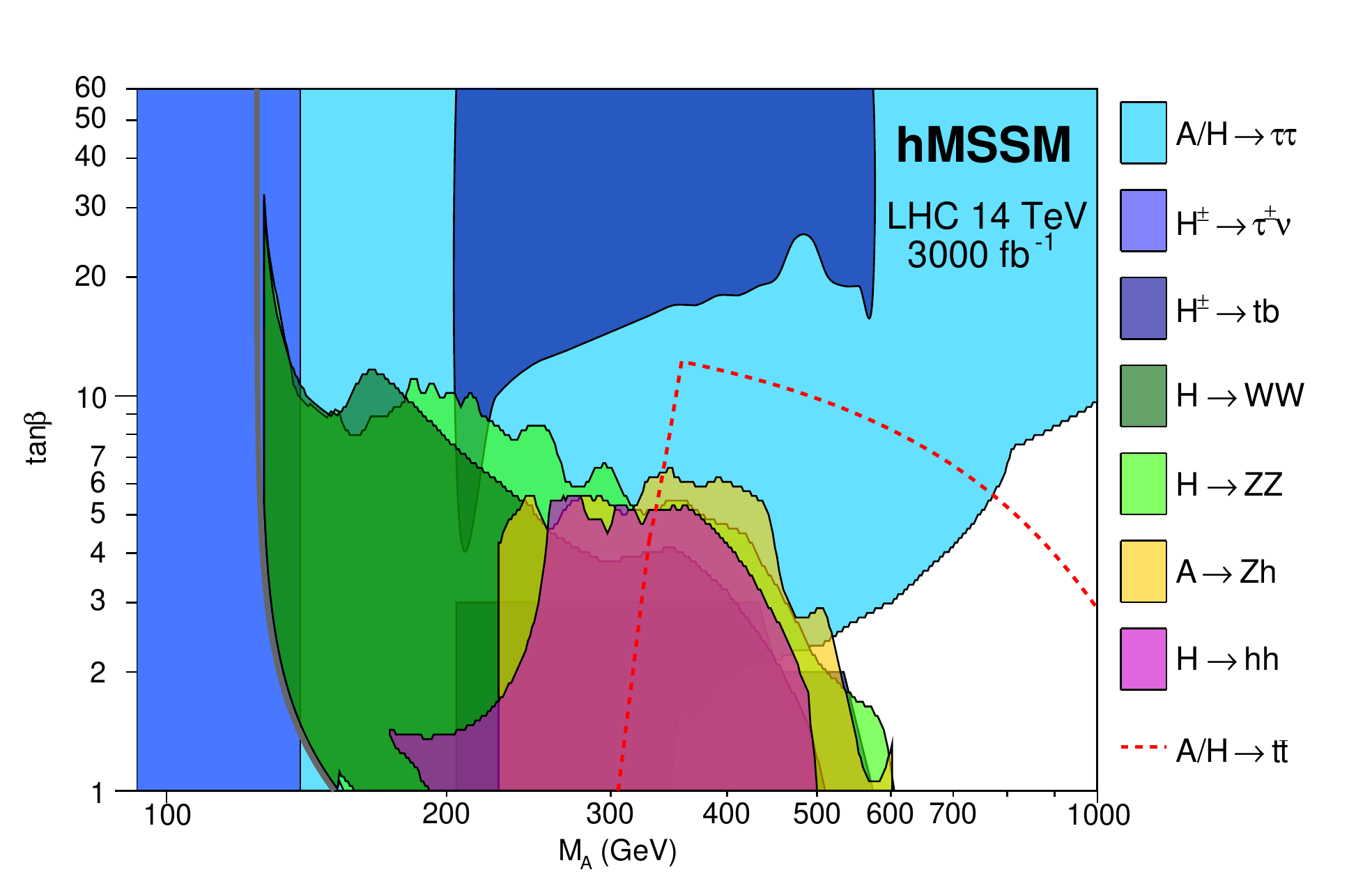}\\[-.2mm]
\includegraphics[scale=0.8]{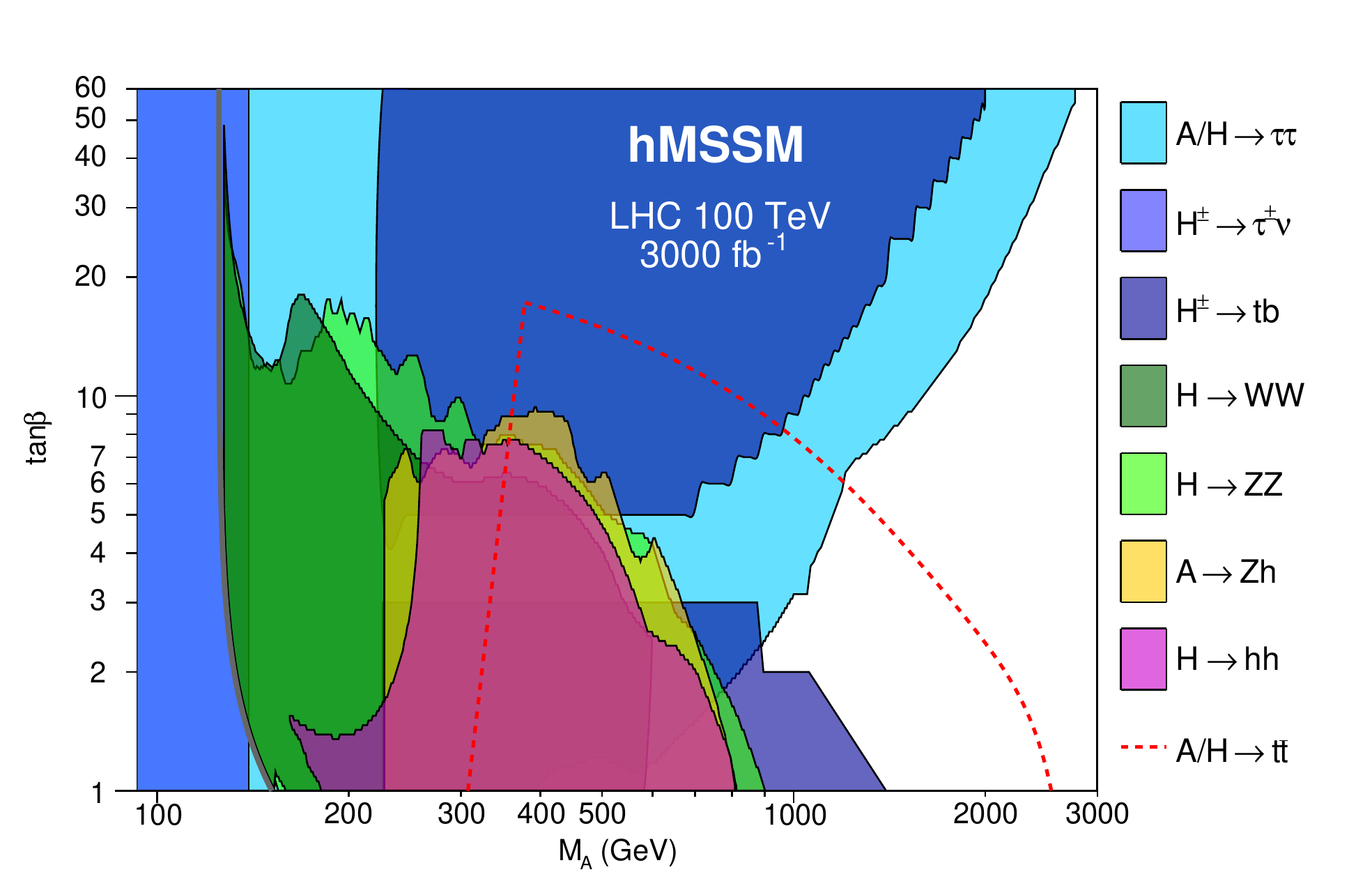}
\vspace*{-3mm}
\caption[]{Projections for the HL--LHC with $\sqrt s=14$ TeV (upper plot) and at $\sqrt{s}=
100$~TeV (lower plot) with $3000$ fb$^{-1}$ data for the $2\sigma$ sensitivity in the 
$h$MSSM $[\tb, M_A]$ plane when ATLAS and CMS searches for the $A/H/H^\pm$ states in 
their fermionic and bosonic decays are combined.}
\label{constraints_hMSSM}
\vspace*{-2cm}
\end{figure}

The output of the projections following this procedure is presented in the $[\tb, M_A]$ 
$h$MSSM plane in the lower part of Fig.~\ref{constraints_hMSSM} in the fermionic and 
bosonic Higgs search channels, including our naive treatment of the $gg\to H/A \to t\bar t$ 
mode. The c.m. energy of $\sqrt s=100$ TeV and a luminosity of 3000 fb$^{-1}$ have
been assumed as well as  the same signal and background efficiencies for the high energy
hadron collider than at the 8 TeV LHC. For the sake of comparison, we also show 
in the upper plot of the figure, the sensitivity at the HL--LHC with $\sqrt s=14$ TeV 
and 3000 fb$^{-1}$ data, that was obtained in Ref.~\cite{Djouadi:2015jea}. 

\begin{figure}[!h]
\vspace*{-3mm}
\centering
\includegraphics[scale=0.8]{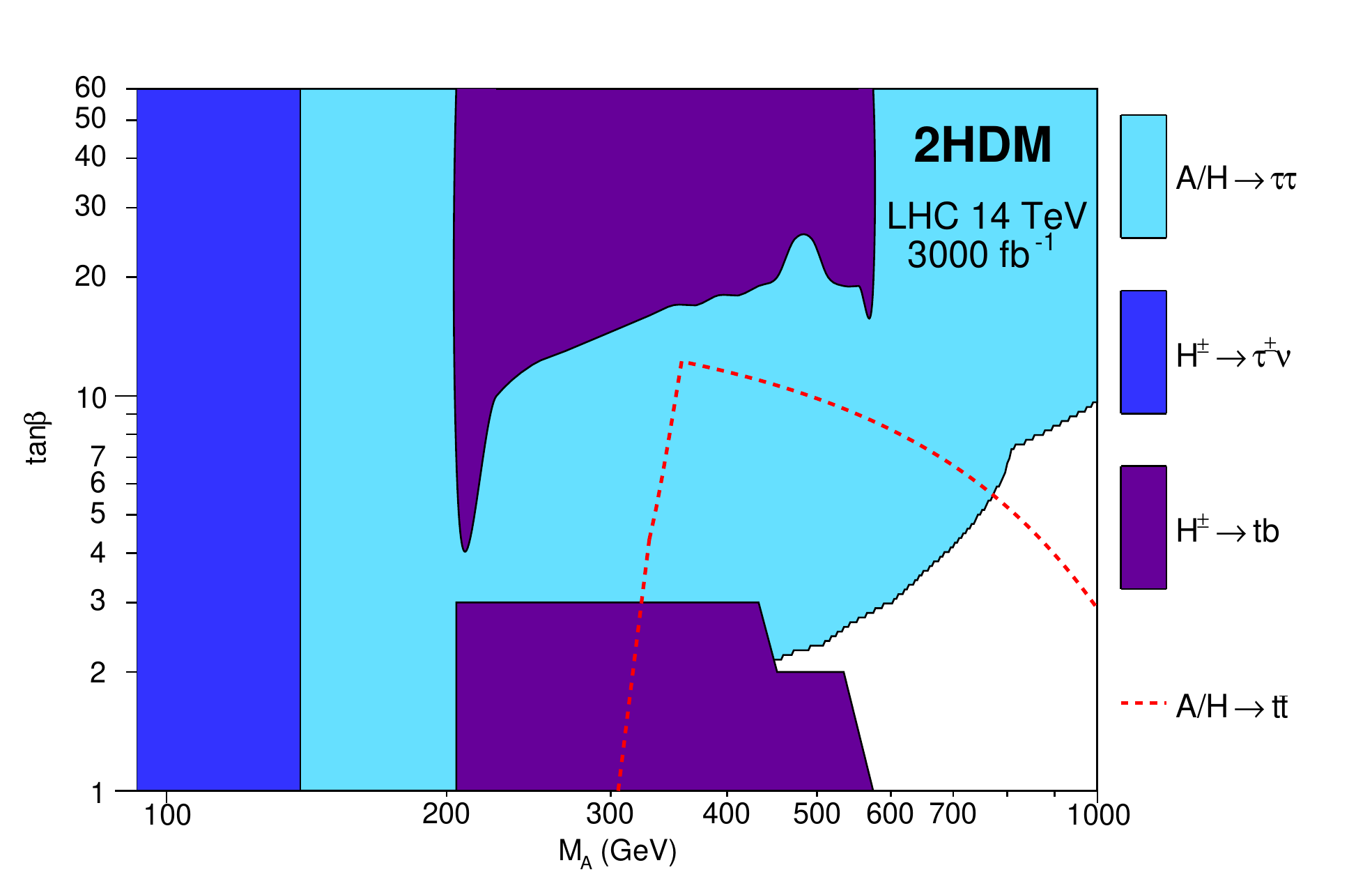}\\[-.2mm]
\includegraphics[scale=0.8]{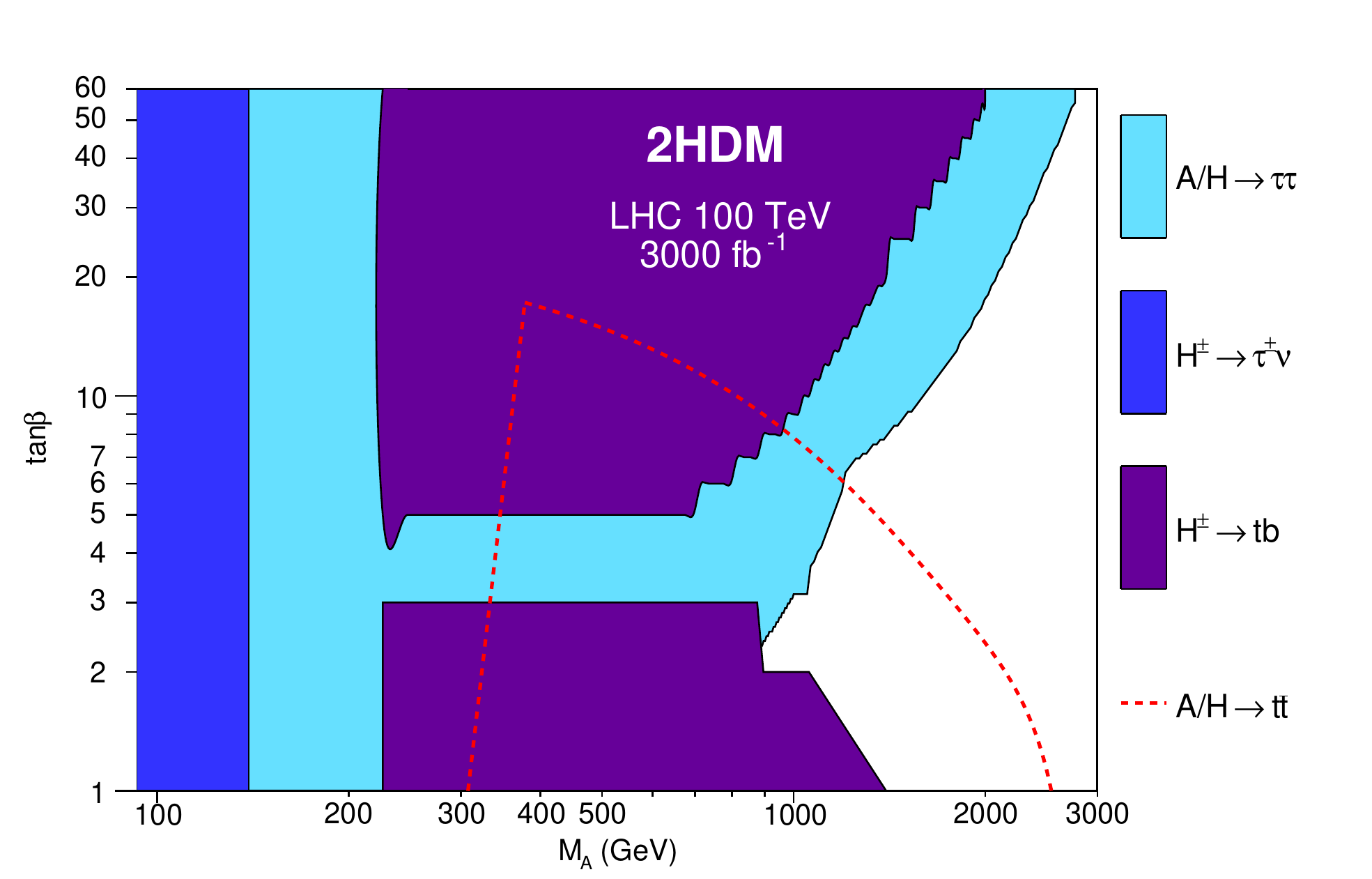}
\vspace*{-3mm}
\caption[]{Projections for the HL--LHC with $\sqrt s=14$ TeV (upper plot) and at $\sqrt{s}=
100$~TeV (lower plot) with $3000$ fb$^{-1}$ data for the $2\sigma$ sensitivity in the 
type II 2HDM  $[\tb, M_\Phi]$ plane (with $M_\Phi=M_A=M_H=M_{H^\pm})$ 
when the ATLAS and CMS searches for the $A/H/H^\pm$ states in their fermionic decays are 
combined in the alignment limit.}
\label{constraints_2HDM}
\vspace*{-2cm}
\end{figure}

In Fig.~\ref{constraints_2HDM}, shown is the same sensitivity in the  $[\tb, M_\Phi ]$ 
plane in our simplified type II 2HDM in which we
assume the alignment limit $\alpha = \beta-\frac12 \alpha$ and 
approximately mass degenerate heavy Higgs bosons $M_A=M_H=M_{H^\pm}=M_\Phi$. 
The same collider parameters as for the $h$MSSM case have been adopted and we also
show for the sake of comparison, the sensitivity in 2HDMs at the HL--LHC.

As can be seen, a vast improvement of the sensitivity to the MSSM and 2HDM parameter
spaces is expected at 100 TeV compared to the HL--LHC with the same luminosity. 
In the very low and very high $\tb$ regions, masses close to 3 TeV can be probed at 
a 100 TeV collider in, respectively the $H/A \to t\bar t$ and $H/A \to \tau^+ \tau^-$ 
modes, compared to only 1.5 TeV at the HL--LHC. The $H/A \to t\bar t$
and $\tau^+\tau$   channels intersect at $M_A=1.5$ TeV for FCC--hh
(instead of $M_A =750$ GeV for HL--LHC), a mass value 
below which the entire $h$MSSM parameter space is fully covered by the searches.   
In the case of the charged Higgs state, the FCC--hh mass reach is lower: $M_{H^\pm}=2$ TeV (1.2 TeV)
at $\tb=60 \, (1)$ but is significantly higher than at HL--LHC. In all cases, the sensitivity 
of a 100 TeV collider in probing the parameter space is twice as large as HL--LHC with the 
same luminosity.

Note that for the fermionic channels, the discussion is qualitatively the
same in our adopted 2HDM than for the $h$MSSM with the difference that the
sensitivity is slightly higher at intermediate $\tb$ values in the former case. 
Indeed, in the $h$MSSM, the branching fractions in these fermionic channel are slightly
suppressed by some bosonic modes such as $H\to hh,WW,ZZ$ and $A\to hZ$ 
that still survive.

 In turn, for these bosonic modes, there is some sensitivity 
only in the $h$MSSM. The modes $H\to hh$ and $A\to hZ$, besides the decays $H\to WW$ 
and $ZZ$, could be observed up to $M_A=1$ TeV (0.5 TeV) for $\tb \approx 7$, a significant
improvement over the LHC at least in the low $\tb$ range.

\clearpage

\clearpage

\section{Conclusions}

We have analysed the prospects of future high--energy proton--proton colliders to
probe the Higgs sectors of the Standard Model and of some of its new physics 
extensions. In the SM context, we have studied the production of the observed Higgs 
particle in the dominant channels and have shown that the one to two orders of 
magnitude increase of the cross sections (depending on the considered channels) at 
energies close to $\sqrt s=100$ TeV would allow for a much more accurate determination 
of the Higgs couplings to gauge bosons and fermions. Some  observables, like the ratio 
of partial widths for Higgs decays into two photons and into four leptons, could be 
then measured at the per--mille level with a few ab$^{-1}$ data. We then analysed the 
various processes for Higgs pair production that allow for the measurement of the 
triple Higgs coupling, probably the only parameter that would remain undetermined after 
the high luminosity option of the LHC. Again, the rates at 100 TeV are
so large that a relatively precise measurement would be possible. The
last SM parameter to be probed would be then the quartic Higgs
coupling which can be accessed only in tripe Higgs production. At
$\sqrt s=100$ TeV,  the rates for the dominant gluon--fusion process
are not completely negligible and a luminosity of a few ten ab$^{-1}$
could  make the formidable challenge of observing three Higgs
particles not entirely hopeless.

In a second step, we have considered the production of the invisible particles that form 
the cosmological dark matter. We have worked in a model--independent effective framework 
in which the DM particle is either a spin--zero, a spin--one or a spin--half Majorana 
fermion that interacts only through the Higgs portal. If the DM particles are heavier 
than $\frac12 M_h$, the only way to observe them would be through Higgs exchange in
the continuum. We have thus evaluated the cross sections in three processes
with missing energy: gluon fusion with an extra jet, vector boson fusion and 
Higgs--strahlung, and shown that at $\sqrt s= 100$ TeV one could probe DM particles with 
a few 100 GeV mass for favourable couplings. 
 
Finally, we have evaluated the potential of a  100 TeV proton collider in probing the
heavy Higgs bosons that are present in extended Higgs scenarios, taking the example of 
two Higgs doublet models and their minimal supersymmetric SM incarnation. We have discussed
in a comprehensive manner the production of the heavier CP--even, the CP--odd and the 
charged Higgs states in all possible channels: single production,
associated production with massive fermions or gauge bosons and pair
production. Taking the examples of a 2HDM
in the alignment limit and the so--called $h$MSSM, we have shown that a collider
with $\sqrt s=100$ TeV and 3 ab$^{-1}$ data could cover the entire parameter space
of the models for Higgs masses up to 1 TeV and that some channels could be observed 
for masses of the additional Higgs bosons up to 3 TeV if their couplings to fermions  
are significant.  

The sensitivity of such a high energy collider in probing the electroweak symmetry 
breaking mechanism is thus far superior than that of the LHC with high luminosity.\bigskip  

\noindent {\bf Acknowledgments:} 

\noindent We thank Marco Zaro for his help in using aMC@NLO and Eleni Vryonidou
for having provided the numbers for $t\bar{t} HH$ production.
AD thanks the CERN Theory Unit for hospitality and the European Research Council for its 
support through the ERC Advanced Grant Higgs@LHC.  JB is supported in
part by the Institutional Strategy of the University of T\"ubingen
(DFG, ZUK~63) and by the DFG Grant JA 1954/1. The work of JQ is
supported by the STFC Grant ST/L000326/1.

\bibliography{H100}

\end{document}